\documentclass[useAMS,usenatbib]{mn2e}
\usepackage{graphicx}
\usepackage{txfonts}
\usepackage[hyphens]{url}
\usepackage{hyperref}
\pdfoutput=1

\usepackage[usenames,dvipsnames]{color}

\newcommand{\Sref}[1]{Section~\ref{#1}}
\newcommand{\Tref}[1]{Table~\ref{#1}}
\newcommand{\Aref}[1]{Appendix~\ref{#1}}
\sfcode`\.=1001\sfcode`\?=1001\sfcode`\!=1001
\newcommand{\Fref}[1]{\ifhmode \ifnum\spacefactor=1001 Figure~\ref{#1}\else Fig.~\ref{#1}\fi \else Figure~\ref{#1}\fi}
\newcommand{\Eref}[1]{\ifhmode \ifnum\spacefactor=1001 Equation~(\ref{#1})\else equation~(\ref{#1})\fi \else Equation~(\ref{#1})\fi}

\newcommand{\cms}{\ensuremath{\textrm{cm\,s}^{-1}}}
\newcommand{\ms}{\ensuremath{\textrm{m\,s}^{-1}}}
\newcommand{\kms}{\ensuremath{\textrm{km\,s}^{-1}}}
\newcommand{\SN}{\ensuremath{\textrm{S/N}}}

\newcommand{\lya}{\ensuremath{\textrm{Ly}\alpha}}

\newcommand{\kNMS}{\ensuremath{k_\textsc{\scriptsize NMS}}}
\newcommand{\kSMS}{\ensuremath{k_\textsc{\scriptsize SMS}}}
\newcommand{\FS}{\ensuremath{F_\textsc{\scriptsize FS}}}
\newcommand{\cm}{\ensuremath{\textrm{cm}^{-1}}}
\newcommand{\GHzamu}{\ensuremath{\textrm{GHz}.\textrm{amu}}}
\newcommand{\MHzfm}{\ensuremath{\textrm{MHz}.\textrm{fm}^{-2}}}

\newcommand{\ion}[2]{\ensuremath{\textrm{#1\,{\scshape{#2}}}}}
\newcommand{\daa}{\ensuremath{\Delta\alpha/\alpha}}

\usepackage{dcolumn}
\newcolumntype{b}{D{(}{\,(}{-1}}  
\usepackage{array}
\newcolumntype{:}{>{\global\let\currentrowstyle\relax}}
\newcolumntype{;}{>{\currentrowstyle}}
\newcommand{\rowstyle}[1]{\gdef\currentrowstyle{#1}%
  #1\ignorespaces
}

\newcommand{\bspsmall}{\vspace{0.5cm}\small\noindent This paper has been typeset
from a \TeX/\LaTeX\ file prepared by the author.\normalsize}

\title[Atomic data for quasar spectroscopy]{Laboratory atomic transition data for precise optical quasar absorption spectroscopy}

\author[M. T. Murphy \& J. C. Berengut]{
  Michael T. Murphy,$^{1}$\thanks{E-mail: mmurphy@swin.edu.au (MTM)} Julian C. Berengut$^{2}$\\
  $^{1}$Centre for Astrophysics and Supercomputing, Swinburne University of Technology, Hawthorn, Victoria 3122, Australia\\
  $^{2}$School of Physics, University of New South Wales, Sydney, NSW 2052, Australia
 }

\begin{document}

\date{Accepted 2013 November 11. Received 2013 November 10; in original form 2013 October 30}

\pagerange{\pageref{firstpage}--\pageref{lastpage}}

\pubyear{2014}

\maketitle

\label{firstpage}

\begin{abstract}
Quasar spectra reveal a rich array of important astrophysical
information about galaxies which intersect the quasar line of
sight. They also enable tests of the variability of fundamental
constants over cosmological time and distance-scales. Key to these
endeavours are the laboratory frequencies, isotopic and hyperfine
structures of various metal-ion transitions. Here we review and
synthesize the existing information about these quantities for
43 transitions which are important for measuring possible changes in the
fine-structure constant, $\alpha$, using optical quasar spectra,
i.e.~those of Na, Mg, Al, Si, Ca, Cr, Mn, Fe, Ni and Zn. We also
summarize the information currently missing that precludes more
transitions being used. We present an up-to-date set of coefficients,
$q$, which define the sensitivity of these transitions to variations
in $\alpha$. New calculations of isotopic structures and $q$
coefficients are performed for \ion{Si}{ii} and \ion{Ti}{ii},
including \ion{Si}{ii} $\lambda$1808 and \ion{Ti}{ii}
$\lambda\lambda$1910.6/1910.9 for the first time. Finally, simulated
absorption-line spectra are used to illustrate the systematic errors
expected if the isotopic/hyperfine structures are omitted from profile
fitting analyses.

To ensure transparency, repeatability and currency of the data and
calculations, we supply a comprehensive database as Supporting
Information. This will be updated as new measurements and calculations
are performed.
\end{abstract}

\begin{keywords}
  atomic data -- line: profiles -- methods: laboratory: atomic -- techniques: spectroscopic
  quasars: absorption lines -- ultraviolet: general
\end{keywords}

\section{Introduction}\label{sec:intro}

The absorption lines observed in the spectra of quasars have proved
powerful probes of the high-redshift Universe. They offer the
possibility for studying the interstellar, circumgalactic and
intergalactic media along the quasar line of sight in considerable
detail, regardless of the existence or brightness of any associated
intervening galaxy
\citep[e.g.][]{Schmidt:1963:1040,Bahcall:1965:1677,Gunn:1965:1633,Rauch:1998:267,Wolfe:2005:861}. Transitions
from ionized metallic species, such as \ion{Si}{ii}, \ion{Mg}{ii}
etc., are particularly useful because they can trace the kinematics of
the intervening gas, i.e.~the typical Doppler broadening of individual
absorption features, $b\la10$\,\kms, is much narrower than the
kinematic spread of gas in typical galaxies ($\ga$30\,\kms), and their
optical depths are sufficiently low that their damping wings do not
swamp the detailed absorption profile
\citep[e.g.][]{Lu:1993:1,Churchill:2001:679,Prochaska:2008:59,Bouche:2013:50}. High-resolution
optical spectroscopy of quasars also enables direct measurements of
the column density of the absorbing ions which, together with that of
the corresponding neutral hydrogen, provides accurate measures of the
metallicity, dust-depletion and ionization characteristics of the
absorbing gas
\citep[e.g.][]{Pettini:1990:48,Prochaska:1996:403,Dessauges-Zavadsky:2007:431}. However,
accurate laboratory data for these transitions are required for
meaningful measurements to be made, particularly their rest-frame
frequencies and oscillator strengths
\citep[e.g.][]{Morton:1991:119,Morton:2003:205}.

The requirement for precise laboratory data is particularly acute
when attempting to constrain changes in the fundamental constants of
Nature using quasar absorption lines. The measured relative velocities
of different metallic transitions in a quasar spectrum are directly
proportional to deviations in the fine-structure constant, $\alpha$,
between the absorbing cloud ($\alpha_z$) and the current laboratory
value \citep[$\alpha_0$,][]{Savedoff:1956:688,Dzuba:1999:888}:
\begin{equation}\label{eq:dadef}
\daa \equiv \frac{\alpha_z-\alpha_0}{\alpha_0} \approx -\frac{\Delta v_i}{c}\frac{\omega_i}{2q_i}\,,
\end{equation}
where the $q$-coefficient (defined in \Aref{app:a}) specifies the
sensitivity of transition $i$'s frequency (or wavenumber, $\omega_i$)
to variations in $\alpha$, and $c$ is the speed of light.
Inaccuracies in the laboratory data for the metallic transitions used
to measure \daa\ therefore translate to possible systematic errors in
the analysis, potentially producing spurious detections of varying
constants. Indeed, evidence has emerged for cosmological variations in
$\alpha$ from large samples of echelle quasar spectra with high
resolving powers
\citep[$R\ga45000$,][]{Webb:1999:884,Murphy:2001:1208,Webb:2001:091301,Murphy:2003:609,Webb:2011:191101,King:2012:3370}
which, so far, studies from smaller samples have neither confirmed or
ruled out
\citep[e.g.][]{Levshakov:2006:L21,Molaro:2008:173,Molaro:2013:A68}. Furthermore,
the first of these studies to use the `Many Multiplet method' -- the
comparison of different transitions from different multiplets, often
incorporating many ionic species -- \citet{Webb:1999:884}, recognised
that the accuracy of the previously-measured laboratory frequencies
for the rest-frame ultraviolet (UV) transitions
($\sim$2300--2800\,\AA) of \ion{Mg}{ii} and \ion{Fe}{ii} was a
limiting factor in measuring \daa\ using quasar spectra. New
laboratory measurements of the \ion{Mg}{ii} frequencies were conducted
using Fourier Transform Spectrometers (FTSs) for that work
\citep{Pickering:1998:131}. It remains important that the laboratory
data -- especially rest-frame frequencies -- for the transitions used
in these studies are re-measured, scrutinized and refined.

An important step, particularly for varying-$\alpha$ studies, was
recently made by \citet{Nave:2011:737} and \citet{Nave:2012:1570}. By
recalibrating the frequency scales of previous FTS measurements, they
placed a large subset of the metal-ion transitions used so far for
varying-$\alpha$ analyses onto a consistent frequency scale. They also
showed that common scale to be consistent with the absolute
frequencies of some transitions established using higher-precision,
frequency comb techniques. It is therefore timely to review all the
current metal-ion transitions used for varying-$\alpha$ studies and,
where relevant and possible, to add further important information,
such as isotopic and hyperfine structure measurements and
calculations.

This paper reviews, compiles and synthesizes the extant measurements
of rest-frame frequencies for the metal-ion transitions useful for
measuring \daa\ in UV, optical and infrared quasar spectra. These are
all electric dipole (E1) transitions from the ground state. We define
``useful for measuring \daa'' to mean that the rest-frame frequency
has been measured with velocity precision $\delta v\la 20$\,\ms,
thereby ensuring that analysis of the quasar spectra should not be
limited by the laboratory data. We consider here transitions of the
following ions:
\ion{Na}{i},
\ion{Mg}{i} and {\sc ii},
\ion{Al}{ii} and {\sc iii},
\ion{Si}{ii} and {\sc iv},
\ion{Ca}{ii},
\ion{Cr}{ii},
\ion{Mn}{ii},
\ion{Fe}{ii},
\ion{Ni}{ii} and
\ion{Zn}{ii}.
Note that these are predominantly singly-ionized species, the dominant
ionization state for these atoms in quasar absorbers with high enough
column densities of hydrogen to be self-shielded from incoming
ionizing UV photons. As such, these species might be expected to be
largely intermixed within such absorbers and to occupy the same
physical region.

The unresolved isotopic and hyperfine structures of the transitions of
interest are also important to consider in varying-$\alpha$
analyses. The isotopic structures could be particularly important
because we generally have no empirical estimates for the relative
isotopic abundances prevailing in quasar absorption systems: if they
differ from the terrestrial isotopic abundances, the absorption
line centroids will differ slightly from that implied by the
terrestrial ratios, potentially leading to systematic errors
in \daa\
\citep{Murphy:2001:1223,Berengut:2003:022502,Ashenfelter:2004:041102}. Theoretical
models for the variations in the isotopic abundances in low
metallicity environments were explored by \citet{Fenner:2005:468},
and show a wide variety of expected trends for different isotopes of
different ionic species. However, to understand the impact of isotopic
abundance variations on varying-$\alpha$ measurements, a more basic
requirement is for the isotopic structures to be known from
measurements, or at least estimated from theory. This paper presents
calculations of the frequencies of the isotopic and hyperfine
components of the transitions of interest and highlights cases of
significant uncertainty.

Finally, this paper provides a self-consistent set of recommended
sensitivity coefficients, $q$, for these metal-ion transitions.
Following \citet{Dzuba:1999:888}, the $q$-coefficients have been
calculated with a variety of methods by several authors, all showing
good consistency. It is important to realise that errors in the
calculated $q$ coefficients cannot produce spuriously non-zero
measurements of \daa; only systematic errors which introduce measured
velocity shifts between transitions can do that -- see
\Eref{eq:dadef}. Nevertheless, detailed comparison of any non-zero
measurements of \daa\ by different groups is facilitated by a
consistent set of $q$ coefficients.

This paper is organised as follows. In \Sref{sec:data} we tabulate the
laboratory atomic data for the transitions of interest. We summarise
the existing measurements and also fully describe our calculations,
where required, to connect the measured quantities with the
frequencies, isotopic and/or hyperfine structures recommended for
analysis of quasar absorption spectra. \Sref{sec:IHF} describes the
systematic effects on measured quantities in quasar absorption
analyses (redshifts, column densities and Doppler broadening
parameters) if the isotopic and hyperfine structures of these
transitions are ignored when analysing quasar spectra. In
\Sref{sec:conc} we summarize the important laboratory transition
information still missing for the ions considered in this paper and
conclude. An appendix describes our new calculations of the
$q$-coefficients and isotopic structures for transitions of
\ion{Si}{ii} and \ion{Ti}{ii}.

\section{Atomic data}\label{sec:data}

Tables \ref{tab:Na}--\ref{tab:Zn} provide the important laboratory
data for quasar absorption line analysis, focussing particularly on
the frequencies recommended for constraining cosmological changes in
$\alpha$. The isotopic and hyperfine structures are illustrated in
Figs.~\ref{fig:Na_IHF}--\ref{fig:Zn_IHF} where measurements or
calculations were available or newly performed. The structure of these
tables and figures is described in \Sref{ssec:tabs}.

\Sref{ssec:calcs} describes the general basis for our calculations of
composite, isotopic and hyperfine frequencies. For each atom, a
subsequent subsection (Sections \ref{ssec:Na}--\ref{ssec:Zn})
describes the laboratory frequency measurements and the composite,
isotopic and/or hyperfine frequency calculations in detail. To
establish the relative reliability of different transitions for
varying-$\alpha$ analyses, particular attention is paid to whether
each transition has been reproduced in more than one experiment and
how its frequency has been calibrated. For the latter, of particular
note is whether the original FTS measurements were calibrated with the
\ion{Ar}{ii} transition frequencies of \citet{Norlen:1973:249} or
those of \citet{Whaling:1995:1}. \citet{Nave:2011:737} and
\citet{Nave:2012:1570} determined reliable conversion factors between
these two calibration scales, with the \citeauthor{Whaling:1995:1}
scale being consistent with more precise, absolute frequency
measurements using frequency combs. Therefore, all frequencies quoted
in Tables \ref{tab:Na}--\ref{tab:Zn} are either on the absolute
(frequency comb) or \citeauthor{Whaling:1995:1} frequency scales, with
conversions from \citeauthor{Norlen:1973:249} calibrations described
in Sections \ref{ssec:Na}--\ref{ssec:Zn} where necessary.

To ensure the accuracy and transparency of all the information provided
in Tables \ref{tab:Na}--\ref{tab:Zn} and calculations described in
Sections \ref{ssec:calcs} and \ref{ssec:Na}--\ref{ssec:Zn}, we also
provide a comprehensive `Atomic Data spreadsheet' in the online
Supporting Information. This allows future incorporation of
new/alternative measurements and/or calculations.

\subsection{Data tables and isotopic/hyperfine structure figures}\label{ssec:tabs}

In Tables \ref{tab:Na}--\ref{tab:Zn} the transitions are named by
truncating their vacuum wavelengths to the nearest {\AA}ngstr\"om
according to the usual convention in quasar absorption work. They are
further identified by their lower and upper state electronic
configurations, and ``ID'' labels. The latter act as a `short-hand'
notation to assist specifying a set of transitions analysed in a
particular absorption system
\citep[e.g.][]{Murphy:2001:1223,King:2012:3370}. The ID's superscript
is the ionization stage and its subscript is the transition's rank
among those falling redwards of H{\sc \,i} Lyman-$\alpha$ for that ion
in order of decreasing oscillator strength. Hyperfine components are
also identified by the quantum number, $F$, which represents the
vector sum of the nuclear and electron angular momenta. The relative
atomic mass, $A$, is provided for each ion; these determine the
relative widths of quasar absorption lines from different metallic
species when the gas is dominated by thermal broadening. The mass
number identifies individual isotopic or hyperfine components under
the same ``$A$'' column. Rows pertaining to isotopic and hyperfine
components are shown in italics. The tables also provide the relevant
ionization potentials for each ion, i.e.~the energy required to create
the ion from the next lowest ionization stage (IP$^-$) and that
required to reach the next stage (IP$^+$).

The laboratory frequency, $f_0$, is given as the wavenumber,
$\omega_0$, in the tables: $\omega_0\equiv f_0/c=1/\lambda_0$ for
$\lambda_0$ the wavelength in vacuum. For convenience, the tables also
provide $\lambda_0$ for each transition (converted from the measured
frequency). For some transitions, individual isotopic and/or hyperfine
structure wavenumbers were measured in the laboratory experiments. For
others, only the (relative) splitting between isotopic/hyperfine
components was measured or, more often, only calculated, and the
centroid of the `composite' transition -- the weighted sum of its
components -- was measured. Our approach for inferring the unmeasured
quantities in these cases is described below in \Sref{ssec:calcs}, and
the details for each transition are given in Sections
\ref{ssec:Na}--\ref{ssec:Zn}. To summarize that information in Tables
\ref{tab:Na}--\ref{tab:Zn}, we provide a flag, $X$, defined as
follows:
\begin{equation}\label{eq:X}
  X=\left\{
  \begin{array}{l l l}
    0 & \mbox{Measured frequency};\\
    1 & \mbox{Inferred from component frequencies};\\
    2 & \mbox{Inferred from measured composite frequency and}\\
      & \mbox{~~~~~~measured component splitting};\\
    3 & \mbox{Inferred from measured composite frequency and}\\
      & \mbox{~~~~~~calculated component splitting};\\
    4 & \mbox{Inferred from measured component frequency and}\\
      & \mbox{~~~~~~measured component splitting};\\
    5 & \mbox{Inferred from measured component frequency and}\\
      & \mbox{~~~~~~calculated component splitting}.\\
  \end{array}
  \right.
\end{equation}

The variety and challenges of UV FTS laboratory experiments, plus the
significant improvement in precision and accuracy offered by frequency
comb techniques, result in a wide range of frequency uncertainties for
different transitions and their isotopic/hyperfine components. For
convenience in quasar absorption-line work, and particularly for
varying-$\alpha$ analyses, Tables \ref{tab:Na}--\ref{tab:Zn} present
these uncertainties in velocity space, $\delta v$. These range from
just 0.04\,\ms\ for the frequency comb-calibrated \ion{Mg}{ii} isotopes
(see \Tref{tab:Mg}), up to $\approx$20\,\ms\ for the \ion{Ni}{ii}
transitions (see \Tref{tab:Ni}), the maximum measurement uncertainty
allowed under our definition of ``useful for \daa''.

An important parameter for accurate modelling of quasar absorption
lines is the oscillator strength of each transition, $f$. The values
provided in Tables \ref{tab:Na}--\ref{tab:Zn} are those compiled by
\citet{Morton:2003:205}. For isotopic components, the relative
terrestrial isotopic abundance from \citet{Rosman:1998:1275} is
presented as a percentage. This fraction is further broken down for
hyperfine components according to the relative level populations
expected under the assumption of local thermodynamic equilibrium
(LTE). For the relevant transitions discussed in this paper -- those
of $^{23}$\ion{Na}{i}, $^{25}$\ion{Mg}{ii}, $^{27}$\ion{Al}{iii},
$^{43}$\ion{Ca}{ii}, $^{55}$\ion{Mn}{ii} and $^{67}$\ion{Zn}{ii} --
photo-excitation from the cosmic microwave background is sufficient to
make this a very good approximation\footnote{The largest ground-state
  hyperfine energy splitting considered in this paper is
  $8\times10^{-5}$\,eV for $^{55}$\ion{Mn}{ii} $\lambda$2606. This is
  an order of magnitude smaller than the present-day cosmic microwave
  background photon energy of $\approx\,7\times10^{-4}$\,eV, so we can
  expect all hyperfine levels in the ground state to be equally
  populated at all redshifts.}.

Finally, the tables also provide the recommended $q$ coefficient for
each transition, i.e.~its sensitivity to variations in $\alpha$.  For
single-valence-electron ions (i.e.~\ion{Na}{i}, \ion{Mg}{ii},
\ion{Al}{iii}, \ion{Si}{iv}, and \ion{Ca}{ii}) our recommended $q$
coefficients were calculated by \citet{Dzuba:2007:062510} using the
all-order coupled-cluster single--double method with added third-order
many-body perturbation theory (MBPT) and Breit corrections. Note that
the quoted uncertainties are not statistical and should not be
considered or used as such. Instead, the errors quoted are indicative
of the (small) discrepancy with other methods: second-order MBPT
\citep{Dzuba:1999:230}, second-order Brueckner orbitals
\citep{Berengut:2004:064101}, or second-order MBPT with Breit
corrections \citep{Savukov:2008:042501}.  Note that Breit corrections
to the $q$ coefficients are very small in all the transitions
considered. For all the many-valence-electron ions, at a minimum the
$q$ coefficients are calculated using configuration interaction
(CI). In many cases, we present results obtained using the atomic
structure package AMBiT, an implementation of the combination of CI
and MBPT described in \citet{Berengut:2006:012504} \citep[see
also][]{Dzuba:1996:3948,Berengut:2005:044501}. More details of this
method are presented in \Aref{app:a} where we also present new
calculations for the transitions of \ion{Si}{ii} and \ion{Ti}{ii}
shown in Tables \ref{tab:Si} and \ref{tab:Ti}, respectively.
References for the origin of the other many-valence-electron ions are
given in the relevant subsection for each ion, but in all cases the
results we present in this work show a high level of consistency
across methods and authors.

Figures \ref{fig:Na_IHF}--\ref{fig:Zn_IHF} illustrate the isotopic
and/or hyperfine structures for the transitions studied here, where
available. Each panel of each Figure depicts a single transition, with
the black vertical lines indicating the velocity-space positions and
relative strengths (`intensities') of the individual components from
Tables \ref{tab:Na}--\ref{tab:Zn}. The velocity zero point (vertical
cyan line) corresponds to the composite wavelength of the
transition. The ID letter for each transition is provided in the top
right-hand corner of its panel and the components are labelled by
their mass number, $A$. The blue curve in each panel represents the
sum of Gaussian functions centred on, and with the relative amplitudes
of, the isotopic/hyperfine components. The Gaussian width is set to a
full-width-at-half-maximum (FWHM) of 0.3\,\kms, equivalent to a
resolving power of $R=1\times10^{6}$, for all components in all the
Figures. This aids the eye in quickly determining how important the
isotopic/hyperfine structure is for any given transition (assuming
terrestrial isotopic abundances and LTE hyperfine level
populations). For example, one can immediately see that the isotopic
structure of the Mg transitions (\Fref{fig:Mg_IHF}) could be important
for correctly modelling quasar absorption lines, whereas it will be
less important for the \ion{Fe}{ii} transitions (\Fref{fig:Fe_IHF}),
assuming that, for the latter, the theoretical estimates of the
isotopic structure are approximately correct.

\subsection{Composite, isotopic and hyperfine frequency calculations}\label{ssec:calcs}

With few exceptions, the transitions in Tables
\ref{tab:Na}--\ref{tab:Zn} generally comprise multiple isotopic and/or
hyperfine components. At the resolving powers used in most quasar
absorption line work ($R\la100000$), and given the typical Doppler
broadening prevailing in most quasar absorption clouds
($b\sim1$--10\,\kms), this multi-component structure is not important
to model in most studies. However, \Sref{ssec:Nandb} below
demonstrates that care must be taken when modelling \ion{Al}{iii} and
\ion{Mn}{ii} absorption profiles because their hyperfine structures
are particularly broad and can significantly affect column density and
Doppler broadening estimates. For varying-$\alpha$ work, the velocity
centroid of an absorption line is the key measurement, so modelling
the unresolved structure of each transition can be very
important. This is demonstrated explicitly in \Sref{ssec:velshifts}.

Unfortunately, laboratory measurements with sufficient resolution and
suppression of Doppler broadening to resolve apart the
isotopic/hyperfine components of these transitions are few. Thus,
direct measurements of their isotopic/hyperfine component frequencies,
or even the relative splittings, are not available. However, in many
cases the splittings have been calculated, or can be derived from
other measured quantities with sufficient accuracy. In these cases, we
define the relationship between the composite and component
wavenumbers as
\begin{equation}\label{eq:comp}
  \omega_{\rm comp} = \sum_i^{N} a_i\omega_i\,,
\end{equation}
where the coefficient $a_i$ is the relative isotopic abundance of the
$i$th component multiplied by its relative hyperfine intensity,
normalized such that $\sum_i^{N_{\rm comp}}a_i=1$ over the $N$
components. If, as is most usual, the composite wavenumber has been
measured and the isotopic and/or hyperfine component separations,
$\Delta\omega_i \equiv \omega_j - \omega_i$, are known, or can be
estimated, then the wavenumber for the $j$th component is calculated
as
\begin{equation}\label{eq:omega0}
  \omega_j = a_j^{-1}\left(\omega_{\rm comp} - \sum_{i\ne j}^{N} a_i\Delta\omega_i\right)\,.
\end{equation}

For calculating isotopic structures, the frequency shift,
$\delta\nu^{A^\prime\!,\,A}\equiv\nu^{A^\prime}-\nu^A$, for a given
transition in an isotope with mass number $A^\prime$ relative to the
same transition in one with mass number $A$, may be expressed as
\citep[e.g.][]{Berengut:2003:022502}
\begin{equation}\label{eq:iso}
  \delta\nu^{A^\prime\!,\,A} = (\kNMS+\kSMS)\left(\frac{1}{A^\prime}-\frac{1}{A}\right) + \FS\delta\langle r^2\rangle^{A^\prime\!,\,A}\,,
\end{equation}
where $\kNMS\equiv-\nu^A m_{\rm e}$ is the normal mass shift constant
with $m_{\rm e}$ the electron mass in atomic mass units, \kSMS\ is the
specific mass shift constant, \FS\ is the field shift constant and
$\delta\langle r^2\rangle^{A^\prime\!,\,A}$ is the difference between
the root-mean-square nuclear charge radii of the two isotopes. In
Appendix \ref{app:a} we detail new calculations of \kSMS\ and \FS\ for
the \ion{Si}{ii} and \ion{Ti}{ii} transitions shown in Tables
\ref{tab:Si} and \ref{tab:Ti} respectively. Previously published
values for \kSMS\ and \FS\ are used to calculate isotopic structures
using \Eref{eq:iso} for some transitions of \ion{Mg}{ii}, \ion{Fe}{ii} and \ion{Zn}{ii} in
Sections \ref{ssec:Mg}, \ref{ssec:Fe} and \ref{ssec:Zn} respectively.

The hyperfine energy shift for given nuclear and electron states can
be expressed, to second order, as \citep[e.g.][]{Sur:2005:25}
\begin{equation}\label{eq:hyp1}
  \Delta E_{\rm hyp} = \Delta E_{\rm M1} + \Delta E_{\rm E2}\,,
\end{equation}
where the magnetic dipole and electric quadrupole terms are, respectively,
\begin{equation}\label{eq:hyp2}
  \Delta E_{\rm M1} = \frac{A}{2}K{\rm ~~and~~} \Delta E_{\rm E2} = \frac{B}{2}C\,,
\end{equation}
where $A$ and $B$ are the magnetic dipole and electric quadrupole
hyperfine constants, respectively, with
\begin{equation}\label{eq:hyp3}
K\equiv F(F+1) - I(I+1) - J(J+1)
\end{equation}
and
\begin{equation}\label{eq:hyp4}
C\equiv \frac{3K(K+1) - 4I(I+1)J(J+1)}{2I(2I-1)2J(2J-1)}\,.
\end{equation}
Here, $I$, $J$ and $F$ are the quantum numbers representing the total
angular momentum of the nuclear state, {\boldmath{$I$}}, the electron
ground state, {\boldmath{$J$}}, and their vector sum,
{\boldmath{$F$}}. The relative probabilities for the transitions to the excited state hyperfine levels, defined by the electron and total angular momenta, $J^\prime$ and $F^\prime$, is given by the product of the level degeneracies and the Wigner 6$J$-symbol,
\begin{equation}\label{eq:hyp5}
\addtolength{\arraycolsep}{-0.3em} 
S = (2F+1)\,(2F^\prime+1)\,\left\{\begin{array}{ccc}
J       & F      & I \\
F^\prime & J^\prime & 1
\addtolength{\arraycolsep}{0.3em}
\end{array}\right\}^2\,.
\end{equation}

Hyperfine structures are calculated using equations
(\ref{eq:hyp1})--(\ref{eq:hyp5}) with previously measured or
calculated values of $A$ and $B$ for transitions of
$^{23}$\ion{Na}{i}, $^{25}$\ion{Mg}{ii}, $^{43}$\ion{Ca}{ii} and
$^{67}$\ion{Zn}{ii} in Sections \ref{ssec:Na}, \ref{ssec:Mg},
\ref{ssec:Ca} and \ref{ssec:Zn} respectively. In those cases, and for
the measured hyperfine structures of $^{27}$\ion{Al}{iii} and
$^{55}$\ion{Mn}{ii}, the splitting of the excited states is found to
be negligibly small, so only the hyperfine structure of the ground
states is important to consider for quasar absorption line
studies. This is discussed further in the relevant sections below and
explicitly reflected in Tables \ref{tab:Na}, \ref{tab:Mg},
\ref{tab:Al}, \ref{tab:Ca}, \ref{tab:Mn} and \ref{tab:Zn}. We have not
derived the hyperfine structure for other isotopic species with odd
nucleon numbers ($^{29}$\ion{Si}{ii}, $^{47,49}$\ion{Ti}{ii},
$^{53}$\ion{Cr}{ii}, $^{57}$\ion{Fe}{ii}), either because the relevant
measurements/estimates of $A$ and $B$ are not available in the
literature, or their magnetic moments are relatively small, leading us
to expect small values for $A$ and $B$.

All calculations of composite, isotopic and hyperfine component
wavenumbers are presented in detail, and can be trivially modified to
accommodate new or changed information, in the comprehensive `Atomic
Data spreadsheet' in the online Supporting Information.

\newcommand\oldtabcolsep{\tabcolsep}
\setlength{\tabcolsep}{0.45em}

\subsection{Sodium}\label{ssec:Na}

\addtolength{\tabcolsep}{0.05em}
\begin{table*}
\begin{center}
\caption{
Laboratory data for transitions of Na of interest for quasar absorption-line varying-$\alpha$ studies described in \Sref{ssec:Na}. See \Sref{ssec:tabs} for full descriptions of each column.
}
\label{tab:Na}\vspace{-0.5em}
{\footnotesize
\begin{tabular}{:l;l;c;l;c;l;c;l;l;c;c;c;c}\hline
\multicolumn{1}{c}{Ion}&
\multicolumn{1}{c}{Tran.}&
\multicolumn{1}{c}{$A$}&
\multicolumn{1}{c}{$\omega_0$}&
\multicolumn{1}{c}{$X$}&
\multicolumn{1}{c}{$\lambda_0$}&
\multicolumn{1}{c}{$\delta v$}&
\multicolumn{1}{c}{Lower state}&
\multicolumn{1}{c}{Upper state}&
\multicolumn{1}{c}{ID}&
\multicolumn{1}{c}{IP$^-$, IP$^+$}&
\multicolumn{1}{c}{$f$ or {\it \%}}&
\multicolumn{1}{c}{$q$}\\
&
&
&
\multicolumn{1}{c}{[cm$^{-1}$]}&
&
\multicolumn{1}{c}{[\AA]}&
\multicolumn{1}{c}{[m\,s$^{-1}$]}&
&
&
&
\multicolumn{1}{c}{[eV]}&
&
\multicolumn{1}{c}{[cm$^{-1}$]}\\
\hline
                    Na{\sc \,i}   & 5891   & 22.9898   & 16973.366206(43)$^{}$            & 1 &   5891.583248(15)  & 0.76 & $3\rm{s}~^2\rm{S}_{1/2}                  $ & $3\rm{p}~^2\rm{P}_{3/2}^{\rm o}          $ & Na$^1_{1}$  & ---, 5.14    & 0.6408    & $   62^{a}(2)  $\\
\rowstyle{\itshape}               &        & 23        & 16973.401853$^{b,c,d}$           & 4 &   5891.570875      &      & $F=1                                     $ & $F=0,1,2                                 $ &             &              & 37.5\%    & $     ^{}     $\\
\rowstyle{\itshape}               &        & 23        & 16973.344818$^{b,c,d}$           & 4 &   5891.590672      &      & $F=2                                     $ & $F=1,2,3                                 $ &             &              & 62.5\%    & $     ^{}     $\\
                                  & 5897   & 22.9898   & 16956.170247(43)$^{}$            & 1 &   5897.558148(15)  & 0.76 & $                                        $ & $3\rm{p}~^2\rm{P}_{1/2}^{\rm o}          $ & Na$^1_{2}$  &              & 0.3201    & $   45^{a}(2)  $\\
\rowstyle{\itshape}               &        & 23        & 16956.208494$^{c,e,d}$           & 4 &   5897.544845      &      & $F=1                                     $ & $F=1,2                                   $ &             &              & 37.5\%    & $     ^{}     $\\
\rowstyle{\itshape}               &        & 23        & 16956.147299$^{c,e,d}$           & 4 &   5897.566130      &      & $F=2                                     $ & $F=1,2                                   $ &             &              & 62.5\%    & $     ^{}     $\\
\hline
\end{tabular}
}
{\footnotesize References:
$^{a}$\citet{Dzuba:2007:062510};
$^{b}$\citet{Beckmann:1974:173};
$^{c}$\citet{Juncar:1981:77};
$^{d}$\citet{Yei:1993:1909};
$^{e}$\citet{Wijngaarden:1994:67}.}
\end{center}
\end{table*}

\addtolength{\tabcolsep}{-0.05em}
\begin{figure}
\begin{center}
\includegraphics[width=0.55\columnwidth,bb = 18 322 242 706]{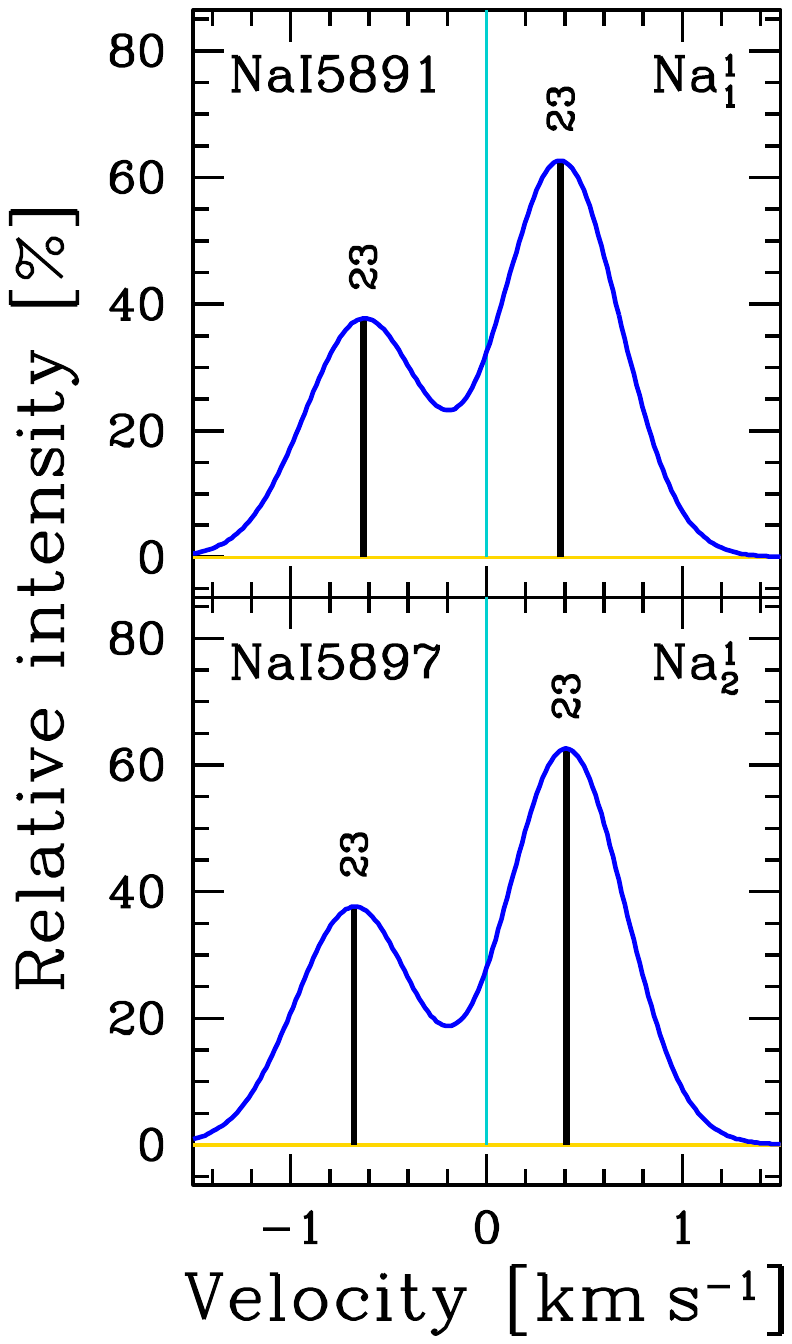}\vspace{-0.5em}
\caption{Na transition hyperfine structures in \Tref{tab:Na} described
  in \Sref{ssec:Na}. See \Sref{ssec:tabs} for Figure details.}
\label{fig:Na_IHF}
\end{center}
\end{figure}

\Tref{tab:Na} provides the atomic data for the two strongest
\ion{Na}{i} transitions which are common, though often not
particularly strong, in quasar absorption systems. \Fref{fig:Na_IHF}
illustrates the hyperfine structures for these transitions. The
\ion{Na}{i} $\lambda\lambda$5891/5897 doublet, or D2 and D1
transitions are observable at redshifts $z\la0.6$ in optical quasar
absorption spectra. However, they have not been used so far in
varying-$\alpha$ studies (to our knowledge) for three main reasons:
(i) they are typically much weaker than the ubiquitous Mg and
\ion{Fe}{ii} transitions observable at the same redshifts; (ii) they
often fall in spectral regions containing significant telluric
absorption; and (iii) their frequencies are relatively insensitive to
variations in $\alpha$. Nevertheless, being the strongest \ion{Na}{i}
transitions from the ground state by a factor of $\ga$30
\citep{Morton:2003:205}, they are the only \ion{Na}{i} transitions
that are suitable for varying-$\alpha$ work. Na has only one stable
isotope ($^{23}$Na), so it is immune to systematic effects in
varying-$\alpha$ analyses related to isotopic abundance evolution
(along with Al and Mn). However, the D1/D2 lines do have hyperfine
structure, which is dominated by the $\sim$1\,\kms\ splitting of the
ground state, and these have been measured to high precision (see
below).

To our knowledge, the most recent and most precise absolute frequency
measurements for the \ion{Na}{i} D lines are those of
\citet{Juncar:1981:77}. Using a saturated-absorption technique, they
locked a tunable dye laser to well-studied hyperfine components of
iodine gas (I$_2$) transitions near the `cross-over' (i.e.~average)
frequencies of particular hyperfine components of the \ion{Na}{i}
transitions. Within the final measurement uncertainties
they present, their I$_2$ calibration scale is consistent with modern
frequency comb-calibrated measurements of the same I$_2$ reference
lines \citep[e.g.][]{Balling:2008:3}, so we have not corrected their
\ion{Na}{i} frequency measurements for calibration differences.

Starting from the basic `cross-over' frequency measurements of
\citet{Juncar:1981:77}, we used modern measurements of the ground and
excited state hyperfine splittings to determine absolute frequencies
for all hyperfine components. The magnetic dipole hyperfine constant
($A$) for the ground state was taken from \citet{Beckmann:1974:173}
while those for the D1 and D2 excited states were taken from
\citet{Wijngaarden:1994:67} and \citet{Yei:1993:1909}, respectively.
The electric quadrupole hyperfine constant ($B$) for the D2 line was
also taken from the latter. The hyperfine structure was then
determined via equations (\ref{eq:hyp1})--(\ref{eq:hyp5}). In
\Tref{tab:Na} we average over the hyperfine splitting of the excited
states because it is negligible ($\sim$60\,\ms) compared to the large
ground-state splitting ($\sim$1\,\kms). Finally, the composite
frequencies were determined from the hyperfine component frequencies
via \Eref{eq:comp}. We note that the composite wavelengths in
\Tref{tab:Na} differ slightly (up to 1.1$\sigma$,
i.e.~$1.5\times10^{-5}$\,\AA, or 0.8\,\ms) from those reported by
\citet{Juncar:1981:77}, a difference we suspect is attributable to the
updated excited-state hyperfine structures we employed.

The final hyperfine splittings of the \ion{Na}{i} D lines illustrated
in \Fref{fig:Na_IHF} are relatively broad ($\sim$1\,\kms) and render
the final line shapes significantly asymmetric. Therefore, absorption
line modelling of modern, high-resolution quasar spectra, could be
inaccurate if the hyperfine isotopic structure is not taken into
account. This is considered further in \Sref{sec:IHF}.

\subsection{Magnesium}\label{ssec:Mg}

\begin{table*}
\begin{center}
\caption{
Laboratory data for transitions of Mg of interest for quasar absorption-line varying-$\alpha$ studies described in \Sref{ssec:Mg}. See \Sref{ssec:tabs} for full descriptions of each column.
}
\label{tab:Mg}\vspace{-0.5em}
{\footnotesize
\begin{tabular}{:l;l;c;l;c;l;c;l;l;c;c;c;c}\hline
\multicolumn{1}{c}{Ion}&
\multicolumn{1}{c}{Tran.}&
\multicolumn{1}{c}{$A$}&
\multicolumn{1}{c}{$\omega_0$}&
\multicolumn{1}{c}{$X$}&
\multicolumn{1}{c}{$\lambda_0$}&
\multicolumn{1}{c}{$\delta v$}&
\multicolumn{1}{c}{Lower state}&
\multicolumn{1}{c}{Upper state}&
\multicolumn{1}{c}{ID}&
\multicolumn{1}{c}{IP$^-$, IP$^+$}&
\multicolumn{1}{c}{$f$ or {\it \%}}&
\multicolumn{1}{c}{$q$}\\
&
&
&
\multicolumn{1}{c}{[cm$^{-1}$]}&
&
\multicolumn{1}{c}{[\AA]}&
\multicolumn{1}{c}{[m\,s$^{-1}$]}&
&
&
&
\multicolumn{1}{c}{[eV]}&
&
\multicolumn{1}{c}{[cm$^{-1}$]}\\
\hline
                    Mg{\sc \,i}   & 2026   & 24.3050   & 49346.772620(36)$^{}$            & 1 &  2026.4749788(15)  &  0.2 & $3\rm{s}^2~^1\rm{S}_0                    $ & $3\rm{s}4\rm{p}~^1\rm{P}_1^{\rm o}       $ & Mg$^1_{2}$  & ---, 7.65    & 0.113     & $   87^{a,b}(7)  $\\
\rowstyle{\itshape}               &        & 26        & 49346.854173(40)$^{c}$           & 0 &  2026.4716298(16)  &  0.2 & $                                        $ & $                                        $ &             &              & 11.01\%   & $     ^{}     $\\
\rowstyle{\itshape}               &        & 25        & 49346.807724(40)$^{c}$           & 0 &  2026.4735372(16)  &  0.2 & $                                        $ & $                                        $ &             &              & 10.00\%   & $     ^{}     $\\
\rowstyle{\itshape}               &        & 24        & 49346.756809(35)$^{c}$           & 0 &  2026.4756281(14)  &  0.2 & $                                        $ & $                                        $ &             &              & 78.99\%   & $     ^{}     $\\
                                  & 2852   & 24.3050   & 35051.28076(19)$^{}$             & 1 &   2852.962797(15)  &  1.6 & $                                        $ & $3\rm{s}3\rm{p}~^1\rm{P}_1^{\rm o}       $ & Mg$^1_{1}$  &              & 1.83      & $   90^{a,b}(10) $\\
\rowstyle{\itshape}               &        & 26        & 35051.32015(25)$^{d}$            & 0 &   2852.959591(20)  &  2.1 & $                                        $ & $                                        $ &             &              & 11.01\%   & $     ^{}     $\\
\rowstyle{\itshape}               &        & 25        & 35051.29784(25)$^{d}$            & 0 &   2852.961407(20)  &  2.1 & $                                        $ & $                                        $ &             &              & 10.00\%   & $     ^{}     $\\
\rowstyle{\itshape}               &        & 24        & 35051.27311(17)$^{d}$            & 0 &   2852.963420(14)  &  1.5 & $                                        $ & $                                        $ &             &              & 78.99\%   & $     ^{}     $\\
                    Mg{\sc \,ii}  & 2796   & 24.3050   & 35760.85414(20)$^{}$             & 1 &   2796.353790(16)  &  1.7 & $3\rm{s}~^2\rm{S}_{1/2}                  $ & $3\rm{p}~^2\rm{P}_{3/2}^{\rm o}          $ & Mg$^2_{1}$  & 7.65, 15.04  & 0.6155    & $  212^{e}(2)  $\\
\rowstyle{\itshape}               &        & 26        & 35760.9403866(53)$^{f}$          & 0 & 2796.34704565(42)  & 0.04 & $                                        $ & $                                        $ &             &              & 11.01\%   & $     ^{}     $\\
\rowstyle{\itshape}               &        & 25        & 35760.85850(64)$^{f,g,h}$        & 3 &   2796.353449(50)  &  5.4 & $F=2                                     $ & $F=1,2,3                                 $ &             &              & 4.17\%    & $     ^{}     $\\
\rowstyle{\itshape}               &        & 25        & 35760.91502(64)$^{f,g,h}$        & 3 &   2796.349030(50)  &  5.4 & $F=3                                     $ & $F=2,3,4                                 $ &             &              & 5.83\%    & $     ^{}     $\\
\rowstyle{\itshape}               &        & 24        & 35760.8373967(53)$^{f}$          & 0 & 2796.35509903(42)  & 0.04 & $                                        $ & $                                        $ &             &              & 78.99\%   & $     ^{}     $\\
                                  & 2803   & 24.3050   & 35669.30440(20)$^{}$             & 1 &   2803.530982(16)  &  1.7 & $                                        $ & $3\rm{p}~^2\rm{P}_{1/2}^{\rm o}          $ & Mg$^2_{2}$  &              & 0.3058    & $  121^{e}(2)  $\\
\rowstyle{\itshape}               &        & 26        & 35669.3905712(53)$^{f}$          & 0 & 2803.52420938(42)  & 0.04 & $                                        $ & $                                        $ &             &              & 11.01\%   & $     ^{}     $\\
\rowstyle{\itshape}               &        & 25        & 35669.30492(64)$^{f,g,h}$        & 3 &   2803.530941(50)  &  5.4 & $F=2                                     $ & $F=2,3                                   $ &             &              & 4.17\%    & $     ^{}     $\\
\rowstyle{\itshape}               &        & 25        & 35669.36798(64)$^{f,g,h}$        & 3 &   2803.525985(50)  &  5.4 & $F=3                                     $ & $F=2,3                                   $ &             &              & 5.83\%    & $     ^{}     $\\
\rowstyle{\itshape}               &        & 24        & 35669.2876697(53)$^{f}$          & 0 & 2803.53229720(42)  & 0.04 & $                                        $ & $                                        $ &             &              & 78.99\%   & $     ^{}     $\\
\hline
\end{tabular}
}
{\footnotesize References:
$^{a}$\citet{Berengut:2005:044501};
$^{b}$\citet{Savukov:2008:042501};
$^{c}$\citet{Hannemann:2006:012505};
$^{d}$\citet{Salumbides:2006:L41};
$^{e}$\citet{Dzuba:2007:062510};
$^{f}$\citet{Batteiger:2009:022503};
$^{g}$\citet{Itano:1981:1364};
$^{h}$\citet{Sur:2005:25}.}
\end{center}
\end{table*}

\begin{figure}
\begin{center}
\includegraphics[width=0.95\columnwidth,bb = 18 322 411 706]{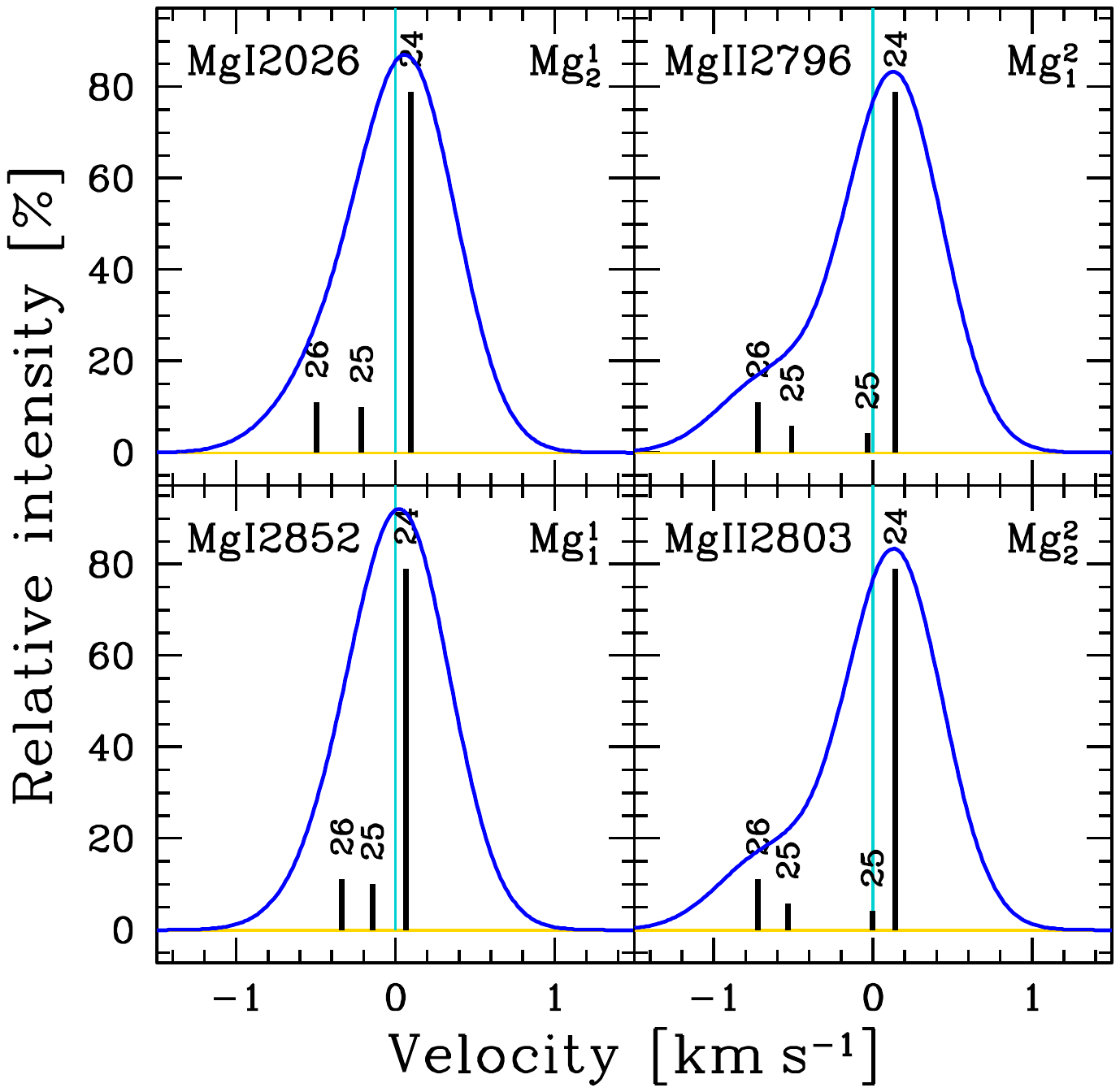}\vspace{-0.5em}
\caption{Mg transition isotopic and hyperfine structures in
  \Tref{tab:Mg} described in \Sref{ssec:Mg}. See \Sref{ssec:tabs} for
  Figure details.}
\label{fig:Mg_IHF}
\end{center}
\end{figure}

\Tref{tab:Mg} provides the atomic data for the two strongest
\ion{Mg}{i} transitions and the \ion{Mg}{ii} doublet which is very
prominent in quasar absorption systems. \Fref{fig:Mg_IHF} illustrates
the isotopic and hyperfine structures for these
transitions. \ion{Mg}{i} $\lambda$2852 is the strongest \ion{Mg}{i}
transition by more than an order of magnitude and is commonly
observed, while \ion{Mg}{i} $\lambda$2026 is often blended with the
\ion{Zn}{ii} $\lambda$2026 transition 50\,\kms\ bluewards of it and,
typically, of similar optical depth. The \ion{Mg}{ii}
$\lambda\lambda$2796/2803 doublet is one of the strongest and,
therefore, easily-identifiable metallic features of quasar absorbers.
The ubiquity of \ion{Mg}{ii} $\lambda\lambda$2796/2803 in quasar
spectra make these transitions prominent in varying-$\alpha$
studies. Only two other \ion{Mg}{ii} transitions fall redwards of the
Lyman-$\alpha$ forest, the $\lambda\lambda$1239/1240 doublet, and
these are 3 orders of magnitude weaker \citep{Morton:2003:205}, so
they are not usually suitable for varying-$\alpha$ studies.

For Mg{\sc \,i}, \citet{Hannemann:2006:012505} and
\citet{Salumbides:2006:L41} measured the frequencies of the 24, 25 and
26 isotopic components of the $\lambda$2026 and $\lambda$2852
transitions, respectively, using laser frequency comb techniques. The
frequencies are therefore on an absolute scale and the uncertainties
are very small, corresponding to velocity uncertainties of
$\sim$0.2\,\ms\ for the $\lambda$2026 transition and $\sim$2\,\ms\ for
$\lambda$2852. The hyperfine structure of the 25 isotope was not
resolved in the measurements of either transition; for $\lambda$2026
the separation between the $^{\rm 25}$Mg{\sc \,i} hyperfine components
is just $\sim$20\,Hz, or $\sim$4\,\ms. Therefore, we do not represent
the $^{\rm 25}$Mg{\sc \,i} hyperfine structure in \Tref{tab:Mg}. The
composite wavelengths for the two Mg{\sc \,i} transitions were
calculated according to \Eref{eq:comp}.

The absolute frequencies of the 24 and 26 isotopic components of the
Mg{\sc \,ii} $\lambda\lambda$2796/2803 doublet have also been measured
with high precision using frequency comb techniques by
\citet{Batteiger:2009:022503}. Their uncertainties correspond to just
$\sim$4\,\cms. However, the transitions of the 25 isotope could not be
produced in this experiment. Instead, \citet{Batteiger:2009:022503}
used \Eref{eq:iso} to calculate the $^{25,24}$Mg{\sc \,ii} isotope
shifts, drawing on calculations of the field shift constants (\FS)
from \citet{Safronova:2001:052501}, \citet{Berengut:2003:022502} and
\citet{Tupitsyn:2003:022511} and the nuclear charge radii
($\delta\langle r^2\rangle^{25,24}$) measurements of
\citet{Angeli:2004:185}, finding $\delta\nu^{25,24}=1621$ and
1620\,MHz for the $\lambda$2796 and $\lambda$2803 transitions,
respectively. The uncertainty of 19\,MHz in each of these estimates,
which is dominated by the measurement uncertainty in $\delta\langle
r^2\rangle^{25,24}$, ultimately limits the precision of the 25 isotope
and also composite frequencies quoted in \Tref{tab:Mg}.

To calculate the hyperfine splitting in the $^{25}$Mg{\sc \,ii}
$\lambda\lambda$2796/2803 transitions we used equations
(\ref{eq:hyp1})--(\ref{eq:hyp5}) with the magnetic dipole hyperfine
constants ($A$) measured for the ground state by
\citet{Itano:1981:1364} and calculated for the excited states by
\citet{Sur:2005:25}; we also use \citet{Sur:2005:25}'s calculated
electric quadrupole hyperfine constant ($B$). Mg has a nuclear spin of
$I=5/2$ so that the $^{25}$Mg{\sc \,ii} $\lambda$2796 ($\lambda$2803)
transition has 6 (4) allowed transitions, three (two) from each of the
two ground states ($F=2$ and 3). The hyperfine splitting is dominated
by that in the ground states, which is itself less than a tenth of the
$^{26,24}$Mg{\sc \,ii} isotopic splitting in these
transitions. Therefore, as illustrated in \Fref{fig:Mg_IHF}, we
represent the $^{25}$Mg{\sc \,ii} $\lambda\lambda$2796/2803 hyperfine
structures with just two components each. All details of these
calculations are provided in the online Supporting Information.

The isotopic structures for the Mg transitions, particularly those of
\ion{Mg}{ii}, illustrated in \Fref{fig:Mg_IHF} are notably broad and
asymmetric, with the 24--26 isotopic component separation being up to
$\sim$0.85\,\kms\ and with the lightest isotope ($^{24}$Mg) being the
most abundant. These facts, combined with the ubiquity of the
\ion{Mg}{ii} $\lambda\lambda$2796/2803 doublet, imply that absorption
line modelling could be inaccurate if the isotopic structure is not
taken into account or the relative Mg isotopic abundances in quasar
absorbers are significantly different to those in the terrestrial
environment. This was recognised as a potential source of significant
errors for varying-$\alpha$ studies by \citet{Webb:1999:884} and
\citet{Murphy:2001:1208,Murphy:2001:1223} \citep[see also,
e.g.,][]{Murphy:2003:609,Ashenfelter:2004:041102,Fenner:2005:468}.

The $q$ coefficients for \ion{Mg}{i}, a two-valence-electron atom,
have been calculated with the same CI+MBPT method described in
\Aref{app:a} by \citep{Berengut:2005:044501} and by a different
implementation that includes the Breit interaction by
\citet{Savukov:2008:042501}. The values recommended in \Tref{tab:Mg}
for \ion{Mg}{i} are the simple average from these two works. The
errors quoted for these $q$ coefficients reflect the differences
between these results, which are largely due to the Breit interaction.

\subsection{Aluminium}\label{ssec:Al}

\begin{table*}
\begin{center}
\caption{
Laboratory data for transitions of Al of interest for quasar absorption-line varying-$\alpha$ studies described in \Sref{ssec:Al}. See \Sref{ssec:tabs} for full descriptions of each column.
}
\label{tab:Al}\vspace{-0.5em}
{\footnotesize
\begin{tabular}{:l;l;c;l;c;l;c;l;l;c;c;c;c}\hline
\multicolumn{1}{c}{Ion}&
\multicolumn{1}{c}{Tran.}&
\multicolumn{1}{c}{$A$}&
\multicolumn{1}{c}{$\omega_0$}&
\multicolumn{1}{c}{$X$}&
\multicolumn{1}{c}{$\lambda_0$}&
\multicolumn{1}{c}{$\delta v$}&
\multicolumn{1}{c}{Lower state}&
\multicolumn{1}{c}{Upper state}&
\multicolumn{1}{c}{ID}&
\multicolumn{1}{c}{IP$^-$, IP$^+$}&
\multicolumn{1}{c}{$f$ or {\it \%}}&
\multicolumn{1}{c}{$q$}\\
&
&
&
\multicolumn{1}{c}{[cm$^{-1}$]}&
&
\multicolumn{1}{c}{[\AA]}&
\multicolumn{1}{c}{[m\,s$^{-1}$]}&
&
&
&
\multicolumn{1}{c}{[eV]}&
&
\multicolumn{1}{c}{[cm$^{-1}$]}\\
\hline
                    Al{\sc \,ii}  & 1670   & 26.9815   & 59851.976(4)$^{a}$               & 0 &    1670.78861(11)  & 20.0 & $3\rm{s}^2~^1\rm{S}_0                    $ & $3\rm{s}3\rm{p}~^1\rm{P}_1^{\rm o}       $ & Al$^2_{1}$  & 5.99, 18.83  & 1.74      & $  270^{b,c}(10) $\\
                    Al{\sc \,iii} & 1854   & 26.9815   & 53916.5480(11)$^{}$              & 1 &   1854.718146(39)  &  6.3 & $3\rm{s}~^2\rm{S}_{1/2}                  $ & $3\rm{p}~^2\rm{P}_{3/2}^{\rm o}          $ & Al$^3_{1}$  & 18.83, 28.45 & 0.559     & $  458^{d}(6)  $\\
\rowstyle{\itshape}               &        & 27        & 53916.8149(8)$^{a}$              & 0 &   1854.708966(28)  &  4.4 & $F=2                                     $ & $F=1,2,3                                 $ &             &              & 41.67\%   & $     ^{}     $\\
\rowstyle{\itshape}               &        & 27        & 53916.3574(8)$^{a}$              & 0 &   1854.724704(28)  &  4.4 & $F=3                                     $ & $F=2,3,4                                 $ &             &              & 58.33\%   & $     ^{}     $\\
                                  & 1862   & 26.9815   & 53682.8884(11)$^{}$              & 1 &   1862.790974(39)  &  6.3 & $                                        $ & $3\rm{p}~^2\rm{P}_{1/2}^{\rm o}          $ & Al$^3_{2}$  &              & 0.278     & $  224^{d}(8)  $\\
\rowstyle{\itshape}               &        & 27        & 53683.1953(8)$^{a}$              & 0 &   1862.780325(28)  &  4.5 & $F=2                                     $ & $F=2,3                                   $ &             &              & 41.67\%   & $     ^{}     $\\
\rowstyle{\itshape}               &        & 27        & 53682.6692(8)$^{a}$              & 0 &   1862.798581(28)  &  4.5 & $F=3                                     $ & $F=2,3                                   $ &             &              & 58.33\%   & $     ^{}     $\\
\hline
\end{tabular}
}
{\footnotesize References:
$^{a}$\citet{Griesmann:2000:L113};
$^{b}$\citet{Angstmann:2004:014102};
$^{c}$\citet{Savukov:2008:042501};
$^{d}$\citet{Dzuba:2007:062510}.}
\end{center}
\end{table*}

\begin{figure}
\begin{center}
\includegraphics[width=0.55\columnwidth,bb = 18 322 242 706]{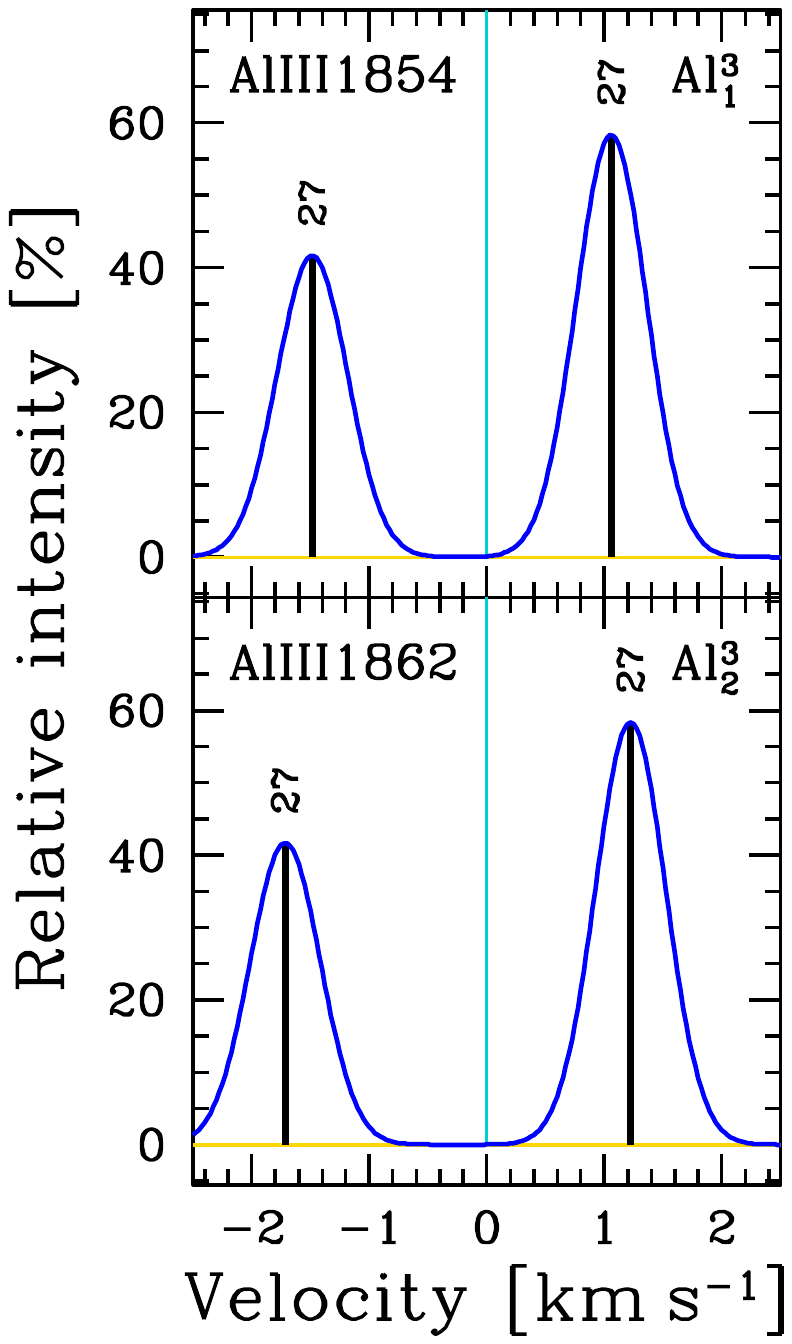}\vspace{-0.5em}
\caption{Al transition hyperfine structures in \Tref{tab:Al} described
  in \Sref{ssec:Al}. See \Sref{ssec:tabs} for Figure details.}
\label{fig:Al_IHF}
\end{center}
\end{figure}

\Tref{tab:Al} provides the atomic data for the single strong Al{\sc
  \,ii} transition and the Al{\sc \,iii} doublet. \Fref{fig:Al_IHF}
illustrates the hyperfine structures for the latter. Al{\sc \,ii}
$\lambda$1670 is one of only 3 transitions of Al{\sc \,ii} known in
the UV and it is the strongest by almost a factor a 1000
\citep{Morton:2003:205}. It is an important ``anchor'' line for
many-multiplet studies of $\alpha$ variation: it is a strong
transition from a relatively abundant ion, so it is detected even in
low column-density absorbers, and it has a very low sensitivity to
$\alpha$ variation (i.e.~a small $q$-coefficient).  Si{\sc \,ii}
$\lambda$1526 is the only similar case in the far UV (i.e.~observed at
high redshift). The \ion{Al}{iii} $\lambda\lambda$1854/1862 doublet
comprises the only known UV transitions of \ion{Al}{iii}
\citep{Morton:2003:205} but it is commonly observed in quasar
absorbers despite \ion{Al}{ii} being the dominant ion in self-shielded
gas. One concern for varying-$\alpha$ studies is whether the
\ion{Al}{ii} and \ion{Al}{iii} absorption profiles are expected to
trace the same velocity structures, i.e.~whether the doubly-ionised
transitions can be analysed together with singly-ionised transitions
in the many-multiplet method. However, the empirical evidence so far
suggests no systematic problem in this regard
\citep[e.g.][]{Murphy:2001:1208}. Importantly, Al has only one stable
isotope ($^{27}$Al), making it one of only three elements discussed in
this paper which is immune to systematic effects in varying-$\alpha$
analyses related to isotopic abundance evolution (the others being Na
and Mn).

The Al{\sc \,ii} $\lambda$1670 laboratory wavenumber has been measured
with 20\,\ms\ precision by \citet{Griesmann:2000:L113} using a FTS and
Penning discharge lamp. Although $^{27}$Al has a relatively large
magnetic dipole moment ($\mu=+3.6$\,nuclear magnetons), no hyperfine
structure was resolved by \citeauthor{Griesmann:2000:L113} for the
spin-zero Al{\sc \,ii} $\lambda$1670 transition. The wavenumber was
calibrated using the Ar{\sc \,ii} scale of \citet{Whaling:1995:1},
placing it, in principle, on the same calibration scale as the other
transitions in this paper \citep[i.e.~that of][]{Nave:2012:1570}. We
are not aware of any similarly-precise measurements of Al{\sc \,ii}
$\lambda$1670.

The wavenumbers for the hyperfine components of Al{\sc \,iii}
$\lambda\lambda$1854/1862 were measured using a FTS and hollow cathode
lamp (HCL) by \citet{Griesmann:2000:L113}. With a nuclear spin of
$I=5/2$, 6 hyperfine components are expected in each transition,
i.e.~3 from each of the two ground states ($F=2$ and 3). Only the
ground state hyperfine structures were resolved by
\citeauthor{Griesmann:2000:L113} so we average over any splitting in
the excited states and represent each of the Al{\sc \,iii}
$\lambda\lambda$1854/1862 transitions by two hyperfine components in
\Tref{tab:Al} and \Fref{fig:Al_IHF}. For consistency with transitions
of other ions considered here, composite wavelengths are also
presented in \Tref{tab:Al}, derived from the measured hyperfine
wavenumbers using \Eref{eq:omega0}. However, as illustrated in
\Fref{fig:Al_IHF}, the hyperfine structure is very widely-spaced --
$>$2.5\,\kms\ -- so, for interpreting high-resolution ($R\ga30000$, or
${\rm FWHM}\la10$\,\kms) quasar absorption spectra, we recommend using
the individual hyperfine components, not treating them as a single,
composite line. Indeed, section \ref{sec:IHF} shows that the effects
of ignoring this hyperfine structure in quasar absorption studies are
relatively large. Like Al{\sc \,ii} $\lambda$1670, the Al{\sc \,iii}
wavenumbers measured by \citeauthor{Griesmann:2000:L113} were placed
on the \citet{Whaling:1995:1} calibration scale and have not, to our
knowledge, been reproduced to similar precision in subsequent
experiments.

The $q$ coefficients for \ion{Al}{ii}, a two-valence-electron ion,
have been calculated using a CI+MBPT method, similar to that described
in \Aref{app:a}, by \citet{Angstmann:2004:014102} and
\citet{Savukov:2008:042501}, with consistent results. We adopt the
value of $q$ from the latter in \Tref{tab:Al} and the quoted error was
derived from the spread in the two theoretical values.

\subsection{Silicon}\label{ssec:Si}

\begin{table*}
\begin{center}
\caption{
Laboratory data for transitions of Si of interest for quasar absorption-line varying-$\alpha$ studies described in \Sref{ssec:Si}. See \Sref{ssec:tabs} for full descriptions of each column.
}
\label{tab:Si}\vspace{-0.5em}
{\footnotesize
\begin{tabular}{:l;l;c;l;c;l;c;l;l;c;c;c;c}\hline
\multicolumn{1}{c}{Ion}&
\multicolumn{1}{c}{Tran.}&
\multicolumn{1}{c}{$A$}&
\multicolumn{1}{c}{$\omega_0$}&
\multicolumn{1}{c}{$X$}&
\multicolumn{1}{c}{$\lambda_0$}&
\multicolumn{1}{c}{$\delta v$}&
\multicolumn{1}{c}{Lower state}&
\multicolumn{1}{c}{Upper state}&
\multicolumn{1}{c}{ID}&
\multicolumn{1}{c}{IP$^-$, IP$^+$}&
\multicolumn{1}{c}{$f$ or {\it \%}}&
\multicolumn{1}{c}{$q$}\\
&
&
&
\multicolumn{1}{c}{[cm$^{-1}$]}&
&
\multicolumn{1}{c}{[\AA]}&
\multicolumn{1}{c}{[m\,s$^{-1}$]}&
&
&
&
\multicolumn{1}{c}{[eV]}&
&
\multicolumn{1}{c}{[cm$^{-1}$]}\\
\hline
                    Si{\sc \,ii}  & 1526   & 28.0855   & 65500.4538(7)$^{a}$              & 0 &   1526.706980(16)  &  3.2 & $3\rm{s}^23\rm{p}~^2\rm{P}_{1/2}^{\rm o} $ & $3\rm{s}^24\rm{s}~^2\rm{S}_{1/2}         $ & Si$^2_{2}$  & 8.15, 16.35  & 0.133     & $   47^{b}(4)  $\\
\rowstyle{\itshape}               &        & 30        & 65500.42585$^{b}$                & 3 &  1526.7076312      &      & $                                        $ & $                                        $ &             &              & 3.087\%   & $     ^{}     $\\
\rowstyle{\itshape}               &        & 29        & 65500.43980$^{b}$                & 3 &  1526.7073061      &      & $                                        $ & $                                        $ &             &              & 4.683\%   & $     ^{}     $\\
\rowstyle{\itshape}               &        & 28        & 65500.45545$^{b}$                & 3 &  1526.7069415      &      & $                                        $ & $                                        $ &             &              & 92.230\%  & $     ^{}     $\\
                                  & 1808   & 28.0855   & 55309.3404(4)$^{a}$              & 0 &   1808.012883(13)  &  2.2 & $                                        $ & $3\rm{s}3\rm{p}^2~^2\rm{D}_{3/2}         $ & Si$^2_{4}$  &              & 0.00208   & $  526^{b}(16) $\\
\rowstyle{\itshape}               &        & 30        & 55309.49010$^{b}$                & 3 &  1808.0079895      &      & $                                        $ & $                                        $ &             &              & 3.087\%   & $     ^{}     $\\
\rowstyle{\itshape}               &        & 29        & 55309.41403$^{b}$                & 3 &  1808.0104763      &      & $                                        $ & $                                        $ &             &              & 4.683\%   & $     ^{}     $\\
\rowstyle{\itshape}               &        & 28        & 55309.33165$^{b}$                & 3 &  1808.0131691      &      & $                                        $ & $                                        $ &             &              & 92.230\%  & $     ^{}     $\\
                    Si{\sc \,iv}  & 1393   & 28.0855   & 71748.355(2)$^{a}$               & 0 &   1393.760177(39)  &  8.4 & $2\rm{p}^63\rm{s}~^2\rm{S}_{1/2}         $ & $2\rm{p}^63\rm{p}~^2\rm{P}_{3/2}^{\rm o} $ & Si$^4_{1}$  & 33.49, 45.14 & 0.513     & $  823^{c}(2)  $\\
\rowstyle{\itshape}               &        & 30        & 71748.5517$^{d}$                 & 3 &  1393.7563570      &      & $                                        $ & $                                        $ &             &              & 3.087\%   & $     ^{}     $\\
\rowstyle{\itshape}               &        & 29        & 71748.4513$^{d}$                 & 3 &  1393.7583075      &      & $                                        $ & $                                        $ &             &              & 4.683\%   & $     ^{}     $\\
\rowstyle{\itshape}               &        & 28        & 71748.3435$^{d}$                 & 3 &  1393.7604003      &      & $                                        $ & $                                        $ &             &              & 92.230\%  & $     ^{}     $\\
                                  & 1402   & 28.0855   & 71287.376(2)$^{a}$               & 0 &   1402.772912(39)  &  8.4 & $                                        $ & $2\rm{p}^63\rm{p}~^2\rm{P}_{1/2}^{\rm o} $ & Si$^4_{2}$  &              & 0.254     & $  361^{c}(2)  $\\
\rowstyle{\itshape}               &        & 30        & 71287.5743$^{d}$                 & 3 &  1402.7690089      &      & $                                        $ & $                                        $ &             &              & 3.087\%   & $     ^{}     $\\
\rowstyle{\itshape}               &        & 29        & 71287.4731$^{d}$                 & 3 &  1402.7710015      &      & $                                        $ & $                                        $ &             &              & 4.683\%   & $     ^{}     $\\
\rowstyle{\itshape}               &        & 28        & 71287.3644$^{d}$                 & 3 &  1402.7731393      &      & $                                        $ & $                                        $ &             &              & 92.230\%  & $     ^{}     $\\
\hline
\end{tabular}
}
{\footnotesize References:
$^{a}$\citet{Griesmann:2000:L113};
$^{b}$This work;
$^{c}$\citet{Dzuba:2007:062510};
$^{d}$\citet{Berengut:2003:022502}.}
\end{center}
\end{table*}

\begin{figure}
\begin{center}
\includegraphics[width=0.95\columnwidth,bb = 18 322 411 706]{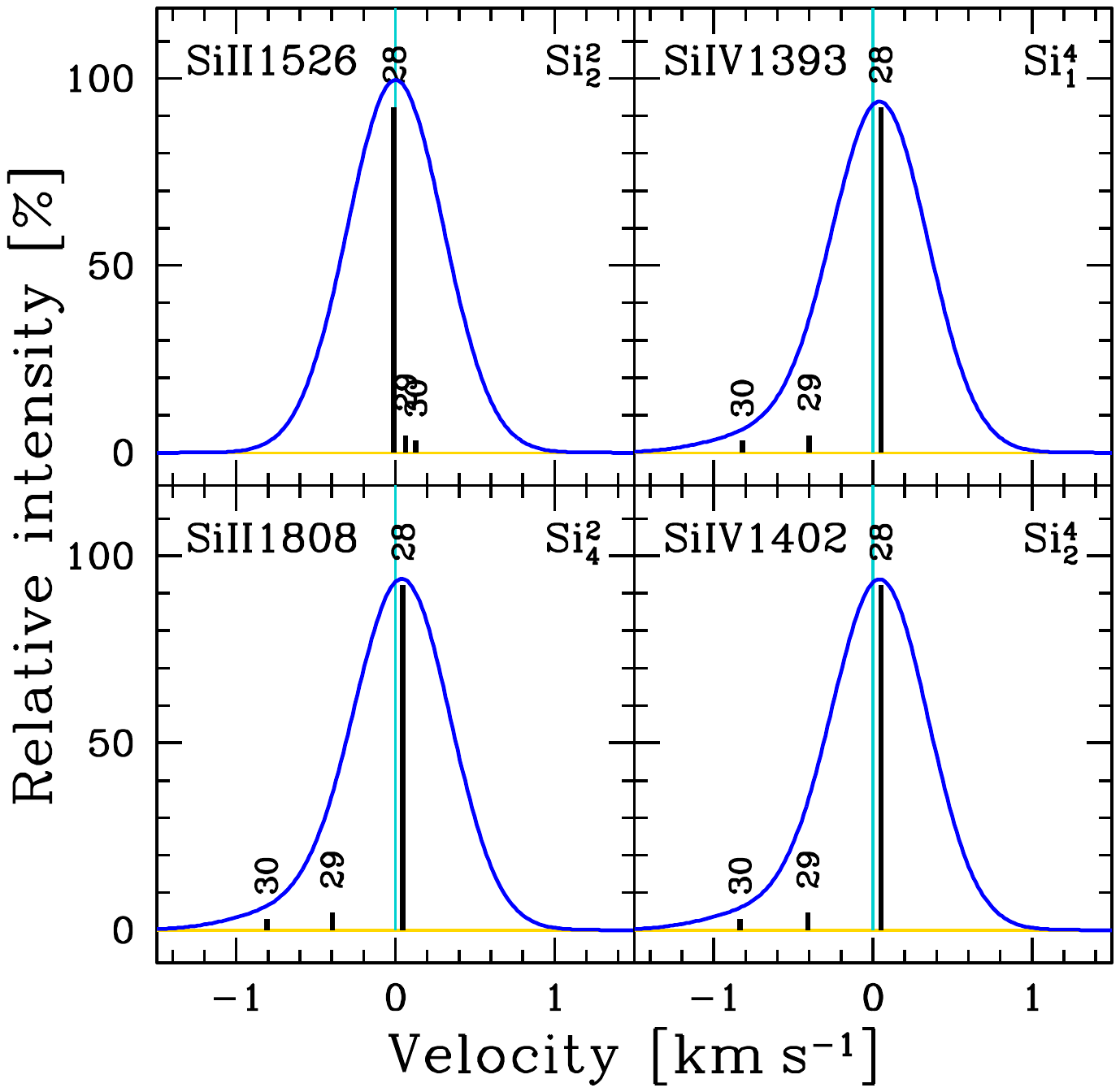}\vspace{-0.5em}
\caption{Si transition isotopic structures in \Tref{tab:Si} described
  in \Sref{ssec:Si}. See \Sref{ssec:tabs} for Figure details.}
\label{fig:Si_IHF}
\end{center}
\end{figure}

\Tref{tab:Si} summarises the atomic data for the two Si{\sc \,ii}
transitions with precisely-measured laboratory wavelengths and the
Si{\sc \,iv} doublet which is very prominent in quasar absorption
spectra. \Fref{fig:Si_IHF} illustrates the isotopic structures
for these transitions.

While many ground-state transitions of Si{\sc \,ii} are known in the
far UV, only 4 of the detectably-strong ones occur redwards of H{\sc
  \,i} Lyman-$\alpha$ and are, therefore, useful for varying-$\alpha$
studies. However, the two bluest Si{\sc \,ii} transitions,
$\lambda$1260 and $\lambda$1304, do not have precisely-measured
laboratory wavelengths (to our knowledge). The two remaining useful
transitions, Si{\sc \,ii} $\lambda$1526 and $\lambda$1808, form a very
useful pair for varying-$\alpha$ studies due to their very different
oscillator strengths: when Si{\sc \,ii} $\lambda$1526 begins to
saturate, $\lambda$1808 becomes detectable, thereby ensuring that a
useful ``anchor'' line (small $q$-coefficient) is nearly always
available in the far UV.

Si{\sc \,ii} and {\sc iv} likely arise in different astrophysical
environments: given the ionization potentials shown in \Tref{tab:Si},
Si{\sc \,ii} will be the dominant ion in self-shielded, predominantly
neutral gas while Si{\sc \,iv} will be much more prominent in more
diffuse, fairly highly-ionized gas. The highest optical depth velocity
components in a Si{\sc \,ii} absorption profile are therefore unlikely
to also have relatively high optical depth in Si{\sc \,iv}. They may
not even be detectable. Therefore, Si{\sc \,iv} absorption lines
generally cannot be fit with the same velocity structure as the many
singly-ionized species in varying-$\alpha$ studies. Instead, the
Si{\sc \,iv} doublet is used in the ``alkali doublet method'' for
constraining variations in $\alpha$
\citep[e.g.][]{Cowie:1995:596,Varshalovich:1996:6,Murphy:2001:1237}. The
ubiquity of the Si{\sc \,iv} doublet transitions and the relatively
large difference between their $q$-coefficients makes this is one of
the best doublets for this purpose.

The laboratory wavenumbers for the Si{\sc \,ii} and {\sc iv}
transitions in \Tref{tab:Si} were measured by
\citet{Griesmann:2000:L113} using the same FTS and Penning discharge
lamp setup and scans in which they measured Al{\sc \,ii}
$\lambda$1670. That is, the Si{\sc \,ii}/{\sc iv} and Al{\sc \,ii}
$\lambda$1670 wavenumbers are all on the same wavenumber scale which
was calibrated with the \citet{Whaling:1995:1} Ar{\sc \,ii}
measurements. The velocity precision achieved is nevertheless better
than for Al{\sc \,ii} $\lambda$1670: $\sim3$\,\ms\ for the Si{\sc
  \,ii} transitions and $\sim8$\,\ms\ even for the highly excited
Si{\sc \,iv} lines. We are not aware of any subsequent,
similarly-precise measurements of these Si{\sc \,ii} and {\sc iv}
transitions.

There exist no measurements of the isotopic structures for the
\ion{Si}{ii} and {\sc iv} transitions in \Tref{tab:Si}, to our
knowledge. \citet{Berengut:2003:022502} calculated the specific mass
shift (\kSMS) for \ion{Si}{ii} $\lambda$1526 and the \ion{Si}{iv}
doublet, finding that for the former it opposed the normal mass shift
(\kNMS), nearly cancelling it out. Therefore, it is important to
estimate the field shift (\FS) to better understand the likely
isotopic structure of \ion{Si}{ii} $\lambda$1526. We are not aware of
a previous calculation of \ion{Si}{ii} $\lambda$1808's isotopic
structure. As a rough approximation, previous varying-$\alpha$ studies
\citep[e.g.][]{Murphy:2001:1208,Murphy:2003:609} estimated it by
scaling the specific mass shift of the \ion{Mg}{ii} doublet. By
comparison, \ion{Si}{iv}'s isotopic structure is simpler to calculate,
with the expectation that the results are more reliable. Thus, we have
used the results of \citet{Berengut:2003:022502} for the \ion{Si}{iv}
doublet in \Tref{tab:Si}. Note the similarity of the \ion{Si}{iv}
structures in \Fref{fig:Si_IHF}, implying that alkali doublet method
varying-$\alpha$ constraints using it are very insensitive to isotopic
abundance variations between the laboratory and quasar absorption
clouds.

To remedy the situation for \ion{Si}{ii} we have undertaken new
\emph{ab initio} calculations of isotopic structures for a variety of
its transitions in \Aref{app:a}. The same calculations also yield new
$q$ coefficients, derived with a similar and consistent approach as
that used for the $q$ coefficients of the other species studied
here. The detailed results for \ion{Si}{ii} are presented in
\Tref{tab:SiII_final} and the new isotopic structures and $q$
coefficients are provided in \Tref{tab:Si} and illustrated in
\Fref{fig:Si_IHF}. These calculations confirm that the field shift for
\ion{Si}{ii} $\lambda$1526 is a relatively small effect, leaving the
expected near-cancelation between the normal and specific mass shifts
to provide a very narrow isotopic splitting. The \ion{Si}{ii}
$\lambda$1808 isotopic splitting, $c\delta\nu^{30,28}/\nu=0.86$\,\kms\
is also very similar to that assumed by, e.g.,
\citet{Murphy:2003:609}, $0.63$\,\kms.

\subsection{Calcium}\label{ssec:Ca}
\begin{table*}
\begin{center}
\caption{
Laboratory data for transitions of Ca of interest for quasar absorption-line varying-$\alpha$ studies described in \Sref{ssec:Ca}. See \Sref{ssec:tabs} for full descriptions of each column.
}
\label{tab:Ca}\vspace{-0.5em}
{\footnotesize
\begin{tabular}{:l;l;c;l;c;l;c;l;l;c;c;c;c}\hline
\multicolumn{1}{c}{Ion}&
\multicolumn{1}{c}{Tran.}&
\multicolumn{1}{c}{$A$}&
\multicolumn{1}{c}{$\omega_0$}&
\multicolumn{1}{c}{$X$}&
\multicolumn{1}{c}{$\lambda_0$}&
\multicolumn{1}{c}{$\delta v$}&
\multicolumn{1}{c}{Lower state}&
\multicolumn{1}{c}{Upper state}&
\multicolumn{1}{c}{ID}&
\multicolumn{1}{c}{IP$^-$, IP$^+$}&
\multicolumn{1}{c}{$f$ or {\it \%}}&
\multicolumn{1}{c}{$q$}\\
&
&
&
\multicolumn{1}{c}{[cm$^{-1}$]}&
&
\multicolumn{1}{c}{[\AA]}&
\multicolumn{1}{c}{[m\,s$^{-1}$]}&
&
&
&
\multicolumn{1}{c}{[eV]}&
&
\multicolumn{1}{c}{[cm$^{-1}$]}\\
\hline
                    Ca{\sc \,ii}  & 3934   & 40.0780   & 25414.416438(86)$^{}$            & 1 &   3934.774589(13)  &  1.0 & $4\rm{s}~^2\rm{S}_{1/2}                  $ & $4\rm{p}~^2\rm{P}_{3/2}^{\rm o}          $ & Ca$^2_{1}$  & 6.11, 11.87  & 0.6267    & $  446^{a}(6)  $\\
\rowstyle{\itshape}               &        & 48        & 25414.4735(42)$^{b,c,d}$         & 2 &    3934.76575(65)  &   49 & $                                        $ & $                                        $ &             &              & 0.187\%   & $     ^{}     $\\
\rowstyle{\itshape}               &        & 46        & 25414.4583(33)$^{b,c,d}$         & 2 &    3934.76811(51)  &   38 & $                                        $ & $                                        $ &             &              & 0.004\%   & $     ^{}     $\\
\rowstyle{\itshape}               &        & 44        & 25414.4436(23)$^{b,c,d}$         & 2 &    3934.77038(35)  &   27 & $                                        $ & $                                        $ &             &              & 2.086\%   & $     ^{}     $\\
\rowstyle{\itshape}               &        & 43        & 25414.38165(73)$^{e,b,c,f}$      & 2 &    3934.77998(11)  &  8.6 & $F=3                                     $ & $F=2,3,4                                 $ &             &              & 0.059\%   & $     ^{}     $\\
\rowstyle{\itshape}               &        & 43        & 25414.48235(73)$^{e,b,c,f}$      & 2 &    3934.76438(11)  &  8.6 & $F=4                                     $ & $F=3,4,5                                 $ &             &              & 0.076\%   & $     ^{}     $\\
\rowstyle{\itshape}               &        & 42        & 25414.4301(12)$^{b,c,d}$         & 2 &    3934.77248(18)  &   14 & $                                        $ & $                                        $ &             &              & 0.647\%   & $     ^{}     $\\
\rowstyle{\itshape}               &        & 40        & 25414.415619(17)$^{d}$           & 0 &  3934.7747160(26)  & 0.20 & $                                        $ & $                                        $ &             &              & 96.941\%  & $     ^{}     $\\
                                  & 3969   & 40.0780   & 25191.520729(66)$^{}$            & 1 &   3969.589652(10)  & 0.79 & $                                        $ & $4\rm{p}~^2\rm{P}_{1/2}^{\rm o}          $ & Ca$^2_{2}$  &              & 0.3116    & $  222^{a}(2)  $\\
\rowstyle{\itshape}               &        & 48        & 25191.5794(42)$^{b,c,g}$         & 2 &    3969.58041(66)  &   50 & $                                        $ & $                                        $ &             &              & 0.187\%   & $     ^{}     $\\
\rowstyle{\itshape}               &        & 46        & 25191.5652(33)$^{b,c,g}$         & 2 &    3969.58264(51)  &   39 & $                                        $ & $                                        $ &             &              & 0.004\%   & $     ^{}     $\\
\rowstyle{\itshape}               &        & 44        & 25191.5496(23)$^{b,c,g}$         & 2 &    3969.58510(36)  &   27 & $                                        $ & $                                        $ &             &              & 2.086\%   & $     ^{}     $\\
\rowstyle{\itshape}               &        & 43        & 25191.47818(46)$^{e,b,c,f}$      & 2 &    3969.59636(7)   &  5.5 & $F=3                                     $ & $F=3,4                                   $ &             &              & 0.059\%   & $     ^{}     $\\
\rowstyle{\itshape}               &        & 43        & 25191.59225(46)$^{e,b,c,f}$      & 2 &    3969.57838(7)   &  5.5 & $F=4                                     $ & $F=3,4                                   $ &             &              & 0.076\%   & $     ^{}     $\\
\rowstyle{\itshape}               &        & 42        & 25191.5348(12)$^{b,c,g}$         & 2 &    3969.58743(19)  &   14 & $                                        $ & $                                        $ &             &              & 0.647\%   & $     ^{}     $\\
\rowstyle{\itshape}               &        & 40        & 25191.519868(57)$^{g}$           & 0 &  3969.5897875(89)  & 0.67 & $                                        $ & $                                        $ &             &              & 96.941\%  & $     ^{}     $\\
\hline
\end{tabular}
}
{\footnotesize References:
$^{a}$\citet{Dzuba:2007:062510};
$^{b}$\citet{Maleki:1992:524};
$^{c}$\citet{Martensson-Pendrill:1992:4675};
$^{d}$\citet{Wolf:2009:223901};
$^{e}$\citet{Arbes:1994:27};
$^{f}$\citet{Nortershauser:1998:33};
$^{g}$\citet{Wolf:2008:032511}.}
\end{center}
\end{table*}

\begin{figure}
\begin{center}
\includegraphics[width=0.55\columnwidth,bb = 18 322 242 706]{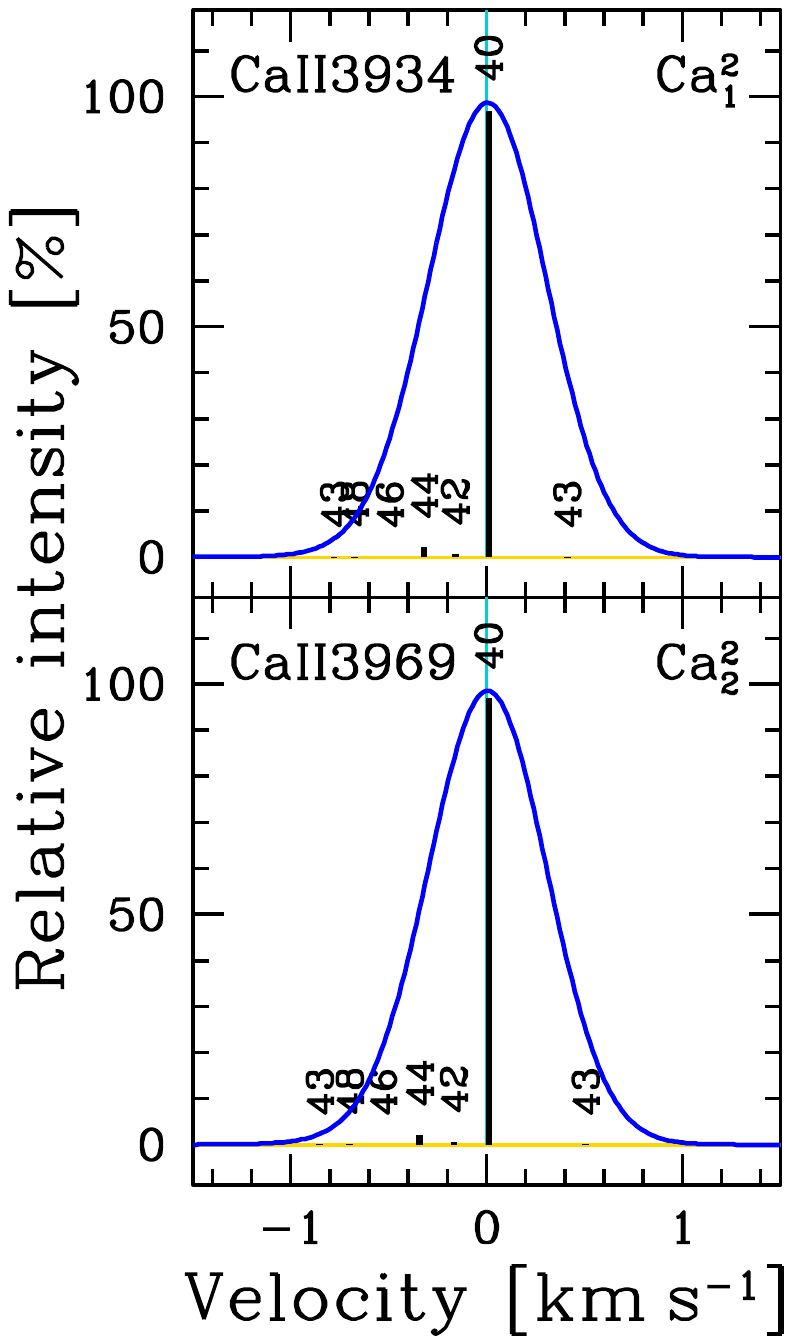}\vspace{-0.5em}
\caption{Ca transition isotopic and hyperfine structures in
  \Tref{tab:Ca} described in \Sref{ssec:Ca}. See \Sref{ssec:tabs} for
  Figure details.}
\label{fig:Ca_IHF}
\end{center}
\end{figure}

\Tref{tab:Ca} shows the atomic data for the two strongest \ion{Ca}{ii}
transitions which are common, though often not particularly strong, in
quasar absorption systems
\citep[e.g.][]{Zych:2009:1429}. \Fref{fig:Ca_IHF} illustrates the
isotopic and hyperfine structures for these transitions. The
\ion{Ca}{ii} $\lambda\lambda$3934/3969 doublet, or K and H transitions
are observable at redshifts $z\la1.5$ in optical quasar absorption
spectra. However, like the \ion{Na}{i} $\lambda\lambda$5891/5897
doublet, they have not been used so far in varying-$\alpha$ studies
(to our knowledge) for the same two main reasons: weak absorption
compared to the ubiquitous Mg and \ion{Fe}{ii} transitions at the same
$z$ and, often, significant telluric contamination. They are
nevertheless $\ga$100 times stronger than other known \ion{Ca}{ii}
transitions \citep{Morton:2003:205}, so they are the only ones
suitable for varying-$\alpha$ work. Ca has 6 stable isotopes,
$^{40-48}$Ca, only one of which has an odd nucleon number (43). All
the isotopic splittings, and the significant hyperfine splittings for
transitions of the latter, have been measured with adequate precision
for high-resolution quasar absorption studies (see below).

Absolute frequencies were recently measured for the
$^{40}$\ion{Ca}{ii} H and K lines by \citet{Wolf:2008:032511} and
\citet{Wolf:2009:223901}, respectively, using frequency-comb
spectroscopy techniques. The isotopic splittings were measured by
\citet{Martensson-Pendrill:1992:4675} and \citet{Maleki:1992:524}. We
averaged the two measurements for each isotope and transition,
weighting the average by the inverse-square uncertainties
\citep[including the 2\,MHz per mass unit difference from $^{44}$Ca
systematic error quoted by][]{Martensson-Pendrill:1992:4675}. The
magnetic dipole hyperfine constant ($A$) for the ground state was
measured by \citet{Arbes:1994:27}, while \citet{Nortershauser:1998:33}
measured the $A$ values for the H and K excited states and also the
electric quadrupole hyperfine constant ($B$) for the $\lambda$3934
excited state. The hyperfine structure was then determined using
equations (\ref{eq:hyp1})--(\ref{eq:hyp5}). In \Tref{tab:Ca} we
simplify the hyperfine structure by averaging over the splitting of
the excited states. This is justified for applications in
high-resolution spectroscopy because the excited state splitting
($\sim$200\,\ms) is much smaller than the ground-state splitting
($\sim$1.5\,\kms) and the isotopic abundance of $^{43}$Ca is very
small, just 0.14\,\%. Finally, the composite frequencies were
determined from the isotopic and hyperfine component frequencies via
\Eref{eq:comp}.

Despite the isotopic/hyperfine structures for the \ion{Ca}{ii} H and K
lines illustrated in \Fref{fig:Na_IHF} being relatively broad
($\sim$1\,\kms), the final line shape remains almost symmetric and
centered very close ($\sim$10\,\kms) to the $^{40}$Ca isotopic
components because they have by-far the highest (terrestrial) isotopic
abundance \citep[96.9\,\%; ][]{Rosman:1998:1275}. However, if the
isotopic abundances of Ca are very different in high-redshift quasar
absorption systems, modelling of \ion{Ca}{ii} profiles with the
terrestrial abundances may lead to spurious line-shifts. We are not
aware of any theoretical work that provides basic expectations for Ca
in this regard, unlike the other species considered in this paper
\citep{Fenner:2005:468}.

\subsection{Titanium}\label{ssec:Ti}
\begin{table*}
\begin{center}
\caption{
Laboratory data for transitions of Ti of interest for quasar absorption-line varying-$\alpha$ studies described in \Sref{ssec:Ti}. See \Sref{ssec:tabs} for full descriptions of each column.
}
\label{tab:Ti}\vspace{-0.5em}
{\footnotesize
\begin{tabular}{:l;l;c;l;c;l;c;l;l;c;c;c;c}\hline
\multicolumn{1}{c}{Ion}&
\multicolumn{1}{c}{Tran.}&
\multicolumn{1}{c}{$A$}&
\multicolumn{1}{c}{$\omega_0$}&
\multicolumn{1}{c}{$X$}&
\multicolumn{1}{c}{$\lambda_0$}&
\multicolumn{1}{c}{$\delta v$}&
\multicolumn{1}{c}{Lower state}&
\multicolumn{1}{c}{Upper state}&
\multicolumn{1}{c}{ID}&
\multicolumn{1}{c}{IP$^-$, IP$^+$}&
\multicolumn{1}{c}{$f$ or {\it \%}}&
\multicolumn{1}{c}{$q$}\\
&
&
&
\multicolumn{1}{c}{[cm$^{-1}$]}&
&
\multicolumn{1}{c}{[\AA]}&
\multicolumn{1}{c}{[m\,s$^{-1}$]}&
&
&
&
\multicolumn{1}{c}{[eV]}&
&
\multicolumn{1}{c}{[cm$^{-1}$]}\\
\hline
                    Ti{\sc \,ii}  & 1910.6 & 47.867    & 52339.240(1)$^{a}$               & 0 &    1910.61238(4)   &  5.7 & $3\rm{d}^24\rm{s}~a^4\rm{F}_{3/2}        $ & $3\rm{d}4\rm{s}4\rm{p}^4\rm{D}_{1/2}^{\rm o}$ & Ti$^2_{4}$  & 6.82, 13.58  & 0.104     & $-1414^{b}(170)$\\
\rowstyle{\itshape}               &        & 50        & 52339.1552$^{b}$                 & 3 &   1910.615477      &      & $                                        $ & $                                        $ &             &              & 5.18\%    & $     ^{}     $\\
\rowstyle{\itshape}               &        & 49        & 52339.1943$^{b}$                 & 3 &   1910.614051      &      & $                                        $ & $                                        $ &             &              & 5.41\%    & $     ^{}     $\\
\rowstyle{\itshape}               &        & 48        & 52339.2361$^{b}$                 & 3 &   1910.612524      &      & $                                        $ & $                                        $ &             &              & 73.72\%   & $     ^{}     $\\
\rowstyle{\itshape}               &        & 47        & 52339.2785$^{b}$                 & 3 &   1910.610976      &      & $                                        $ & $                                        $ &             &              & 7.44\%    & $     ^{}     $\\
\rowstyle{\itshape}               &        & 46        & 52339.3233$^{b}$                 & 3 &   1910.609339      &      & $                                        $ & $                                        $ &             &              & 8.25\%    & $     ^{}     $\\
                                  & 1910.9 & 47.867    & 52329.889(1)$^{a}$               & 0 &    1910.95380(4)   &  5.7 & $                                        $ & $3\rm{d}4\rm{s}4\rm{p}^4\rm{F}_{3/2}^{\rm o}$ & Ti$^2_{5}$  &              & 0.0980    & $-1689^{b}(250)$\\
\rowstyle{\itshape}               &        & 50        & 52329.8057$^{b}$                 & 3 &   1910.956836      &      & $                                        $ & $                                        $ &             &              & 5.18\%    & $     ^{}     $\\
\rowstyle{\itshape}               &        & 49        & 52329.8441$^{b}$                 & 3 &   1910.955436      &      & $                                        $ & $                                        $ &             &              & 5.41\%    & $     ^{}     $\\
\rowstyle{\itshape}               &        & 48        & 52329.8852$^{b}$                 & 3 &   1910.953935      &      & $                                        $ & $                                        $ &             &              & 73.72\%   & $     ^{}     $\\
\rowstyle{\itshape}               &        & 47        & 52329.9268$^{b}$                 & 3 &   1910.952416      &      & $                                        $ & $                                        $ &             &              & 7.44\%    & $     ^{}     $\\
\rowstyle{\itshape}               &        & 46        & 52329.9708$^{b}$                 & 3 &   1910.950808      &      & $                                        $ & $                                        $ &             &              & 8.25\%    & $     ^{}     $\\
                                  & 3067   & 47.867    & 32602.6283(14)$^{c,a,d}$         & 0 &    3067.23737(13)  & 13.0 & $                                        $ & $3\rm{d}^24\rm{p~z}^4\rm{D}_{3/2}^{\rm o}$ & Ti$^2_{7}$  &              & 0.0489    & $  855^{b}(80) $\\
\rowstyle{\itshape}               &        & 50        & 32602.65258$^{b}$                & 3 &   3067.235089      &      & $                                        $ & $                                        $ &             &              & 5.18\%    & $     ^{}     $\\
\rowstyle{\itshape}               &        & 49        & 32602.64187$^{b}$                & 3 &   3067.236097      &      & $                                        $ & $                                        $ &             &              & 5.41\%    & $     ^{}     $\\
\rowstyle{\itshape}               &        & 48        & 32602.62930$^{b}$                & 3 &   3067.237279      &      & $                                        $ & $                                        $ &             &              & 73.72\%   & $     ^{}     $\\
\rowstyle{\itshape}               &        & 47        & 32602.61768$^{b}$                & 3 &   3067.238373      &      & $                                        $ & $                                        $ &             &              & 7.44\%    & $     ^{}     $\\
\rowstyle{\itshape}               &        & 46        & 32602.60483$^{b}$                & 3 &   3067.239582      &      & $                                        $ & $                                        $ &             &              & 8.25\%    & $     ^{}     $\\
                                  & 3073   & 47.867    & 32532.3566(7)$^{c,a,d}$          & 0 &    3073.86278(7)   &  6.5 & $                                        $ & $3\rm{d}^24\rm{p~z}^4\rm{D}_{1/2}^{\rm o}$ & Ti$^2_{3}$  &              & 0.121     & $  739^{b}(80) $\\
\rowstyle{\itshape}               &        & 50        & 32532.38075$^{b}$                & 3 &   3073.860495      &      & $                                        $ & $                                        $ &             &              & 5.18\%    & $     ^{}     $\\
\rowstyle{\itshape}               &        & 49        & 32532.37011$^{b}$                & 3 &   3073.861501      &      & $                                        $ & $                                        $ &             &              & 5.41\%    & $     ^{}     $\\
\rowstyle{\itshape}               &        & 48        & 32532.35759$^{b}$                & 3 &   3073.862683      &      & $                                        $ & $                                        $ &             &              & 73.72\%   & $     ^{}     $\\
\rowstyle{\itshape}               &        & 47        & 32532.34604$^{b}$                & 3 &   3073.863774      &      & $                                        $ & $                                        $ &             &              & 7.44\%    & $     ^{}     $\\
\rowstyle{\itshape}               &        & 46        & 32532.33325$^{b}$                & 3 &   3073.864983      &      & $                                        $ & $                                        $ &             &              & 8.25\%    & $     ^{}     $\\
                                  & 3230   & 47.867    & 30958.5871(10)$^{c,d}$           & 0 &    3230.12157(10)  &  9.7 & $                                        $ & $3\rm{d}^24\rm{p~z}^4\rm{F}_{5/2}^{\rm o}$ & Ti$^2_{6}$  &              & 0.0687    & $  721^{b}(70) $\\
\rowstyle{\itshape}               &        & 50        & 30958.61173$^{b}$                & 3 &   3230.119002      &      & $                                        $ & $                                        $ &             &              & 5.18\%    & $     ^{}     $\\
\rowstyle{\itshape}               &        & 49        & 30958.60088$^{b}$                & 3 &   3230.120133      &      & $                                        $ & $                                        $ &             &              & 5.41\%    & $     ^{}     $\\
\rowstyle{\itshape}               &        & 48        & 30958.58816$^{b}$                & 3 &   3230.121461      &      & $                                        $ & $                                        $ &             &              & 73.72\%   & $     ^{}     $\\
\rowstyle{\itshape}               &        & 47        & 30958.57639$^{b}$                & 3 &   3230.122688      &      & $                                        $ & $                                        $ &             &              & 7.44\%    & $     ^{}     $\\
\rowstyle{\itshape}               &        & 46        & 30958.56338$^{b}$                & 3 &   3230.124046      &      & $                                        $ & $                                        $ &             &              & 8.25\%    & $     ^{}     $\\
                                  & 3242   & 47.867    & 30836.4271(10)$^{c,d}$           & 0 &    3242.91785(11)  &  9.7 & $                                        $ & $3\rm{d}^24\rm{p~z}^4\rm{F}_{3/2}^{\rm o}$ & Ti$^2_{2}$  &              & 0.232     & $  563^{b}(30) $\\
\rowstyle{\itshape}               &        & 50        & 30836.45206$^{b}$                & 3 &   3242.915229      &      & $                                        $ & $                                        $ &             &              & 5.18\%    & $     ^{}     $\\
\rowstyle{\itshape}               &        & 49        & 30836.44106$^{b}$                & 3 &   3242.916386      &      & $                                        $ & $                                        $ &             &              & 5.41\%    & $     ^{}     $\\
\rowstyle{\itshape}               &        & 48        & 30836.42817$^{b}$                & 3 &   3242.917742      &      & $                                        $ & $                                        $ &             &              & 73.72\%   & $     ^{}     $\\
\rowstyle{\itshape}               &        & 47        & 30836.41623$^{b}$                & 3 &   3242.918997      &      & $                                        $ & $                                        $ &             &              & 7.44\%    & $     ^{}     $\\
\rowstyle{\itshape}               &        & 46        & 30836.40304$^{b}$                & 3 &   3242.920385      &      & $                                        $ & $                                        $ &             &              & 8.25\%    & $     ^{}     $\\
                                  & 3384   & 47.867    & 29544.4551(10)$^{c,d}$           & 0 &    3384.72988(11)  & 10.1 & $                                        $ & $3\rm{d}^24\rm{p~z}^4\rm{G}_{5/2}^{\rm o}$ & Ti$^2_{1}$  &              & 0.358     & $  408^{b}(30) $\\
\rowstyle{\itshape}               &        & 50        & 29544.48132$^{b}$                & 3 &   3384.726877      &      & $                                        $ & $                                        $ &             &              & 5.18\%    & $     ^{}     $\\
\rowstyle{\itshape}               &        & 49        & 29544.46970$^{b}$                & 3 &   3384.728208      &      & $                                        $ & $                                        $ &             &              & 5.41\%    & $     ^{}     $\\
\rowstyle{\itshape}               &        & 48        & 29544.45618$^{b}$                & 3 &   3384.729757      &      & $                                        $ & $                                        $ &             &              & 73.72\%   & $     ^{}     $\\
\rowstyle{\itshape}               &        & 47        & 29544.44358$^{b}$                & 3 &   3384.731201      &      & $                                        $ & $                                        $ &             &              & 7.44\%    & $     ^{}     $\\
\rowstyle{\itshape}               &        & 46        & 29544.42969$^{b}$                & 3 &   3384.732792      &      & $                                        $ & $                                        $ &             &              & 8.25\%    & $     ^{}     $\\
\hline
\end{tabular}
}
{\footnotesize References:
$^{a}$\citet{Ruffoni:2010:424};
$^{b}$This work;
$^{c}$\citet{Aldenius:2009:014008};
$^{d}$\citet{Nave:2012:1570}.}
\end{center}
\end{table*}

\begin{figure*}
\begin{center}
\includegraphics[width=0.60\textwidth,bb = 18 153 580 706]{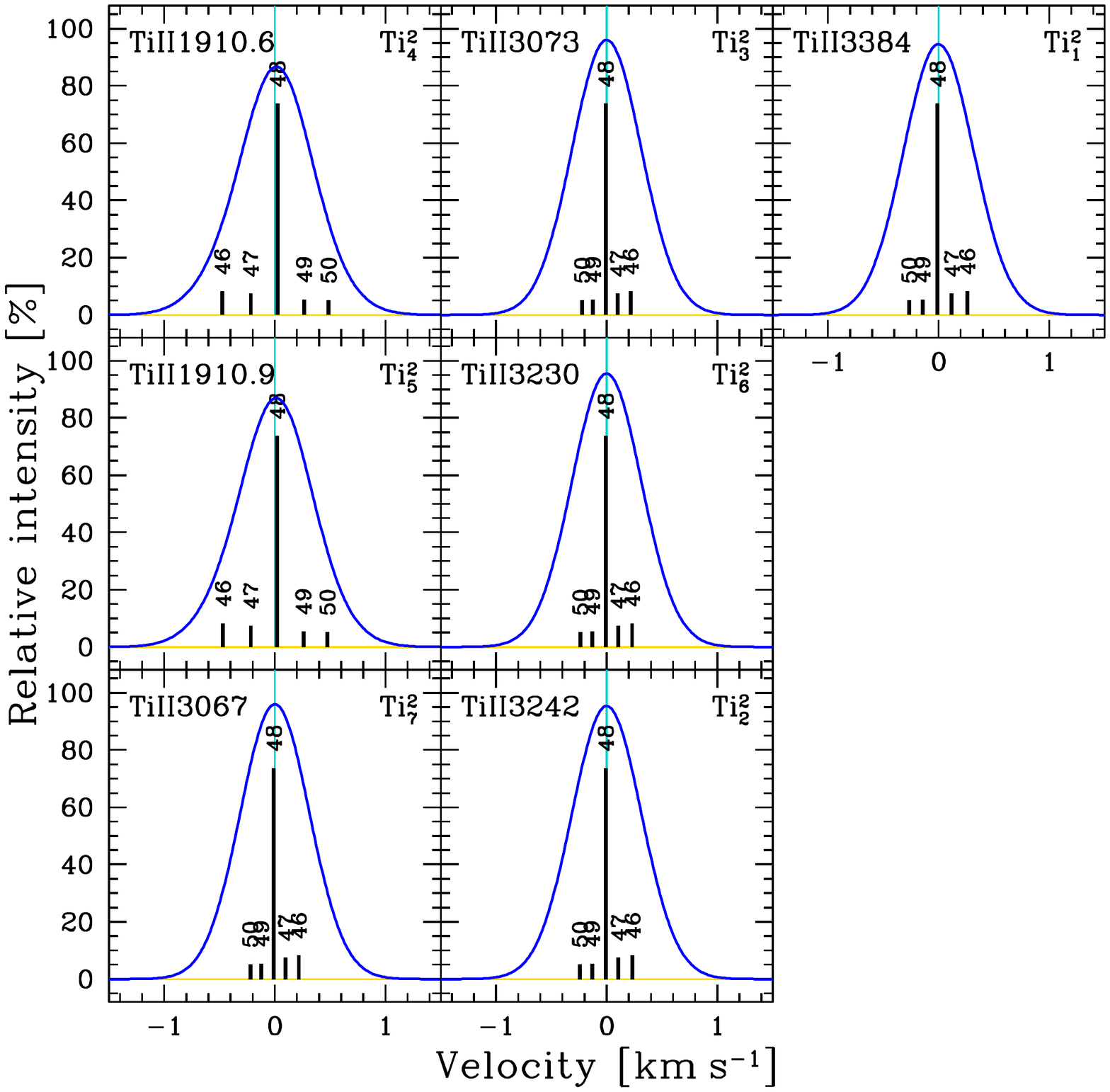}\vspace{-0.5em}
\caption{Ti transition isotopic structures in \Tref{tab:Ti} described
  in \Sref{ssec:Ti}. See \Sref{ssec:tabs} for Figure details.}
\label{fig:Ti_IHF}
\end{center}
\end{figure*}

\Tref{tab:Ti} summarises the atomic data for the seven Ti{\sc \,ii}
transitions with precisely-measured laboratory wavelengths and
\Fref{fig:Ti_IHF} illustrates their isotopic structures. To our
knowledge, Ti{\sc \,ii} has not yet been utilized in varying-$\alpha$
studies, mainly because of two factors: the frequencies of the
transitions in \Tref{tab:Ti} have only recently been measured
precisely enough \citep{Aldenius:2006:444,Ruffoni:2010:424} and Ti has
a low abundance in general, some 2.5 orders of magnitude below that of
Fe \citep[e.g.][]{Asplund:2009:481}, so it is only detected in quasar
absorbers with the strongest metal absorption, and those are the
rarest. Nevertheless, the Ti{\sc \,ii} transitions in \Tref{tab:Ti}
are a particularly interesting combination for varying-$\alpha$
studies because they have similar oscillator strengths and, most
importantly, the pair near 1910\,\AA\ have large, negative
$q$-coefficients whereas the 5 transitions redwards of 3000\,\AA\
have moderate, positive $q$-coefficients. That is, within the same
ion, transitions are available (with precisely measured laboratory
frequencies) which shift in opposite directions to each other as
$\alpha$ varies. This property is shared only by Fe{\sc \,ii} -- see
section \ref{ssec:Fe}. However, aside from identifying absorption
systems with strong enough metal absorption to allow high-quality
spectra of the Ti{\sc \,ii} transitions to be obtained, another
challenge is that the negative- and positive-$q$ transitions lie so
far apart. This could mean that any long-range wavelength
miscalibrations of the quasar spectra may play an important role and
it limits the number of quasar absorbers where all the Ti{\sc \,ii}
transitions in \Tref{tab:Ti} can be detected outside the
Lyman-$\alpha$ forest.

The wavenumbers for the 5 Ti{\sc \,ii} transitions redwards of
3000\,\AA\ were measured with $\sim$10\,\ms\ precision by
\citet{Aldenius:2006:444} \citep[see also][]{Aldenius:2009:014008}
using a FTS and HCL. The wavenumbers were calibrated using the Ar{\sc
  \,ii} scale of \citet{Whaling:1995:1}. However, following the
comprehensive re-calibration analysis of \citet{Nave:2011:737}, the
wavenumbers of \citet{Aldenius:2009:014008} were increased by 3.7
parts per 10$^{8}$ by \citet{Nave:2012:1570}. For the Ti{\sc \,ii}
$\lambda$3230, $\lambda$3243 and $\lambda$3384 transitions, these
increased wavenumbers are the composite values reported in
\Tref{tab:Ti}.

The wavenumbers for the pair of Ti{\sc \,ii} transitions at 1910\,\AA\
and the $\lambda\lambda$3067/3073 doublet were measured by
\citet{Ruffoni:2010:424} with $\sim$10--20\,\ms\ precision using a FTS
and HCL. Again, the wavenumbers were calibrated using the Ar{\sc \,ii}
scale of \citet{Whaling:1995:1}. However, the Mg{\sc \,i}
$\lambda$2026 and $\lambda$2852 transitions discussed in section
\ref{ssec:Mg} were recorded simultaneously as calibration
standards. \citeauthor{Ruffoni:2010:424} found that their Ar{\sc
  \,ii}-calibrated Mg{\sc \,i} wavenumbers agreed well with the
frequency-comb based, absolute measurements of
\citet{Hannemann:2006:012505} and
\citet{Salumbides:2006:L41}. \citeauthor{Ruffoni:2010:424}'s Ti{\sc \,ii}
$\lambda\lambda$3067/3073 doublet wavenumbers also agree with the
corrected \citet{Aldenius:2009:014008} values reported by
\citet{Nave:2012:1570}. Therefore, we adopt a simple weighted mean
wavenumber for the composite values reported in \Tref{tab:Ti} for
the $\lambda\lambda$3067/3073 doublet. For the $\lambda$1910.6 and
$\lambda$1910.9 transitions, we adopt \citeauthor{Ruffoni:2010:424}'s
wavenumbers.

To our knowledge, there exists no measurements of the isotopic
structures for the \ion{Ti}{ii} transitions in
\Tref{tab:Ti}. \citet{Berengut:2008:235702} calculated the total
isotopic shifts for the 5 transitions redwards of 3000\,\AA\ but no
previous estimates are available for the pair of transitions at
1910\,\AA. We have therefore undertaken new \emph{ab initio} isotopic
structure and $q$ coefficient calculations for all 7 \ion{Ti}{ii}
transitions of interest in \Aref{app:a}. The detailed results are
presented in \Tref{tab:TiII_final} and the new isotopic structures and
$q$ coefficients are provided in \Tref{tab:Ti} and illustrated in
\Fref{fig:Ti_IHF}. The new calculations were very similar to those of
\citet{Berengut:2008:235702} and we find very similar results for the
5 transitions redwards of 3000\,\AA. For the pair of transitions at
1910\,\AA, we find that the isotopic structures are reversed, with the
heavier isotopes appearing at lower frequencies. This reversal
correlates with the reversed sign of the $q$ coefficients for these
transitions. This is very important for varying-$\alpha$ studies: even
if the isotopic structures in \Fref{fig:Ti_IHF} are fit to the
quasar absorption profiles, any significant deviations from the
terrestrial Ti isotopic abundances in the quasar absorbers will lead
to spurious line-shift measurements which will mimic a variation in
$\alpha$.

\subsection{Chromium}\label{ssec:Cr}

\begin{table*}
\begin{center}
\caption{
Laboratory data for transitions of Cr of interest for quasar absorption-line varying-$\alpha$ studies described in \Sref{ssec:Cr}. See \Sref{ssec:tabs} for full descriptions of each column.
}
\label{tab:Cr}\vspace{-0.5em}
{\footnotesize
\begin{tabular}{:l;l;c;l;c;l;c;l;l;c;c;c;c}\hline
\multicolumn{1}{c}{Ion}&
\multicolumn{1}{c}{Tran.}&
\multicolumn{1}{c}{$A$}&
\multicolumn{1}{c}{$\omega_0$}&
\multicolumn{1}{c}{$X$}&
\multicolumn{1}{c}{$\lambda_0$}&
\multicolumn{1}{c}{$\delta v$}&
\multicolumn{1}{c}{Lower state}&
\multicolumn{1}{c}{Upper state}&
\multicolumn{1}{c}{ID}&
\multicolumn{1}{c}{IP$^-$, IP$^+$}&
\multicolumn{1}{c}{$f$ or {\it \%}}&
\multicolumn{1}{c}{$q$}\\
&
&
&
\multicolumn{1}{c}{[cm$^{-1}$]}&
&
\multicolumn{1}{c}{[\AA]}&
\multicolumn{1}{c}{[m\,s$^{-1}$]}&
&
&
&
\multicolumn{1}{c}{[eV]}&
&
\multicolumn{1}{c}{[cm$^{-1}$]}\\
\hline
                    Cr{\sc \,ii}  & 2056   & 51.9961   & 48632.0597(14)$^{a,b,c}$         & 0 &   2056.256728(60)  &  8.7 & $3\rm{d}^5~^6\rm{S}_{5/2}                $ & $3\rm{d}^44\rm{p}~^6\rm{P}_{7/2}^{\rm o} $ & Cr$^2_{1}$  & 6.77, 16.50  & 0.103     & $-1061^{d}(70) $\\
\rowstyle{\itshape}               &        & 54        & 48631.98144$^{d}$                & 3 &  2056.2600379      &      & $                                        $ & $                                        $ &             &              & 2.365\%   & $     ^{}     $\\
\rowstyle{\itshape}               &        & 53        & 48632.02086$^{d}$                & 3 &  2056.2583712      &      & $                                        $ & $                                        $ &             &              & 9.501\%   & $     ^{}     $\\
\rowstyle{\itshape}               &        & 52        & 48632.06174$^{d}$                & 3 &  2056.2566427      &      & $                                        $ & $                                        $ &             &              & 83.789\%  & $     ^{}     $\\
\rowstyle{\itshape}               &        & 50        & 48632.14853$^{d}$                & 3 &  2056.2529732      &      & $                                        $ & $                                        $ &             &              & 4.345\%   & $     ^{}     $\\
                                  & 2062   & 51.9961   & 48491.0583(14)$^{a,b,c}$         & 0 &   2062.235873(60)  &  8.7 & $                                        $ & $3\rm{d}^44\rm{p}~^6\rm{P}_{5/2}^{\rm o} $ & Cr$^2_{2}$  &              & 0.0759    & $-1280^{d}(70) $\\
\rowstyle{\itshape}               &        & 54        & 48490.98199$^{d}$                & 3 &  2062.2391193      &      & $                                        $ & $                                        $ &             &              & 2.365\%   & $     ^{}     $\\
\rowstyle{\itshape}               &        & 53        & 48491.02049$^{d}$                & 3 &  2062.2374821      &      & $                                        $ & $                                        $ &             &              & 9.501\%   & $     ^{}     $\\
\rowstyle{\itshape}               &        & 52        & 48491.06028$^{d}$                & 3 &  2062.2357899      &      & $                                        $ & $                                        $ &             &              & 83.789\%  & $     ^{}     $\\
\rowstyle{\itshape}               &        & 50        & 48491.14471$^{d}$                & 3 &  2062.2321991      &      & $                                        $ & $                                        $ &             &              & 4.345\%   & $     ^{}     $\\
                                  & 2066   & 51.9961   & 48398.8729(14)$^{a,b,c}$         & 0 &   2066.163819(60)  &  8.8 & $                                        $ & $3\rm{d}^44\rm{p}~^6\rm{P}_{3/2}^{\rm o} $ & Cr$^2_{3}$  &              & 0.0512    & $-1421^{d}(70) $\\
\rowstyle{\itshape}               &        & 54        & 48398.79495$^{d}$                & 3 &  2066.1671454      &      & $                                        $ & $                                        $ &             &              & 2.365\%   & $     ^{}     $\\
\rowstyle{\itshape}               &        & 53        & 48398.83418$^{d}$                & 3 &  2066.1654706      &      & $                                        $ & $                                        $ &             &              & 9.501\%   & $     ^{}     $\\
\rowstyle{\itshape}               &        & 52        & 48398.87486$^{d}$                & 3 &  2066.1637338      &      & $                                        $ & $                                        $ &             &              & 83.789\%  & $     ^{}     $\\
\rowstyle{\itshape}               &        & 50        & 48398.96123$^{d}$                & 3 &  2066.1600466      &      & $                                        $ & $                                        $ &             &              & 4.345\%   & $     ^{}     $\\
\hline
\end{tabular}
}
{\footnotesize References:
$^{a}$\citet{Aldenius:2009:014008};
$^{b}$\citet{Pickering:2000:163};
$^{c}$\citet{Nave:2012:1570};
$^{d}$\citet{Berengut:2011:052520}.}
\end{center}
\end{table*}

\begin{figure}
\begin{center}
\includegraphics[width=0.55\columnwidth,bb = 18 153 242 706]{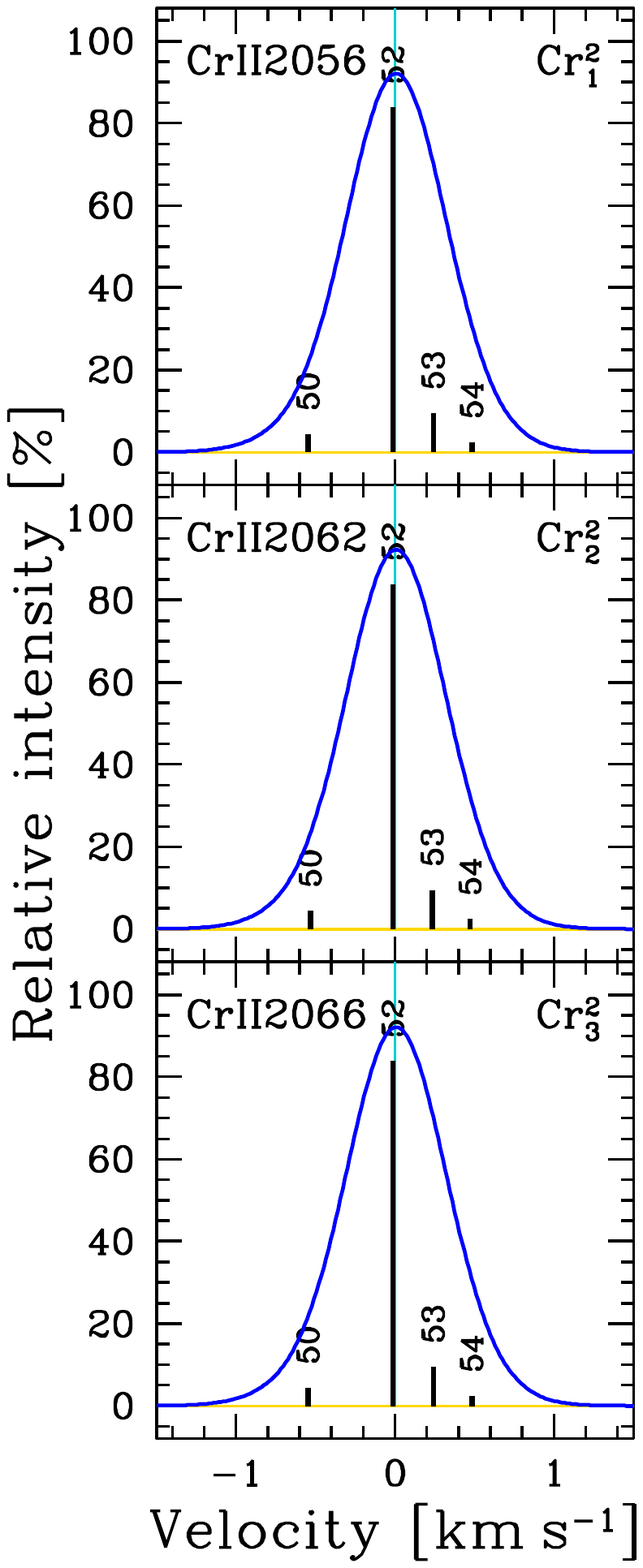}\vspace{-0.5em}
\caption{Cr transition isotopic structures in \Tref{tab:Cr} described
  in \Sref{ssec:Cr}. See \Sref{ssec:tabs} for Figure details.}
\label{fig:Cr_IHF}
\end{center}
\end{figure}

\Tref{tab:Cr} summarises the atomic data for the Cr{\sc \,ii} triplet
near 2060\,\AA\ and \Fref{fig:Cr_IHF} illustrates their isotopic
structures. While several other ground-state transitions of Cr{\sc
  \,ii} lie in the UV, redwards of H{\sc \,i}
Lyman-$\alpha$,\footnote{Notably Cr{\sc \,ii} $\lambda$2026 which
  falls very close to the Mg{\sc \,i} and Zn{\sc \,ii} $\lambda$2026
  transitions discussed in sections \ref{ssec:Mg} and \ref{ssec:Zn},
  respectively.} they are all $>$30 times weaker than the Cr{\sc \,ii}
triplet near 2060\,\AA. These three Cr{\sc \,ii} lines are therefore
the most commonly observed in quasar absorption spectra. Nevertheless,
they are usually very weak because of the comparatively low abundance
of Cr. Thus, even though the Cr{\sc \,ii} triplet has been used in
some varying-$\alpha$ studies, it has generally not provided the
dominant constraints
\citep[e.g.][]{Murphy:2001:1208,King:2012:3370}. This is despite the
remarkable feature that they have fairly large but, most-importantly,
negative $q$-coefficients.

The Cr{\sc \,ii} triplet wavenumbers have been measured in two
independent analyses with FTSs and HCLs. The wavenumbers measured by
\citet{Pickering:2000:163} were calibrated using the older Ar{\sc
  \,ii} scale of \citet{Norlen:1973:249}, while those measured by
\citet{Aldenius:2006:444} \citep[see also][]{Aldenius:2009:014008}
were calibrated using the \citet{Whaling:1995:1}
scale. \citet{Nave:2012:1570} placed both sets of measurements on a
common calibration scale consistent with that of absolute standards
derived from frequency comb measurements and we adopt
\citeauthor{Nave:2012:1570}'s corrections here. The final composite
wavenumbers shown in \Tref{tab:Cr} are a simple weighted mean of the
corrected values from \citet{Pickering:2000:163} and
\citet{Aldenius:2009:014008}.

There are no measurements of the isotopic structure of the Cr{\sc
  \,ii} triplet transitions, to our
knowledge. \citet{Berengut:2011:052520} calculated the isotopic
component separations (their table 6) and we have used these and
\Eref{eq:omega0} to report the isotopic structures in
\Tref{tab:Cr}. The resulting isotopic splittings are reasonably wide
for a relatively heavy ion like \ion{Cr}{ii}, up to
$\sim$0.55\,\kms. Therefore, if the isotopic abundances in quasar
absorbers do not reflect those in the terrestrial environment,
spurious line-shifts could be measured for the \ion{Cr}{ii}
transitions. \citet{Fenner:2005:468} explored the isotopic abundances
expected in lower-than-solar metallicity environments using a chemical
evolution model of a Milky Way-like disc galaxy, finding that the
sub-dominant Cr isotopes should be even less abundant at the
metallicities typical of the highest column density quasar
absorbers. With the composite frequencies and those of the dominant
isotope ($^{52}$Cr) being so similar in the \ion{Cr}{ii} triplet, this
would imply little systematic effect on measurements of $\alpha$ in
quasar absorbers. However, while this might represent the basic
theoretical expectation, it is important to recognise that no current
constraints on the actual Cr isotopic abundances in quasar absorbers
exist.

Our recommended $q$ coefficients for \ion{Cr}{ii} in \Tref{tab:Cr} are
taken from the CI+MBPT calculations of
\citet{Berengut:2011:052520}. These calculations are in good agreement
with previous CI calculations \citep{Dzuba:2002:022501}.

\subsection{Manganese}\label{ssec:Mn}

\begin{table*}
\begin{center}
\caption{
Laboratory data for transitions of Mn of interest for quasar absorption-line varying-$\alpha$ studies described in \Sref{ssec:Mn}. See \Sref{ssec:tabs} for full descriptions of each column.
}
\label{tab:Mn}\vspace{-0.5em}
{\footnotesize
\begin{tabular}{:l;l;c;l;c;l;c;l;l;c;c;c;c}\hline
\multicolumn{1}{c}{Ion}&
\multicolumn{1}{c}{Tran.}&
\multicolumn{1}{c}{$A$}&
\multicolumn{1}{c}{$\omega_0$}&
\multicolumn{1}{c}{$X$}&
\multicolumn{1}{c}{$\lambda_0$}&
\multicolumn{1}{c}{$\delta v$}&
\multicolumn{1}{c}{Lower state}&
\multicolumn{1}{c}{Upper state}&
\multicolumn{1}{c}{ID}&
\multicolumn{1}{c}{IP$^-$, IP$^+$}&
\multicolumn{1}{c}{$f$ or {\it \%}}&
\multicolumn{1}{c}{$q$}\\
&
&
&
\multicolumn{1}{c}{[cm$^{-1}$]}&
&
\multicolumn{1}{c}{[\AA]}&
\multicolumn{1}{c}{[m\,s$^{-1}$]}&
&
&
&
\multicolumn{1}{c}{[eV]}&
&
\multicolumn{1}{c}{[cm$^{-1}$]}\\
\hline
                    Mn{\sc \,ii}  & 2576   & 54.9380   & 38806.6923(17)$^{a,b,c}$         & 0 &    2576.87512(11)  & 12.9 & $3\rm{d}^54\rm{s~a}^7\rm{S}_3            $ & $3\rm{d}^54\rm{p~z}^7\rm{P}_4^{\rm o}    $ & Mn$^2_{1}$  & 7.44, 15.64  & 0.361     & $ 1276^{d}(150)$\\
\rowstyle{\itshape}               &        & 55        & 38806.45473$^{}$                 & 3 &   2576.890898      &      & $F=0.5,1.5                               $ & $F=1.5,2.5                               $ &             &              & 28.571\%  & $     ^{}     $\\
\rowstyle{\itshape}               &        & 55        & 38806.62838$^{}$                 & 3 &   2576.879368      &      & $F=2.5                                   $ & $F=1.5,2.5,3.5                           $ &             &              & 23.801\%  & $     ^{}     $\\
\rowstyle{\itshape}               &        & 55        & 38806.77173$^{}$                 & 3 &   2576.869849      &      & $F=3.5                                   $ & $F=2.5,3.5,4.5                           $ &             &              & 19.030\%  & $     ^{}     $\\
\rowstyle{\itshape}               &        & 55        & 38806.88249$^{}$                 & 3 &   2576.862494      &      & $F=4.5                                   $ & $F=3.5,4.5,5.5                           $ &             &              & 14.286\%  & $     ^{}     $\\
\rowstyle{\itshape}               &        & 55        & 38806.97756$^{}$                 & 3 &   2576.856181      &      & $F=5.5                                   $ & $F=4.5,5.5,6.5                           $ &             &              & 14.312\%  & $     ^{}     $\\
                                  & 2594   & 54.9380   & 38543.1250(17)$^{a,b,c}$         & 0 &    2594.49643(11)  & 12.9 & $                                        $ & $3\rm{d}^54\rm{p~z}^7\rm{P}_3^{\rm o}    $ & Mn$^2_{2}$  &              & 0.280     & $ 1030^{d}(150)$\\
\rowstyle{\itshape}               &        & 55        & 38542.89261$^{}$                 & 3 &   2594.512068      &      & $F=0.5,1.5                               $ & $F=0.5,1.5,2.5                           $ &             &              & 28.579\%  & $     ^{}     $\\
\rowstyle{\itshape}               &        & 55        & 38543.06316$^{}$                 & 3 &   2594.500587      &      & $F=2.5                                   $ & $F=1.5,2.5,3.5                           $ &             &              & 23.841\%  & $     ^{}     $\\
\rowstyle{\itshape}               &        & 55        & 38543.20275$^{}$                 & 3 &   2594.491191      &      & $F=3.5                                   $ & $F=2.5,3.5,4.5                           $ &             &              & 19.078\%  & $     ^{}     $\\
\rowstyle{\itshape}               &        & 55        & 38543.31105$^{}$                 & 3 &   2594.483901      &      & $F=4.5                                   $ & $F=3.5,4.5,5.5                           $ &             &              & 14.289\%  & $     ^{}     $\\
\rowstyle{\itshape}               &        & 55        & 38543.40454$^{}$                 & 3 &   2594.477608      &      & $F=5.5                                   $ & $F=4.5,5.5                               $ &             &              & 14.213\%  & $     ^{}     $\\
                                  & 2606   & 54.9380   & 38366.2313(17)$^{a,b,c}$         & 0 &    2606.45877(11)  & 13.0 & $                                        $ & $3\rm{d}^54\rm{p~z}^7\rm{P}_2^{\rm o}    $ & Mn$^2_{3}$  &              & 0.198     & $  869^{d}(150)$\\
\rowstyle{\itshape}               &        & 55        & 38365.94424$^{}$                 & 3 &   2606.478271      &      & $F=0.5,1.5                               $ & $F=0.5,1.5,2.5                           $ &             &              & 28.563\%  & $     ^{}     $\\
\rowstyle{\itshape}               &        & 55        & 38366.15463$^{}$                 & 3 &   2606.463977      &      & $F=2.5                                   $ & $F=1.5,2.5,3.5                           $ &             &              & 23.793\%  & $     ^{}     $\\
\rowstyle{\itshape}               &        & 55        & 38366.32705$^{}$                 & 3 &   2606.452264      &      & $F=3.5                                   $ & $F=2.5,3.5,4.5                           $ &             &              & 19.052\%  & $     ^{}     $\\
\rowstyle{\itshape}               &        & 55        & 38366.46082$^{}$                 & 3 &   2606.443176      &      & $F=4.5                                   $ & $F=3.5,4.5                               $ &             &              & 14.282\%  & $     ^{}     $\\
\rowstyle{\itshape}               &        & 55        & 38366.57520$^{}$                 & 3 &   2606.435406      &      & $F=5.5                                   $ & $F=4.5                                   $ &             &              & 14.310\%  & $     ^{}     $\\
\hline
\end{tabular}
}
{\footnotesize References:
$^{a}$\citet{Aldenius:2009:014008};
$^{b}$\citet{Blackwell-Whitehead:2005:705};
$^{c}$\citet{Nave:2012:1570};
$^{d}$\citet{Berengut:2004:064101}.}
\end{center}
\end{table*}

\begin{figure}
\begin{center}
\includegraphics[width=0.55\columnwidth,bb = 18 153 242 706]{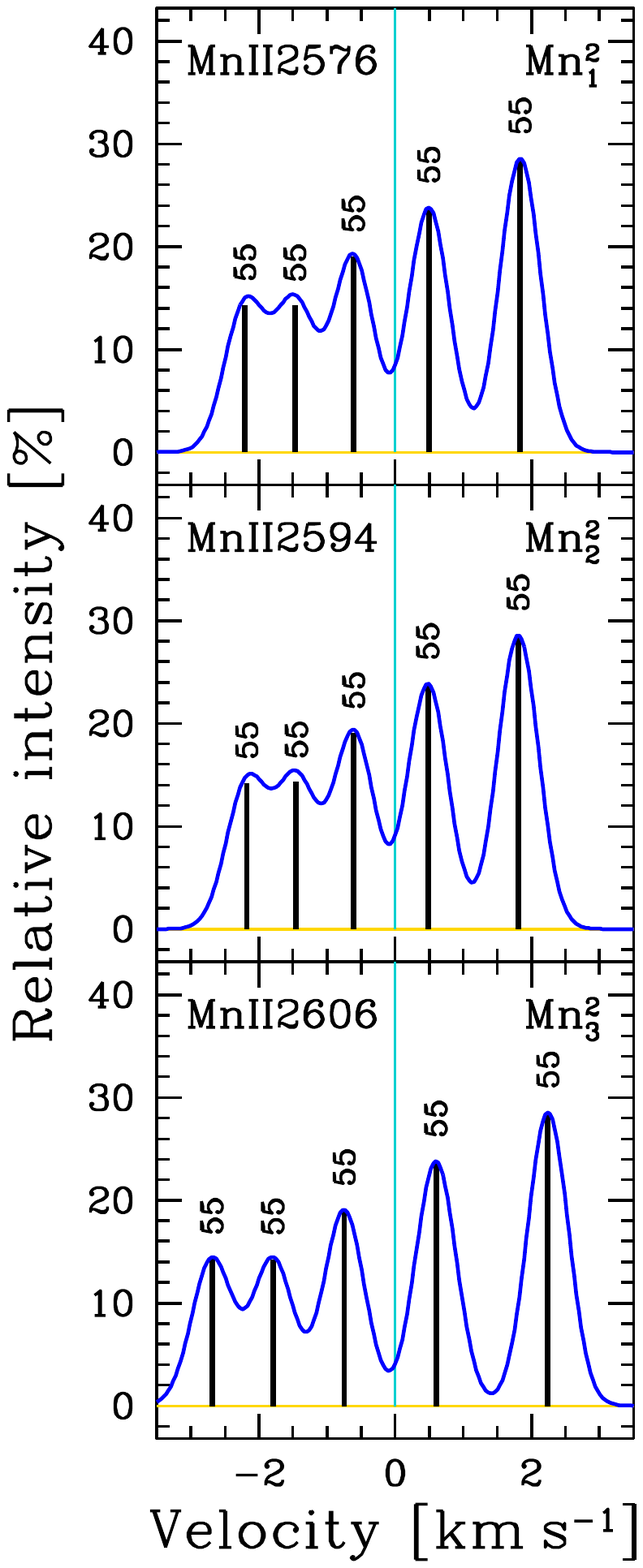}\vspace{-0.5em}
\caption{Mn transition hyperfine structures in \Tref{tab:Mn} described
  in \Sref{ssec:Mn}. See \Sref{ssec:tabs} for Figure details.}
\label{fig:Mn_IHF}
\end{center}
\end{figure}

\Tref{tab:Mn} summarises the atomic data for Mn{\sc \,ii} triplet
near 2595\,\AA\ and \Fref{fig:Mn_IHF} illustrates their hyperfine
structures. Only 8 transitions of Mn{\sc \,ii} lie redwards of H{\sc
  \,i} Lyman-$\alpha$ \citep{Morton:2003:205} and the Mn{\sc \,ii}
triplet are by-far the strongest; they are more than two orders of
magnitude stronger than the next strongest transition. Despite
having a similar terrestrial abundance to Cr, their higher absolute
oscillator strengths (of order 0.2--0.3) make these three Mn{\sc \,ii}
lines more frequently observed in quasar absorption spectra, and more
often usable in varying-$\alpha$ analyses. Nevertheless, their
laboratory wavenumbers were only measured precisely enough for the
first time in 2005, and, to our knowledge, only one study has utilized
them to constrain possible variations in $\alpha$
\citep{King:2012:3370}. Like Al, Mn has only one stable isotope
($^{55}$Mn), making it immune to systematic effects related to
isotopic abundance evolution.

The Mn{\sc \,ii} triplet wavenumbers have been measured in two
independent analyses using FTSs with HCLs. The wavenumbers measured by
\citet{Blackwell-Whitehead:2005:705} were calibrated using the older
Ar{\sc \,ii} scale of \citet{Norlen:1973:249}, while those measured by
\citet{Aldenius:2006:444} \citep[see also][]{Aldenius:2009:014008}
were calibrated using the \citet{Whaling:1995:1} scale. Following the
analysis of \citet{Nave:2011:737}, \citet{Nave:2012:1570} concluded
that \citeauthor{Aldenius:2009:014008}' wavenumbers should be
increased by 3.7 parts in 10$^{8}$, but no reference is made to
\citeauthor{Blackwell-Whitehead:2005:705}'s wavenumbers. However, the
calibration procedure of \citeauthor{Blackwell-Whitehead:2005:705} and
\citet{Pickering:2000:163} appear to be the same, and
\citet{Nave:2012:1570} conclude that \citeauthor{Pickering:2000:163}'s
wavenumbers should be increased by 10.6 parts in 10$^{8}$, so we adopt
the same correction here for
\citeauthor{Blackwell-Whitehead:2005:705}'s wavenumbers. With both
sets of Mn{\sc \,ii} corrected to the same wavenumber scale as above
-- that equivalent to the Ar{\sc \,ii} \citet{Whaling:1995:1} scale
consistent with the absolute scale derived from frequency comb
measurements -- we adopt the weighted mean composite wavenumbers in
\Tref{tab:Mn}.

The FTS spectra of both \citet{Blackwell-Whitehead:2005:705} and
\citet{Aldenius:2009:014008} resolve significant hyperfine structure
in the Mn{\sc \,ii} triplet transitions. $^{55}$Mn has a relatively
large magnetic dipole moment ($\mu=+3.5$\,nuclear magnetons), so many
observable Mn transitions in stellar spectra have prominent hyperfine
structure. Indeed, ignoring the hyperfine structure of Mn transitions
can lead to substantial systematic errors in stellar elemental
abundances \citep[e.g.][]{Prochaska:2000:L57}, sometimes as much as
one or two orders of magnitude
\citep{Jomaron:1999:555}.

\citet{Blackwell-Whitehead:2005:705} first calculated the hyperfine
structure for the Mn{\sc \,ii} ground state. With a nuclear spin of
$I=5/2$, the ground state contains 6 hyperfine levels with
half-integer total angular momenta,
$F=\frac{1}{2}$--$\frac{11}{2}$. Given the triplet excited state
electronic configurations and selection rules, the three Mn{\sc \,ii}
transitions have a total of 14--16 hyperfine components
each. \citeauthor{Blackwell-Whitehead:2005:705} determined the
hyperfine structure constants ($A$ and $B$) by fitting their FTS
spectra with their calculated structure, finding that ground state
splitting dominated the excited state splitting. Therefore, in
\Tref{tab:Mn} we group together the hyperfine components from each
ground state and provide their composite wavenumber. We also grouped
together the two lowest-lying ground states ($F=\frac{1}{2}$ \&
$\frac{3}{2}$) because their wavenumbers differed relatively little;
that is, each Mn{\sc \,ii} is represented by 5 grouped hyperfine
components in \Tref{tab:Mn}. We scaled these grouped component
wavenumbers to ensure that their composite value [\Eref{eq:omega0}]
agreed with the weighted mean composite wavenumber from the two
different FTS studies for each transition in \Tref{tab:Mn}.

The recommended $q$ coefficients for \ion{Mn}{ii} in \Tref{tab:Mn} are
taken from the AMBiT CI calculations of
\citet{Berengut:2004:064101}. The older CI calculations of
\citet{Dzuba:1999:230} are consistent within the stated uncertainties.

\subsection{Iron}\label{ssec:Fe}

\begin{table*}
\begin{center}
\caption{
Laboratory data for transitions of Fe of interest for quasar absorption-line varying-$\alpha$ studies described in \Sref{ssec:Fe}. See \Sref{ssec:tabs} for full descriptions of each column.
}
\label{tab:Fe}\vspace{-0.5em}
{\footnotesize
\begin{tabular}{:l;l;c;l;c;l;c;l;l;c;c;c;c}\hline
\multicolumn{1}{c}{Ion}&
\multicolumn{1}{c}{Tran.}&
\multicolumn{1}{c}{$A$}&
\multicolumn{1}{c}{$\omega_0$}&
\multicolumn{1}{c}{$X$}&
\multicolumn{1}{c}{$\lambda_0$}&
\multicolumn{1}{c}{$\delta v$}&
\multicolumn{1}{c}{Lower state}&
\multicolumn{1}{c}{Upper state}&
\multicolumn{1}{c}{ID}&
\multicolumn{1}{c}{IP$^-$, IP$^+$}&
\multicolumn{1}{c}{$f$ or {\it \%}}&
\multicolumn{1}{c}{$q$}\\
&
&
&
\multicolumn{1}{c}{[cm$^{-1}$]}&
&
\multicolumn{1}{c}{[\AA]}&
\multicolumn{1}{c}{[m\,s$^{-1}$]}&
&
&
&
\multicolumn{1}{c}{[eV]}&
&
\multicolumn{1}{c}{[cm$^{-1}$]}\\
\hline
                    Fe{\sc \,ii}  & 1608   & 55.845    & 62171.623(3)$^{a}$               & 0 &   1608.450852(78)  & 14.5 & $3\rm{d}^64\rm{s~a}^6\rm{D}_{9/2}        $ & $3\rm{d}^54\rm{s}4\rm{p~y}^6\rm{P}_{7/2}^{\rm o}$ & Fe$^2_{5}$  & 7.87, 16.18  & 0.0577    & $-1165^{b,c}(300)$\\
\rowstyle{\itshape}               &        & 58        & 62171.57878$^{c}$                & 3 &  1608.4519963      &      & $                                        $ & $                                        $ &             &              & 0.282\%   & $     ^{}     $\\
\rowstyle{\itshape}               &        & 57        & 62171.59948$^{c}$                & 3 &  1608.4514606      &      & $                                        $ & $                                        $ &             &              & 2.119\%   & $     ^{}     $\\
\rowstyle{\itshape}               &        & 56        & 62171.62093$^{c}$                & 3 &  1608.4509059      &      & $                                        $ & $                                        $ &             &              & 91.754\%  & $     ^{}     $\\
\rowstyle{\itshape}               &        & 54        & 62171.66620$^{c}$                & 3 &  1608.4497347      &      & $                                        $ & $                                        $ &             &              & 5.845\%   & $     ^{}     $\\
                                  & 1611   & 55.845    & 62065.527(3)$^{a}$               & 0 &   1611.200369(78)  & 14.5 & $                                        $ & $3\rm{d}^64\rm{p~y}^4\rm{F}_{7/2}^{\rm o}$ & Fe$^2_{10}$ &              & 0.00138   & $ 1330^{b,c}(300)$\\
\rowstyle{\itshape}               &        & 58        & 62065.55558$^{c}$                & 3 &  1611.1996271      &      & $                                        $ & $                                        $ &             &              & 0.282\%   & $     ^{}     $\\
\rowstyle{\itshape}               &        & 57        & 62065.54220$^{c}$                & 3 &  1611.1999744      &      & $                                        $ & $                                        $ &             &              & 2.119\%   & $     ^{}     $\\
\rowstyle{\itshape}               &        & 56        & 62065.52834$^{c}$                & 3 &  1611.2003342      &      & $                                        $ & $                                        $ &             &              & 91.754\%  & $     ^{}     $\\
\rowstyle{\itshape}               &        & 54        & 62065.49909$^{c}$                & 3 &  1611.2010936      &      & $                                        $ & $                                        $ &             &              & 5.845\%   & $     ^{}     $\\
                                  & 2249   & 55.845    & 44446.9044(19)$^{a}$             & 0 &   2249.875472(96)  & 12.8 & $                                        $ & $3\rm{d}^64\rm{p~z}^4\rm{D}_{7/2}^{\rm o}$ & Fe$^2_{9}$  &              & 0.00182   & $ 1604^{d}(200)$\\
\rowstyle{\itshape}               &        & 58        & 44446.93584$^{c}$                & 3 &  2249.8738801      &      & $                                        $ & $                                        $ &             &              & 0.282\%   & $     ^{}     $\\
\rowstyle{\itshape}               &        & 57        & 44446.92112$^{c}$                & 3 &  2249.8746253      &      & $                                        $ & $                                        $ &             &              & 2.119\%   & $     ^{}     $\\
\rowstyle{\itshape}               &        & 56        & 44446.90587$^{c}$                & 3 &  2249.8753970      &      & $                                        $ & $                                        $ &             &              & 91.754\%  & $     ^{}     $\\
\rowstyle{\itshape}               &        & 54        & 44446.87369$^{c}$                & 3 &  2249.8770263      &      & $                                        $ & $                                        $ &             &              & 5.845\%   & $     ^{}     $\\
                                  & 2260   & 55.845    & 44232.5390(19)$^{e,a}$           & 0 &   2260.779108(97)  & 12.9 & $                                        $ & $3\rm{d}^64\rm{p~z}^4\rm{F}_{9/2}^{\rm o}$ & Fe$^2_{8}$  &              & 0.00244   & $ 1435^{d}(150)$\\
                                  & 2344   & 55.845    & 42658.2443(14)$^{e,a}$           & 0 &   2344.212747(76)  &  9.7 & $                                        $ & $3\rm{d}^64\rm{p~z}^6\rm{P}_{7/2}^{\rm o}$ & Fe$^2_{3}$  &              & 0.114     & $ 1375^{b,c}(300)$\\
\rowstyle{\itshape}               &        & 58        & 42658.26949$^{c}$                & 3 &  2344.2113616      &      & $                                        $ & $                                        $ &             &              & 0.282\%   & $     ^{}     $\\
\rowstyle{\itshape}               &        & 57        & 42658.25768$^{c}$                & 3 &  2344.2120103      &      & $                                        $ & $                                        $ &             &              & 2.119\%   & $     ^{}     $\\
\rowstyle{\itshape}               &        & 56        & 42658.24546$^{c}$                & 3 &  2344.2126822      &      & $                                        $ & $                                        $ &             &              & 91.754\%  & $     ^{}     $\\
\rowstyle{\itshape}               &        & 54        & 42658.21964$^{c}$                & 3 &  2344.2141007      &      & $                                        $ & $                                        $ &             &              & 5.845\%   & $     ^{}     $\\
                                  & 2367   & 55.845    & 42237.0563(19)$^{a}$             & 0 &    2367.58924(11)  & 13.5 & $                                        $ & $3\rm{d}^64\rm{p~z}^6\rm{F}_{7/2}^{\rm o}$ & Fe$^2_{15}$ &              & 0.0000216 & $ 1803^{d}(200)$\\
\rowstyle{\itshape}               &        & 58        & 42237.08770$^{c}$                & 3 &   2367.587479      &      & $                                        $ & $                                        $ &             &              & 0.282\%   & $     ^{}     $\\
\rowstyle{\itshape}               &        & 57        & 42237.07300$^{c}$                & 3 &   2367.588303      &      & $                                        $ & $                                        $ &             &              & 2.119\%   & $     ^{}     $\\
\rowstyle{\itshape}               &        & 56        & 42237.05777$^{c}$                & 3 &   2367.589157      &      & $                                        $ & $                                        $ &             &              & 91.754\%  & $     ^{}     $\\
\rowstyle{\itshape}               &        & 54        & 42237.02563$^{c}$                & 3 &   2367.590959      &      & $                                        $ & $                                        $ &             &              & 5.845\%   & $     ^{}     $\\
                                  & 2374   & 55.845    & 42114.8376(14)$^{e,a}$           & 0 &   2374.460064(78)  &  9.8 & $                                        $ & $3\rm{d}^64\rm{p~z}^6\rm{F}_{9/2}^{\rm o}$ & Fe$^2_{6}$  &              & 0.03130   & $ 1625^{b,c}(100)$\\
\rowstyle{\itshape}               &        & 58        & 42114.86890$^{c}$                & 3 &  2374.4582998      &      & $                                        $ & $                                        $ &             &              & 0.282\%   & $     ^{}     $\\
\rowstyle{\itshape}               &        & 57        & 42114.85425$^{c}$                & 3 &  2374.4591258      &      & $                                        $ & $                                        $ &             &              & 2.119\%   & $     ^{}     $\\
\rowstyle{\itshape}               &        & 56        & 42114.83907$^{c}$                & 3 &  2374.4599813      &      & $                                        $ & $                                        $ &             &              & 91.754\%  & $     ^{}     $\\
\rowstyle{\itshape}               &        & 54        & 42114.80704$^{c}$                & 3 &  2374.4617873      &      & $                                        $ & $                                        $ &             &              & 5.845\%   & $     ^{}     $\\
                                  & 2382   & 55.845    & 41968.0674(14)$^{e,a}$           & 0 &   2382.763995(80)  & 10.1 & $                                        $ & $3\rm{d}^64\rm{p~z}^6\rm{F}_{11/2}^{\rm o}$ & Fe$^2_{1}$  &              & 0.320     & $ 1505^{b,c}(100)$\\
\rowstyle{\itshape}               &        & 58        & 41968.09852$^{c}$                & 3 &  2382.7622294      &      & $                                        $ & $                                        $ &             &              & 0.282\%   & $     ^{}     $\\
\rowstyle{\itshape}               &        & 57        & 41968.08396$^{c}$                & 3 &  2382.7630560      &      & $                                        $ & $                                        $ &             &              & 2.119\%   & $     ^{}     $\\
\rowstyle{\itshape}               &        & 56        & 41968.06888$^{c}$                & 3 &  2382.7639122      &      & $                                        $ & $                                        $ &             &              & 91.754\%  & $     ^{}     $\\
\rowstyle{\itshape}               &        & 54        & 41968.03705$^{c}$                & 3 &  2382.7657196      &      & $                                        $ & $                                        $ &             &              & 5.845\%   & $     ^{}     $\\
                                  & 2586   & 55.845    & 38660.0532(13)$^{e,a}$           & 0 &   2586.649312(87)  & 10.0 & $                                        $ & $3\rm{d}^64\rm{p~z}^6\rm{D}_{7/2}^{\rm o}$ & Fe$^2_{4}$  &              & 0.0691    & $ 1515^{b,c}(100)$\\
\rowstyle{\itshape}               &        & 58        & 38660.07931$^{c}$                & 3 &  2586.6475648      &      & $                                        $ & $                                        $ &             &              & 0.282\%   & $     ^{}     $\\
\rowstyle{\itshape}               &        & 57        & 38660.06708$^{c}$                & 3 &  2586.6483830      &      & $                                        $ & $                                        $ &             &              & 2.119\%   & $     ^{}     $\\
\rowstyle{\itshape}               &        & 56        & 38660.05441$^{c}$                & 3 &  2586.6492304      &      & $                                        $ & $                                        $ &             &              & 91.754\%  & $     ^{}     $\\
\rowstyle{\itshape}               &        & 54        & 38660.02768$^{c}$                & 3 &  2586.6510194      &      & $                                        $ & $                                        $ &             &              & 5.845\%   & $     ^{}     $\\
                                  & 2600   & 55.845    & 38458.9926(13)$^{e,a}$           & 0 &   2600.172114(88)  & 10.1 & $                                        $ & $3\rm{d}^64\rm{p~z}^6\rm{D}_{9/2}^{\rm o}$ & Fe$^2_{2}$  &              & 0.239     & $ 1370^{b,c}(100)$\\
\rowstyle{\itshape}               &        & 58        & 38459.01850$^{c}$                & 3 &  2600.1703603      &      & $                                        $ & $                                        $ &             &              & 0.282\%   & $     ^{}     $\\
\rowstyle{\itshape}               &        & 57        & 38459.00636$^{c}$                & 3 &  2600.1711816      &      & $                                        $ & $                                        $ &             &              & 2.119\%   & $     ^{}     $\\
\rowstyle{\itshape}               &        & 56        & 38458.99377$^{c}$                & 3 &  2600.1720322      &      & $                                        $ & $                                        $ &             &              & 91.754\%  & $     ^{}     $\\
\rowstyle{\itshape}               &        & 54        & 38458.96721$^{c}$                & 3 &  2600.1738281      &      & $                                        $ & $                                        $ &             &              & 5.845\%   & $     ^{}     $\\
\hline
\end{tabular}
}
{\footnotesize References:
$^{a}$\citet{Nave:2012:1570};
$^{b}$\citet{Dzuba:2002:022501};
$^{c}$\citet{Porsev:2009:032519};
$^{d}$\citet{Berengut:2006:PhD};
$^{e}$\citet{Aldenius:2009:014008}.}
\end{center}
\end{table*}

\begin{figure*}
\begin{center}
\includegraphics[width=0.60\textwidth,bb = 18 153 580 706]{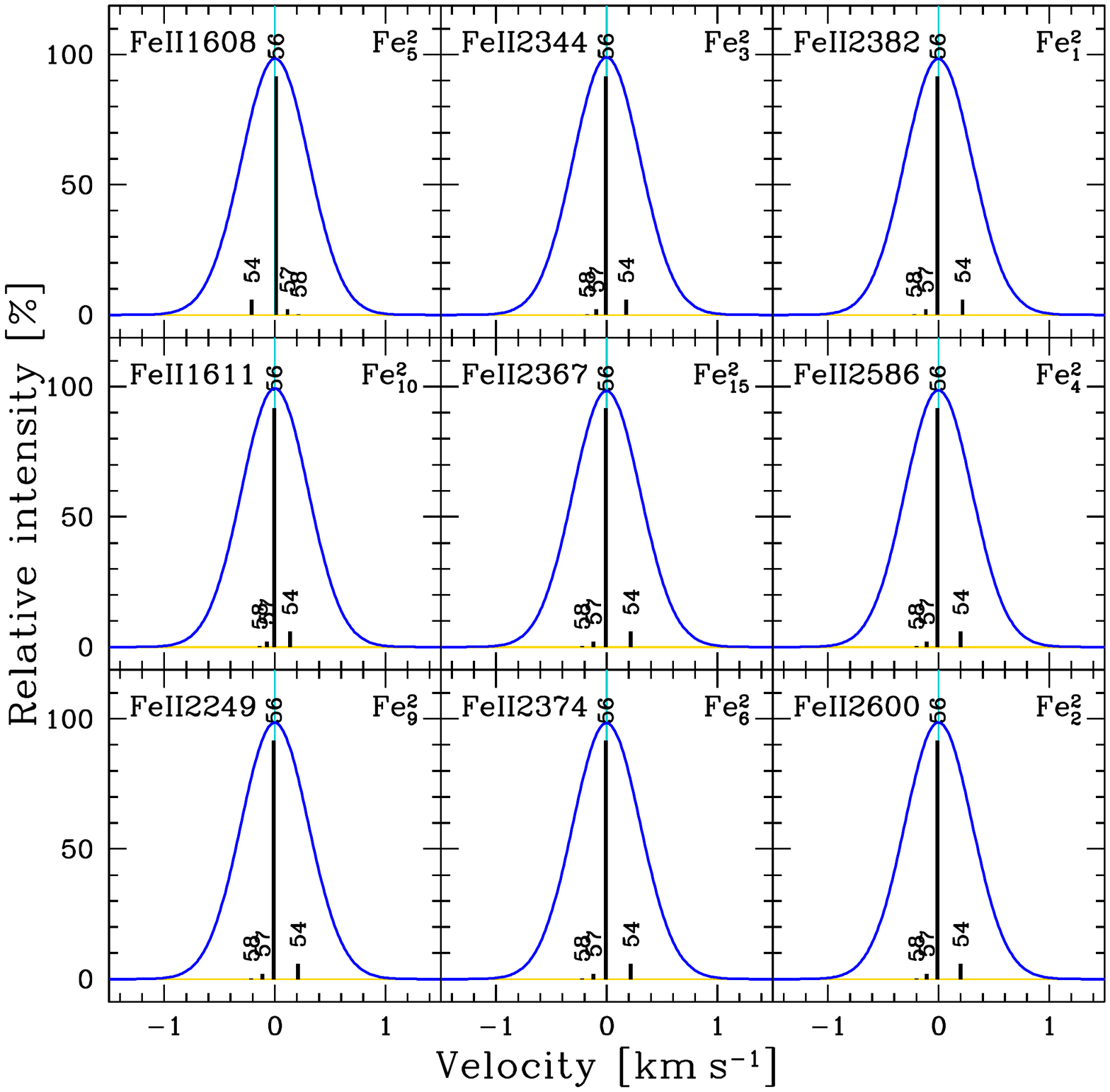}\vspace{-0.5em}
\caption{Fe transition isotopic structures in \Tref{tab:Fe} described
  in \Sref{ssec:Fe}. See \Sref{ssec:tabs} for Figure details.}
\label{fig:Fe_IHF}
\end{center}
\end{figure*}

\Tref{tab:Fe} presents the available atomic data for 10 Fe{\sc \,ii}
lines falling redwards of H{\sc \,i} Lyman-$\alpha$. The isotopic
structures calculated by \citet{Porsev:2009:032519} for nine of these
transitions are illustrated in \Fref{fig:Fe_IHF}. One weak transition
($f=0.024$) with a precisely measured laboratory wavelength,
1260.53557(6)\,\AA\ \citep{Nave:2012:1570}, currently lacks a
calculated $q$ coefficient, so is not shown in \Tref{tab:Fe}. In
practice, that transition is often comprehensively blended with the
very strong Si{\sc \,ii} transition at 1260.4221\,\AA, rendering it
unusable for varying $\alpha$ analyses except when the velocity
structure of the absorber spans $<$27\,\kms. A further 4 very weak
Fe{\sc \,ii} transitions lie redwards of H{\sc \,i} Lyman-$\alpha$
which are stronger than the weakest transition we list in
\Tref{tab:Fe} -- i.e.~with oscillator strengths
$2.5\times10^{-5}<f<1.5\times10^{-4}$ -- and which have
precisely-measured (vacuum) wavelengths: 2234.44624(9),
1901.76075(9), 1620.06106(7) and 1588.68741(6)\,\AA\
\citep{Nave:2013:1}. These transitions also currently lack calculated
$q$ coefficients, so we consider them no further here.

Transitions of \ion{Fe}{i} are very rarely observed in quasar
absorbers because, with its low ionization potential (7.87\,eV), it
has a very low abundance compared with \ion{Fe}{ii}. However, $q$
coefficients for many \ion{Fe}{i} transitions have been calculated
because some were observed towards some bright quasars
\citep{Dzuba:2008:012514,Porsev:2009:032519}. Indeed, of the 13
strongest \ion{Fe}{i} transitions listed by \citet{Morton:2003:205},
spanning a factor of 25 in oscillator strength, 11 have known $q$
coefficients (with up to $\sim$40\,\% uncertainties)\footnote{These 13
  \ion{Fe}{i} transitions' frequencies were presented by
  \citet{Nave:1994:221} and corrected (upwards by 6.7 parts in 10$^8$)
  by \citet{Nave:2011:737} to give the following vacuum wavelengths
  (in decreasing oscillator strength order): 2484.0207(6),
  2523.6081(3), 2167.4532(5), 2719.8328(4), 3021.5185(5),
  2463.3920(3), 2501.8857(3), 2967.7644(4), 3720.9926(7),
  2984.4400(4), 2298.8767(3), 3441.5917(6),
  3861.0055(7)\,\AA.}. However, we do not consider them here because,
unfortunately, their laboratory frequencies are not known with high
enough precision to meet our $\delta v \la 20$\,\kms\ criterion.

Wavenumbers for the six strongest Fe{\sc \,ii} lines in \Tref{tab:Fe},
all at laboratory wavelengths $>$2260\,\AA, were measured by
\citet{Aldenius:2006:444} \citep[see also][]{Aldenius:2009:014008}
using an FTS and HCL. Wavenumbers for all the transitions in
\Tref{tab:Fe} were also previously measured with different FTSs (and
HCLs), as described in \citet{Nave:2012:1570}. \citet{Nave:2011:737}
considered in detail the wavelength calibration scales of all these
measurements and recommended corrections to their measured wavenumbers
to place them on the Ar{\sc \,ii} \citet{Whaling:1995:1} scale which
is consistent with the absolute scale derived from frequency comb
measurements. In particular, they conclude that
\citeauthor{Aldenius:2009:014008}' wavenumbers should be increased by
3.7 parts in 10$^{8}$. \citet{Nave:2012:1570} also used a
comprehensive energy level analysis to derive more precise Ritz
wavenumbers for the Fe{\sc \,ii} transitions in \Tref{tab:Fe}. We
adopt those Ritz composite wavenumbers here. For the six strongest
transitions, we adopt the weighted mean of
\citeauthor{Nave:2012:1570}'s value and the increased
\citet{Aldenius:2009:014008} value as the composite wavenumber in
\Tref{tab:Fe}.

To our knowledge, no measurements exist of the isotopic structure of
the Fe{\sc \,ii} transitions in
\Tref{tab:Fe}. \citet{Porsev:2009:032519} calculated the specific mass
shift constant (\kSMS) for all but the $\lambda$2260
transition. General agreement was found with the earlier calculations
of \citet{Kozlov:2004:062108}, though the systematic uncertainties
were still estimated to $\ga$20\,\%. Assuming the field shift term in
\Eref{eq:iso} to be small, we neglect it and use
\citeauthor{Porsev:2009:032519}'s \kSMS\ values to report the isotopic
structures in \Tref{tab:Fe}. The results are illustrated in
\Fref{fig:Fe_IHF}. Given the considerable uncertainties, and that the
field shifts are ignored here, these structures should only be treated
as representative of what can be expected. Of particular note is that
the calculated isotopic structures are very compact, spanning only
$\sim$400\,\ms, compared with those of other transitions discussed in
this paper. This implies that Fe{\sc \,ii} isotopic structure effects
should not be important for varying $\alpha$ analyses. However, note
that the isotopic structure of the $\lambda$1608 transition is
reversed relative to the others in \Fref{fig:Fe_IHF}; heavier isotopes
have \emph{longer} $\lambda$1608 transition wavelengths. This means
that if the relative Fe isotopic abundances in a quasar absorption
system differ from the terrestrial values used to model the absorption
profiles, the effect on the line centroids will be almost completely
degenerate with a shift in $\alpha$. For example, from
\Fref{fig:Fe_IHF} it is immediately clear that if the
$^{54}$Fe/$^{56}$Fe ratio was inverted in quasar absorbers compared to
the terrestrial value (i.e.~15.7 instead of 0.064), but the
terrestrial ratio was assumed in fits to the Fe{\sc \,ii} transitions,
a spurious velocity shift of $\sim$400\,\ms\ would be measured between
the Fe{\sc \,ii} $\lambda$1608 transition and the other Fe{\sc \,ii}
transitions in \Tref{tab:Fe}. While much less extreme Fe isotopic
abundance differences are expected in quasar absorbers
\citep{Fenner:2005:468}, this example illustrates how using only
Fe{\sc \,ii} transitions in varying $\alpha$ analyses is not immune to
systematic errors from isotopic abundance evolution.

The $q$ coefficients for \ion{Fe}{ii} in \Tref{tab:Fe} were calculated
using CI plus core--valence correlation corrections and Breit
corrections in \citet{Porsev:2007:052507}. On the other hand, the
previous CI calculations by \citet{Dzuba:2002:022501} used
experimental $g$-factors to model level pseudocrossing which, in some
cases, can lead to larger corrections than core--valence
correlations. Both sets of $q$ values are consistent, and our
recommended values are the simple average of the two. The transitions
$\lambda$2260, $\lambda$2249, and $\lambda$2367 were not calculated in
those works: for these we use the values calculated using CI and
experimental $g$-factor fitting presented in
\citet{Berengut:2006:PhD}, which shows good agreement with the other
calculations for the other \ion{Fe}{ii} transitions.

\subsection{Nickel}\label{ssec:Ni}

\begin{table*}
\begin{center}
\caption{
Laboratory data for transitions of Ni of interest for quasar absorption-line varying-$\alpha$ studies described in \Sref{ssec:Ni}. See \Sref{ssec:tabs} for full descriptions of each column.
}
\label{tab:Ni}\vspace{-0.5em}
{\footnotesize
\begin{tabular}{:l;l;c;l;c;l;c;l;l;c;c;c;c}\hline
\multicolumn{1}{c}{Ion}&
\multicolumn{1}{c}{Tran.}&
\multicolumn{1}{c}{$A$}&
\multicolumn{1}{c}{$\omega_0$}&
\multicolumn{1}{c}{$X$}&
\multicolumn{1}{c}{$\lambda_0$}&
\multicolumn{1}{c}{$\delta v$}&
\multicolumn{1}{c}{Lower state}&
\multicolumn{1}{c}{Upper state}&
\multicolumn{1}{c}{ID}&
\multicolumn{1}{c}{IP$^-$, IP$^+$}&
\multicolumn{1}{c}{$f$ or {\it \%}}&
\multicolumn{1}{c}{$q$}\\
&
&
&
\multicolumn{1}{c}{[cm$^{-1}$]}&
&
\multicolumn{1}{c}{[\AA]}&
\multicolumn{1}{c}{[m\,s$^{-1}$]}&
&
&
&
\multicolumn{1}{c}{[eV]}&
&
\multicolumn{1}{c}{[cm$^{-1}$]}\\
\hline
                    Ni{\sc \,ii}  & 1709   & 58.6934   & 58493.0772(40)$^{a,b}$           & 0 &    1709.60402(12)  & 20.5 & $3\rm{d}^9~^2\rm{D}_{5/2}                $ & $3\rm{d}^84\rm{p~z}^2\rm{F}_{5/2}^{\rm o}$ & Ni$^2_{4}$  & 7.64, 18.17  & 0.0324    & $  -20^{c}(250)$\\
                                  & 1741   & 58.6934   & 57420.0191(40)$^{a,b}$           & 0 &    1741.55289(12)  & 20.9 & $                                        $ & $3\rm{d}^84\rm{p~z}^2\rm{D}_{5/2}^{\rm o}$ & Ni$^2_{3}$  &              & 0.0427    & $-1400^{c}(250)$\\
                                  & 1751   & 58.6934   & 57080.3791(40)$^{a,b}$           & 0 &    1751.91549(12)  & 21.0 & $                                        $ & $3\rm{d}^84\rm{p~z}^2\rm{F}_{7/2}^{\rm o}$ & Ni$^2_{6}$  &              & 0.0277    & $ -700^{c}(250)$\\
\hline
\end{tabular}
}
{\footnotesize References:
$^{a}$\citet{Pickering:2000:163};
$^{b}$\citet{Nave:2012:1570};
$^{c}$\citet{Dzuba:2002:022501}.}
\end{center}
\end{table*}

\Tref{tab:Ni} provides the atomic data for the three Ni{\sc \,ii}
transitions redwards of H{\sc \,i} \lya\ with precisely measured
laboratory wavelengths. Several other similarly-strong Ni{\sc \,ii}
transitions exist redwards of $\sim$1370\,\AA\ but all with laboratory
wavelength uncertainties significantly exceeding our $\la$20\,\ms\
criterion for varying $\alpha$ studies. The three Ni{\sc \,ii} lines
in \Tref{tab:Ni} shift in the opposite direction to most transitions
of other species as $\alpha$ varies (i.e.~$q$ is negative), making
them important lines to include in a multi-species fit of a quasar
absorber to constrain $\alpha$-variation. Also of note is the large
range in $q$: $\lambda$1709 is effectively an ``anchor'' line, with a
very small $q$, while $\lambda$1741 has a large, negative
$q$. Therefore, like with Fe{\sc \,ii}, a single ionic species can, in
principle, provide strong constraints on $\alpha$-variation. However,
Ni has several stable isotopes, two of which have high relative
abundances ($^{58}$Ni and $^{60}$Ni), and, to our knowledge, no
isotopic structure measurements or calculations are available for the
transitions in \Tref{tab:Ni}. That is, these Ni{\sc \,ii} transitions
have unknown susceptibility to systematic effects related to isotopic
abundance evolution.

The Ni{\sc \,ii} wavenumbers have been measured only once to high
enough precision for varying $\alpha$ studies by
\citet{Pickering:2000:163} using an FTS with an HCL. Their wavenumbers
were calibrated using the older Ar{\sc \,ii} scale of
\citet{Norlen:1973:249}. \citet{Nave:2012:1570} recommended increasing
\citeauthor{Pickering:2000:163}'s wavenumbers by 10.6 parts in
10$^{8}$ to place them on the newer \citet{Whaling:1995:1} Ar{\sc
  \,ii} calibration scale which is consistent with that of absolute
standards derived from frequency comb measurements. The final
composite wavenumbers shown in \Tref{tab:Ni} are the values from
\citet{Pickering:2000:163} increased accordingly.

The recommended \ion{Ni}{ii} $q$ coefficients in \Tref{tab:Ni} are those
calculated by \citet{Dzuba:2002:022501}.

\subsection{Zinc}\label{ssec:Zn}

\begin{table*}
\begin{center}
\caption{
Laboratory data for transitions of Zn of interest for quasar absorption-line varying-$\alpha$ studies described in \Sref{ssec:Zn}. See \Sref{ssec:tabs} for full descriptions of each column.
}
\label{tab:Zn}\vspace{-0.5em}
{\footnotesize
\begin{tabular}{:l;l;c;l;c;l;c;l;l;c;c;c;c}\hline
\multicolumn{1}{c}{Ion}&
\multicolumn{1}{c}{Tran.}&
\multicolumn{1}{c}{$A$}&
\multicolumn{1}{c}{$\omega_0$}&
\multicolumn{1}{c}{$X$}&
\multicolumn{1}{c}{$\lambda_0$}&
\multicolumn{1}{c}{$\delta v$}&
\multicolumn{1}{c}{Lower state}&
\multicolumn{1}{c}{Upper state}&
\multicolumn{1}{c}{ID}&
\multicolumn{1}{c}{IP$^-$, IP$^+$}&
\multicolumn{1}{c}{$f$ or {\it \%}}&
\multicolumn{1}{c}{$q$}\\
&
&
&
\multicolumn{1}{c}{[cm$^{-1}$]}&
&
\multicolumn{1}{c}{[\AA]}&
\multicolumn{1}{c}{[m\,s$^{-1}$]}&
&
&
&
\multicolumn{1}{c}{[eV]}&
&
\multicolumn{1}{c}{[cm$^{-1}$]}\\
\hline
                    Zn{\sc \,ii}  & 2026   & 65.38     & 49355.00544(76)$^{a,b,c,d}$      & 0 &   2026.136946(31)  &  4.6 & $3\rm{d}^{10}4\rm{s}~^2\rm{S}_{1/2}      $ & $3\rm{d}^{10}4\rm{p}~^2\rm{P}_{3/2}^{\rm o}$ & Zn$^2_{1}$  & 9.39, 17.96  & 0.501     & $ 2470^{e,f}(25) $\\
\rowstyle{\itshape}               &        & 70        & 49355.05268(86)$^{g}$            & 2 &   2026.135007(35)  &  5.2 & $                                        $ & $                                        $ &             &              & 0.62\%    & $     ^{}     $\\
\rowstyle{\itshape}               &        & 68        & 49355.03373(80)$^{g}$            & 2 &   2026.135785(33)  &  4.8 & $                                        $ & $                                        $ &             &              & 18.75\%   & $     ^{}     $\\
\rowstyle{\itshape}               &        & 67        & 49355.1572(50)$^{h,g,i}$         & 3 &    2026.13072(21)  & 30.4 & $F=2                                     $ & $F=1,2,3                                 $ &             &              & 1.71\%    & $     ^{}     $\\
\rowstyle{\itshape}               &        & 67        & 49354.9316(18)$^{h,g,i}$         & 3 &   2026.139979(76)  & 11.2 & $F=3                                     $ & $F=2,3,4                                 $ &             &              & 2.39\%    & $     ^{}     $\\
\rowstyle{\itshape}               &        & 66        & 49355.01138(79)$^{g}$            & 2 &   2026.136702(32)  &  4.8 & $                                        $ & $                                        $ &             &              & 27.90\%   & $     ^{}     $\\
\rowstyle{\itshape}               &        & 64        & 49354.9888(10)$^{g}$             & 2 &   2026.137628(43)  &  6.3 & $                                        $ & $                                        $ &             &              & 48.63\%   & $     ^{}     $\\
                                  & 2062   & 65.38     & 48481.08070(76)$^{a,b,c,d}$      & 0 &   2062.660291(32)  &  4.7 & $                                        $ & $3\rm{d}^{10}4\rm{p}~^2\rm{P}_{1/2}^{\rm o}$ & Zn$^2_{2}$  &              & 0.246     & $ 1563^{e,f}(25) $\\
\rowstyle{\itshape}               &        & 70        & 48481.12899$^{j,g}$              & 3 &   2062.658236      &      & $                                        $ & $                                        $ &             &              & 0.62\%    & $     ^{}     $\\
\rowstyle{\itshape}               &        & 68        & 48481.10970$^{j,g}$              & 3 &   2062.659057      &      & $                                        $ & $                                        $ &             &              & 18.75\%   & $     ^{}     $\\
\rowstyle{\itshape}               &        & 67        & 48481.24812$^{h,j,g,i}$          & 3 &   2062.653167      &      & $F=2                                     $ & $F=2,3                                   $ &             &              & 1.71\%    & $     ^{}     $\\
\rowstyle{\itshape}               &        & 67        & 48480.99644$^{h,j,g,i}$          & 3 &   2062.663876      &      & $F=3                                     $ & $F=2,3                                   $ &             &              & 2.39\%    & $     ^{}     $\\
\rowstyle{\itshape}               &        & 66        & 48481.08685$^{j,g}$              & 3 &   2062.660029      &      & $                                        $ & $                                        $ &             &              & 27.90\%   & $     ^{}     $\\
\rowstyle{\itshape}               &        & 64        & 48481.06363$^{j,g}$              & 3 &   2062.661017      &      & $                                        $ & $                                        $ &             &              & 48.63\%   & $     ^{}     $\\
\hline
\end{tabular}
}
{\footnotesize References:
$^{a}$\citet{Aldenius:2009:014008};
$^{b}$\citet{Pickering:2000:163};
$^{c}$\citet{Ruffoni:2010:424};
$^{d}$\citet{Nave:2012:1570};
$^{e}$\citet{Dzuba:2007:062510};
$^{f}$\citet{Savukov:2008:042501};
$^{g}$\citet{Matsubara:2003:209};
$^{h}$\citet{Campbell:1997:2351};
$^{i}$\citet{Dixit:2008:025001};
$^{j}$\citet{Berengut:2003:022502}.}
\end{center}
\end{table*}

\begin{figure}
\begin{center}
\includegraphics[width=0.55\columnwidth,bb = 18 322 242 706]{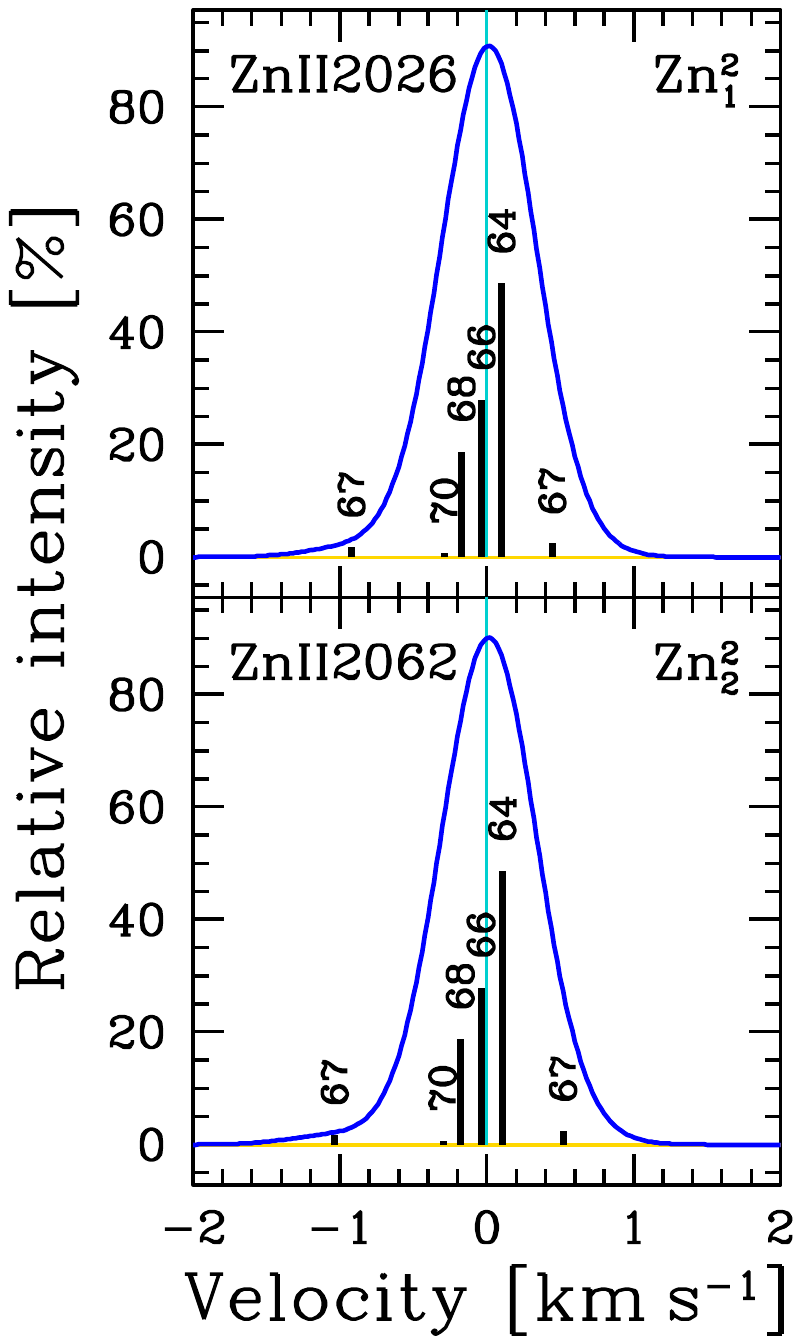}\vspace{-0.5em}
\caption{Zn transition isotopic and hyperfine structures in
  \Tref{tab:Zn} described in \Sref{ssec:Zn}. See \Sref{ssec:tabs} for
  Figure details.}
\label{fig:Zn_IHF}
\end{center}
\end{figure}

\Tref{tab:Zn} summarises the atomic data for Zn{\sc \,ii}
$\lambda\lambda$2026/2062 fine-structure doublet and \Fref{fig:Cr_IHF}
illustrates the isotopic structures. No other Zn{\sc \,ii} transitions
lying redwards of H{\sc \,i} Lyman-$\alpha$ are known
\citep{Morton:2003:205}. The Zn{\sc \,ii} doublet is particularly
important in element abundance studies using quasar absorbers because
Zn is expected to remain predominantly in the gas phase, having an
apparently low affinity for condensing onto (or remaining condensed
on) dust grains \citep{Pettini:1990:48}. Nevertheless, these Zn{\sc
  \,ii} transitions are usually very weak because of the comparatively
low abundance of Zn. Thus, even though the Zn{\sc \,ii} doublet has
been used in some varying-$\alpha$ studies, it has generally not
dominated the constraints
\citep[e.g.][]{Murphy:2001:1208,King:2012:3370}. On the other hand,
they are particularly sensitive to variations in $\alpha$, with
$\lambda$2026 having the largest absolute $q$ value of any transition
considered in this paper.

The composite Zn{\sc \,ii} doublet wavenumbers have been measured with
high enough precision in three separate analyses using FTSs with
HCLs. The wavenumbers measured by \citet{Pickering:2000:163} were
calibrated using the older Ar{\sc \,ii} scale of
\citet{Norlen:1973:249}, while those measured by
\citet{Aldenius:2006:444} \citep[see also][]{Aldenius:2009:014008}
were calibrated using the \citet{Whaling:1995:1}
scale. \citet{Nave:2012:1570} placed both sets of measurements on a
common calibration scale consistent with that of absolute standards
derived from frequency comb measurements and we adopt
\citeauthor{Nave:2012:1570}'s corrections
here. \citet{Ruffoni:2010:424} also recently re-measured the Zn{\sc
  \,ii} transitions simultaneously with the Mg{\sc \,i}/{\sc \,ii} and
Ti{\sc \,ii} transitions discussed in \Sref{ssec:Mg} and
\Sref{ssec:Ti}. Their Ar{\sc \,ii}-calibrated Mg{\sc \,i} wavenumbers
agreed well with the frequency-comb based, absolute measurements of
\citet{Hannemann:2006:012505} and \citet{Salumbides:2006:L41}, and
their Ti{\sc \,ii} $\lambda\lambda$3067/3073 doublet wavenumbers agree
with the corrected \citet{Aldenius:2009:014008} values reported by
\citet{Nave:2012:1570}. Therefore, we apply no correction to
\citeauthor{Ruffoni:2010:424}'s Zn{\sc \,ii} measurements. The final
composite wavenumbers shown in \Tref{tab:Zn} are a simple weighted
mean of the corrected values from \citet{Pickering:2000:163},
\citet{Aldenius:2009:014008} and the original values reported by
\citet{Ruffoni:2010:424}.

\citet{Matsubara:2003:209} have measured the isotopic separations in
the $\lambda$2026 transition between the most abundant isotope,
$^{64}$Zn, and $^{66,68,70}$Zn using laser cooling and Fabry-Perot
spectroscopic techniques. Laser cooling of $^{67}$Zn was not possible in
\citeauthor{Matsubara:2003:209}'s experiment, so we used \Eref{eq:iso}
to estimate its isotopic shift, $\delta\nu^{67,64}$, with
$\kSMS\approx-1365(20)$\,\GHzamu\ taken as a representative value from
the measured $\delta\nu^{68,66}$ in \citeauthor{Matsubara:2003:209}'s
table 1, and multiplying the field shift $\FS\delta\langle
r^2\rangle^{66,\,64}=0.35$\,GHz by the scaling parameter
$\delta\langle r^2\rangle^{67,\,66}/\delta\langle r^2\rangle^{66,\,64}
= 0.19$ from \citet{Campbell:1997:2351}.

There does not appear to be any measurements of the Zn{\sc \,ii}
$\lambda$2062 isotopic separations, so we estimated them based on
\Eref{eq:iso}, the calculations of \citet{Berengut:2003:022502} and
the measurements and calculations above for the $\lambda$2026 line. For
the $\lambda$2062 specific mass shift constant (\kSMS) we scaled the
measured value for $\lambda$2026 adopted above ($-1365$\,\GHzamu) by
the ratio of the calculated values for $\lambda$2062 and $\lambda$2026
from \citeauthor{Berengut:2003:022502} (i.e.~$-1310/-1266$). We
assumed the field shift (\FS) for the $\lambda$2062 transition was the
same as that derived from measurements by \citet{Matsubara:2003:209}
for the $\lambda$2026 transition, an assumption justified by the field
shift estimates of \citeauthor{Berengut:2003:022502} for the Zn{\sc
  \,ii} doublet.

The isotopic splittings illustrated in \Fref{fig:Zn_IHF} are
reasonably narrow. However, the terrestrial isotopic abundances being
distributed fairly evenly amongst three of the isotopes
($^{64,66,68}$Zn). Therefore, the systematic line-shifts induced by
possible isotopic abundance variations between quasar absorbers and
the terrestrial environment could be appreciable. For example,
\citet{Fenner:2005:468} found that the dominant $^{64}$Zn isotope in
the terrestrial environment may have been even more abundant
sub-dominant at the low metallicities typifying the highest column
density quasar absorbers. This could lead to spurious line-shifts in
both \ion{Zn}{ii} transitions of $\sim$100\,\ms\ (i.e.~the separation
between the $^{64}$\ion{Zn}{ii} and composite transition velocities in
\Fref{fig:Zn_IHF}).

To calculate the hyperfine splitting in the $^{67}$Zn{\sc \,ii}
$\lambda\lambda$2026/2062 transitions we used equations
(\ref{eq:hyp1})--(\ref{eq:hyp5}) with the magnetic dipole and electric
quadrupole hyperfine constants ($A$ and $B$) calculated for the ground
and excited states by \citet{Dixit:2008:025001}. Zn has a nuclear spin
of $I=5/2$ so that the $^{67}$Zn{\sc \,ii} $\lambda$2026
($\lambda$2062) transition has 6 (4) allowed transitions, three (two)
from each of the two ground states ($F=2$ and 3). The hyperfine
splitting is dominated by that in the ground states so, as illustrated
in \Fref{fig:Zn_IHF}, we represent the $^{67}$Zn{\sc \,ii}
$\lambda\lambda$2026/2062 hyperfine structures with just two
components each.

\Fref{fig:Zn_IHF} shows that the $^{67}$Zn hyperfine splittings are
quite large, $\sim$1.5\,\kms. This could cause substantial systematic
effects in the measured line velocity if the isotopic abundance
pattern -- where $^{67}$Zn comprises just 4.1\,\% of Zn in the
terrestrial environment \citep{Rosman:1998:1275} -- was very different
in quasar absorbers. The expectation from the simulations of
\citet{Fenner:2005:468} is that $^{67}$Zn should be even less abundant
at the lower metallicities typical of the highest column density
absorbers, but this is nevertheless a concern worth pointing out
because there is currently no observational constraint on the Zn
isotopic abundances in quasar absorbers.

Regarding the $q$ coefficients in \Tref{tab:Zn}, most calculations
treat \ion{Zn}{ii} as a single-valence-electron ion with a closed
$3d^{10}$ shell. While this is generally preferable to treating an 11
electron system with CI, it does make core--valence correlations
large.  Calculations of $q$ coefficients including MBPT and the Breit
interaction have been calculated by \citet{Dzuba:2002:022501} and
\citet{Savukov:2008:042501}, and these are in excellent agreement. On
the other hand, the coupled-cluster single--double method used in
\citet{Dzuba:2007:062510} gives slightly different values. We have
taken a simple average of the values from \citet{Dzuba:2007:062510}
and \citet{Savukov:2008:042501} in this case and assigned indicative
errors in \Tref{tab:Zn} which are large enough to accommodate both.

\setlength{\tabcolsep}{\oldtabcolsep}

\setlength{\tabcolsep}{\oldtabcolsep}

\section{Effects of ignoring isotopic/hyperfine structure in quasar absorption lines}\label{sec:IHF}

Some of the isotopic/hyperfine structures presented in
\Sref{sec:data}, particularly for transitions of \ion{Mg}{i} \& {\sc
  ii}, \ion{Al}{iii} and \ion{Mn}{ii}, are broad and somewhat
asymmetric (see Figs.~\ref{fig:Mg_IHF}, \ref{fig:Al_IHF} \&
\ref{fig:Mn_IHF}). Therefore, if those absorption lines are modelled
using only their composite wavelengths, without the underlying
isotopic/hyperfine structure, we can expect systematic errors in the
fitted parameters -- the redshifts (or velocities), column densities
and Doppler parameters. We explore these systematic effects below.

\subsection{Velocity shifts}\label{ssec:velshifts}

For varying-$\alpha$ studies, systematic effects on the fitted
redshifts (or velocities) of absorption lines are particularly
important. If absorption lines with underlying structure are fitted
with just a single component, velocity shifts will be measured which
may be attributed to a varying $\alpha$, or at least contribute
important systematic errors to varying-$\alpha$ measurements.

\begin{figure}
\begin{center}
\includegraphics[width=0.89\columnwidth,bb = 55 148 300 714]{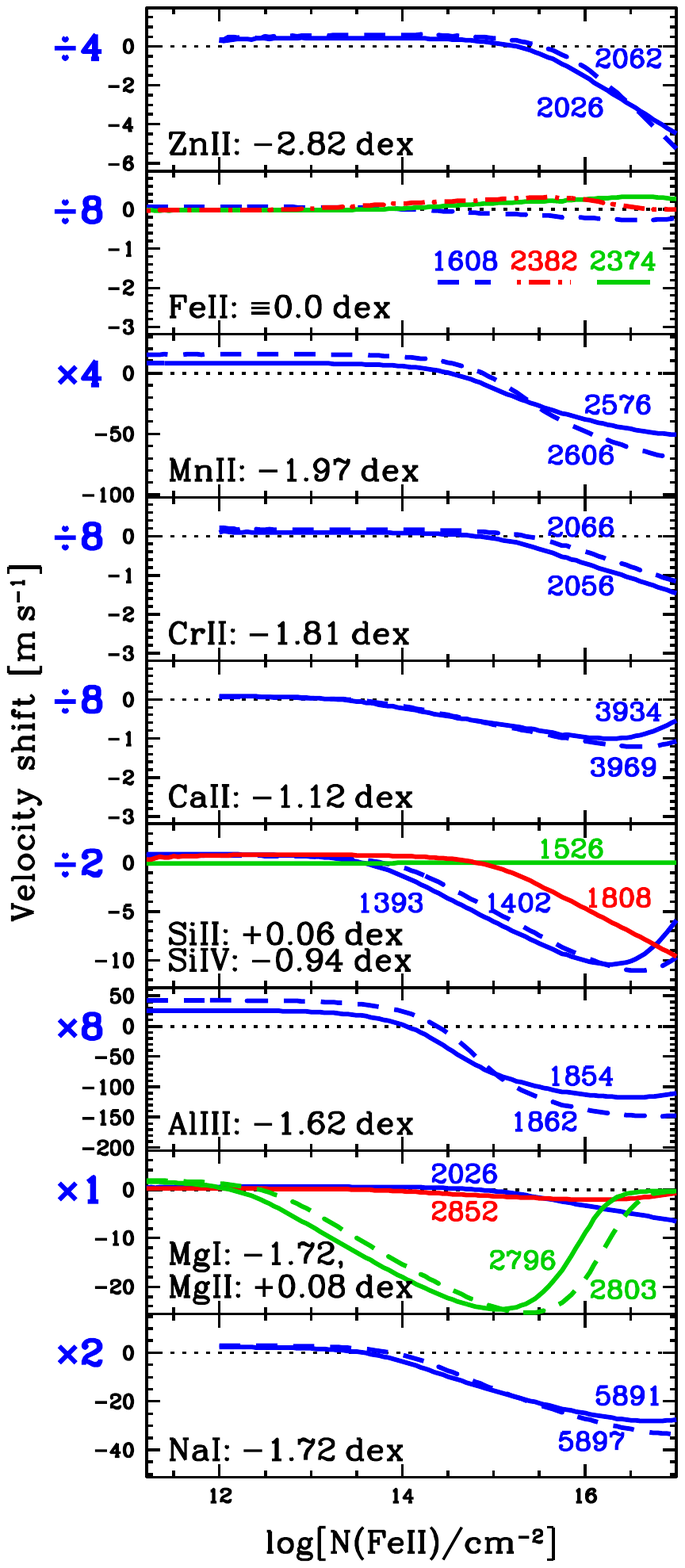}\vspace{-0.5em}
\caption{Spurious velocity shift derived from fitting synthetic absorption line spectra without taking into account the underlying isotopic/hyperfine structures. Selected ionic species are represented in different panels, with each line representing a different transition (labelled with its transition wavelength in \AA). The column density of each ionic species, relative to that of \ion{Fe}{ii}, assumed in the simulated spectra is provided in each panel. The vertical scale of each panel varies; as a quick guide, the
factor by which it differs to that of the Mg panel (bottom panel), is
provided at the left of each panel.}
\label{fig:velshifts_v}
\end{center}
\end{figure}

To explore the magnitude and sign of these spurious velocity shifts as
a function of the strength of the absorption lines, we created
synthetic, high signal-to-noise ratio ($\SN=10^5$ per pixel) spectra
with resolving power $R=75000$ and dispersion 1.3\,\kms\ containing a
single velocity component of all the transitions studied in this paper
and including the isotopic/hyperfine structures illustrated in
Figs.~\ref{fig:Mg_IHF}--\ref{fig:Zn_IHF}. The Doppler broadening
parameter was set to $b=2.5$\,\kms\ for all ionic species. The
relative column densities of the different singly-ionized species were
assumed to follow the meteoritic solar abundance pattern of
\citet[][table 1]{Asplund:2009:481}. To obtain representative column
densities for the sub-dominant ions, \ion{Mg}{i},
\ion{Al}{iii} and \ion{Si}{iv}, we assumed the median
observed values of $\log[N(\ion{Mg}{i})/N(\ion{Mg}{ii})]=-1.8$ by
\citet{Churchill:2001:679}, $\log[N(\ion{Al}{iii})/N(\ion{Al}{ii})]=-0.6$ and
$\log[N(\ion{Si}{iv})/N(\ion{Si}{ii})]=-1.0$ by
\citet{Prochaska:2002:933}. Representative column density ratios for the sub-dominant ions
\ion{Na}{i} and \ion{Ca}{ii} are much more difficult to estimate
because they vary strongly according to the local absorber conditions,
especially the dust content of the gas. Here we adopt the following
rough values for illustration purposes only, guided by the few
available measurements of $\log[N(\ion{Na}{i})/N(\ion{Ca}{ii})]$ and
photoionization modelling conducted by \citet{Richter:2011:A12}:
$\log[N(\ion{Na}{i})/N(\ion{Ca}{ii})]=-0.6$ and
$\log[N(\ion{Ca}{ii})/N(\ion{Mg}{ii})]=-1.2$.

We created synthetic spectra with a range of \ion{Fe}{ii} column
densities, scaling the other species' column densities appropriately,
and fitted them using only the composite transition wavelengths shown
in Tables \ref{tab:Na}--\ref{tab:Zn}. The fits were performed using
the non-linear least squares code {\sc
  vpfit}\footnote{{\urlstyle{rm}\url{http://www.ast.cam.ac.uk/~rfc/vpfit.html}}.}
which was designed for, and has been used in, numerous quasar
absorption line studies, including most previous varying-$\alpha$
analyses.

\Fref{fig:velshifts_v} shows the velocity shift caused by ignoring the
underlying isotopic/hyperfine structure of a single quasar absorption
line. Clearly, the largest systematic effects occur in high column
density absorption lines of \ion{Na}{i}, \ion{Mg}{\sc ii},
\ion{Al}{iii} and \ion{Mn}{ii}. For example, at a \ion{Mg}{ii} column
density of $\log[N(\ion{Mg}{ii})/{\rm
  cm}^{-2}]-0.08=\log[N(\ion{Fe}{ii})/{\rm cm}^{-2}]\sim15.5$, a
spurious velocity shift of $\sim -20$\,\ms\ will be measured in the
\ion{Mg}{ii} doublet transitions, while the broad and asymmetric
hyperfine structures of the \ion{Al}{iii} doublet and \ion{Mn}{ii}
triplet lead to velocity shifts of $\sim -120$ and $\sim -30$\,\ms\
respectively. The reason for the strong dependence of velocity shift
on the absorption line strength [parametrized by $N(\ion{Fe}{ii})$] is
due to `differential isotopic saturation' \citep{Murphy:2001:1223}: as
the column density increases and the strongest isotopic component of
the absorption line saturates, the relative influence of the weaker
components increases, shifting the line centroid towards the simple
mean of the isotopic component velocities. For the strongest
transitions, particularly the \ion{Mg}{ii} doublet, this trend is
overcome and reversed with even higher column density. This is because
the damping wings of the strongest isotopic components begin to
dominate the profile shape, moving the fitted centroid back towards
the component-strength-weighted mean velocity.

Finally, it is also evident in \Fref{fig:velshifts_v} that, for some
transitions, there is a small non-zero velocity shift even at very low
column densities, when the absorption line is optically thin. This is
due to the small non-Voigt shape of the absorption line introduced by
the underlying isotopic/hyperfine components. Thus, it is largest in
transitions of \ion{Al}{iii} and \ion{Mn}{ii} which have the most
extreme hyperfine structures studied here.

\subsection{Column densities and Doppler \boldmath{$b$}-parameters}\label{ssec:Nandb}

\begin{figure}
\begin{center}
\includegraphics[width=0.95\columnwidth,bb = 50 192 580 583]{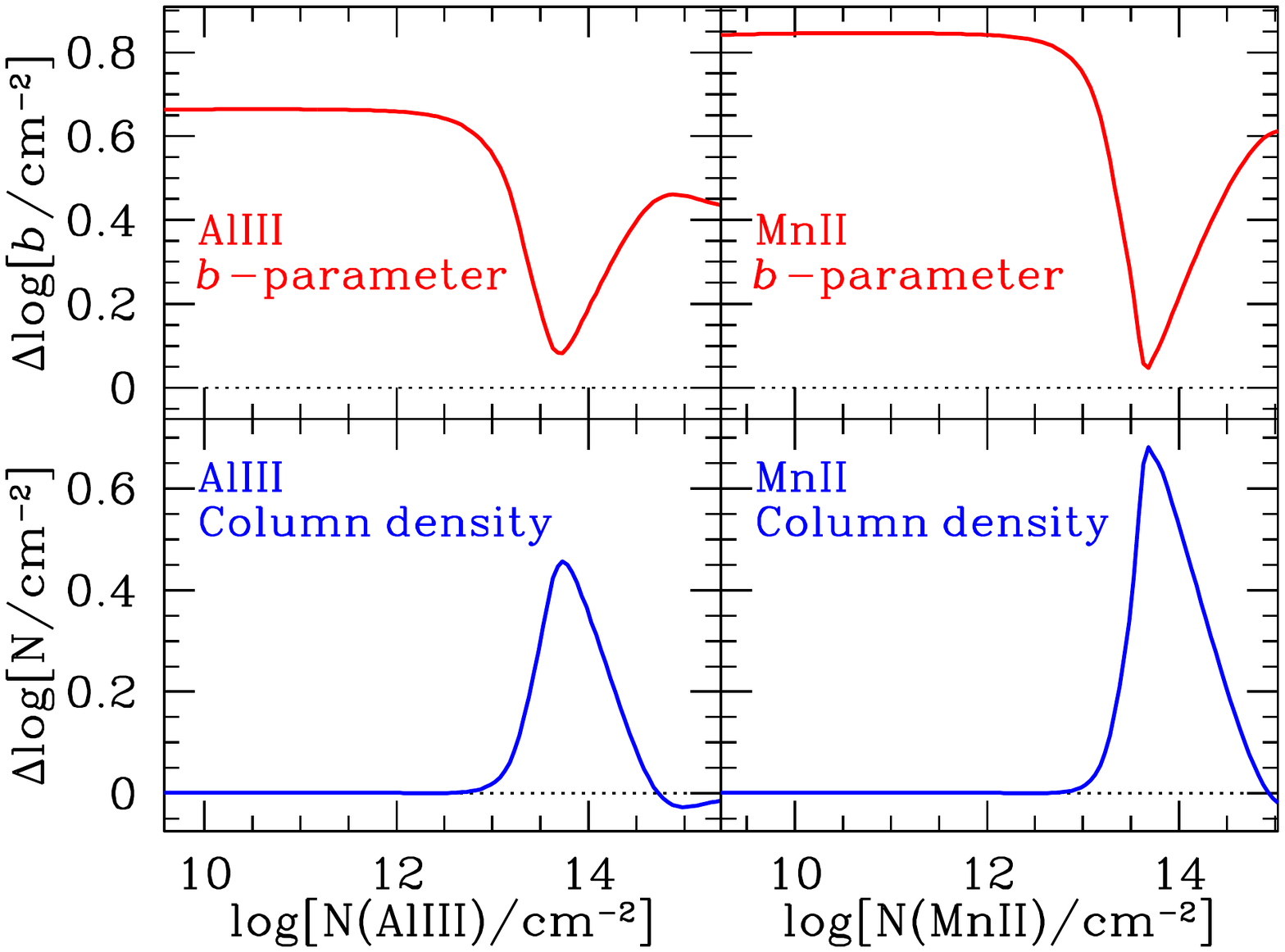}\vspace{-0.5em}
\caption{Systematic errors in the column densities and Doppler
  $b$-parameters of \ion{Al}{iii} and \ion{Mg}{ii} derived from
  simulations with an input $b=2.5$\,\kms. The simulations were
  created with the prominent hyperfine structures of the \ion{Al}{iii}
  and \ion{Mg}{ii} shown in Figs.~\ref{fig:Al_IHF} and
  \ref{fig:Mg_IHF} but were fitted with a model lacking that
  underlying structure.}
\label{fig:velshifts_n}
\end{center}
\end{figure}

\Fref{fig:velshifts_n} shows the systematic effect on the fitted
column density ($N$) and Doppler broadening parameters ($b$) of
\ion{Al}{iii} and \ion{Mn}{ii} when the very pronounced hyperfine
structures of their transitions are ignored (see
Figs.~\ref{fig:Al_IHF} and \ref{fig:Mn_IHF}). The same simulations and
fitting approach used to derive the spurious velocity shifts in
\Sref{ssec:velshifts} were used here. Clearly, the systematic
effects on $N$ and $b$ are anti-correlated, as expected if the fit is
to recover approximately the correct equivalent width of absorption in
each transition. However, it is important to recognise that, at low
$N$ values where the transitions are optically thin
(i.e.~$\log[N/{\rm cm}^{-2}]\la13$), the systematic effect on $N$ is
negligible while that on $b$ is a maximum and $\sim$0.7\,\kms. The
opposite is true when the transitions are saturated, with the effect
on $b$ reducing considerably to only $\sim$0.2\,\kms but that on $N$
increasing to $\sim$0.4--0.6\,dex, or column density overestimates
by factors of $\sim$2.5--4. However, for saturated transitions at more
realistic \SN\ values of $\sim$15--50\,per pixel (cf.~$10^5$ in the
simulations for \Fref{fig:velshifts_n}), the statistical uncertainty
on the column density is very large and altogether
unreliable.

Thus, of the $b$ parameters and column densities, the former are most
affected in practical sense by ignoring the hyperfine structures is
the systematic error on $b$ for optically thin lines. In
varying-$\alpha$ studies, this will manifest itself as an inability to
simultaneously fit the \ion{Al}{iii} and \ion{Mn}{ii} with other
species with much narrower or no isotopic/hyperfine structures:
linking the $b$ parameters between species in the fit may yield poor
(i.e.~statistically unacceptable or unlikely) fits to some/all
transitions.

\section{Conclusions}\label{sec:conc}

We have reviewed and synthesized the existing laboratory atomic data
for metal-line transitions with precisely-known frequencies ($\delta
v\la20$\,\ms). These transitions are of particular interest for quasar
absorption-line studies of possible cosmological variations in the
fine-structure constant, $\alpha$, where the primary concern is the
accuracy of the laboratory frequency and knowledge about any
significant isotopic and/or hyperfine structure. For all transitions
presented here, the composite frequencies should be accurate to within
the stated uncertainties, as should the isotopic/hyperfine frequencies
in transitions where those quantities are measured directly
(\ion{Na}{i}, \ion{Mg}{i} \& {\sc ii}, \ion{Al}{iii}, \ion{Ca}{ii},
\ion{Mn}{ii} and \ion{Zn}{ii} $\lambda$2026). However, for most
transitions -- those of \ion{Si}{ii} \& {\sc iv}, \ion{Ti}{ii},
\ion{Cr}{ii}, \ion{Fe}{ii} and \ion{Zn}{ii} $\lambda$2062 -- the
isotopic/hyperfine splittings are calculated, typically with \emph{ab
  initio} techniques. Indeed, we presented new isotopic structure
calculations of this sort for the \ion{Si}{ii} and \ion{Ti}{ii}
transitions, including \ion{Si}{ii} $\lambda$1808 and \ion{Ti}{ii}
$\lambda\lambda$1910.6/1910.9 for the first time, and found close
agreement with previous estimates for the other
transitions. Nevertheless, it is difficult to estimate how closely any
calculated isotopic structures reflect the real ones; isotopic
structure \emph{measurements} for these transitions are
encouraged. Finally, note that we have no current information about
the isotopic structures of \ion{Fe}{ii} $\lambda$2260 and the
\ion{Ni}{ii} transitions of interest.

Independent, repeated laboratory frequency measurements are critical
for ensuring that tests of the invariance of the fundamental constants
over cosmological time- and distance-scales are reliable. Most
transitions studied here have multiple laboratory frequency
measurements which, following the re-calibration work of
\citet{Nave:2011:737} and \citet{Nave:2012:1570}, agree within the
uncertainties. The exceptions to this are the transitions measured by
\citet{Griesmann:2000:L113} (i.e.~\ion{Al}{ii} \& {\sc iii} and
\ion{Si}{ii} \& {\sc iv}), the bluest and reddest \ion{Ti}{ii}
transitions ($\lambda$1910.6, 1910.9, 3230, 3242 and 3384), some
relatively weak, though nevertheless important, \ion{Fe}{ii}
transitions ($\lambda$1608, 1611, 2249 and 2367), and the \ion{Ni}{ii}
lines. All the transitions presented here are calibrated on frequency
scales that are consistent with the absolute and highly-accurate one
derived in frequency comb measurements. The accuracy of the
frequencies for many transitions relies on the \ion{Ar}{ii}
calibration scale of \citet{Whaling:1995:1} (which is consistent with
frequency comb measurements). However, for the far-UV,
\ion{Ar}{ii}-calibrated transitions without independent, repeated
measurements \citep[e.g.~those of][]{Griesmann:2000:L113},
\citet{Nave:2012:1570} cautions that extending the calibration far
into the UV with confidence is difficult.

\Tref{tab:missing} summarizes the most important missing information
for the ions considered in this paper. It includes the transitions in
Tables \ref{tab:Na}--\ref{tab:Zn} for which repeated laboratory
frequency measurements are desirable and/or we have no current
information (measurement or calculation) about their isotopic
structures. This table also includes ions which did not qualify to
appear in Tables \ref{tab:Na}--\ref{tab:Zn}, either because the
laboratory frequency has not been measured with $\delta v\la20$\,\ms\
accuracy and/or the $q$ coefficient is not known (to our
knowledge). Missing information which prevents ions not considered in
this paper from being useful in varying-$\alpha$ studies is summarized
in \citet{Berengut:2011:9}.

\setlength{\tabcolsep}{0.395em}
\begin{table}
\caption{Summary of the most important missing information for the ions discussed in this paper. For other ions, see \citet{Berengut:2011:9}. The ``Missing'' column specifies the new measurements/calculations that are either required before the transition can be used in varying-$\alpha$ analyses (underlined) or desirable to improve the reliability of such work (not underlined). In particular, the transitions from Tables \ref{tab:Na}--\ref{tab:Zn} for which repeated laboratory measurements do not exist are included here.
}
\label{tab:missing}\vspace{-0.4em}
\setlength{\extrarowheight}{0.7pt}
{\footnotesize
\begin{tabular}{:l;l;l;l;c;l}\hline
\multicolumn{1}{c}{Ion}&
\multicolumn{1}{c}{Tran.}&
\multicolumn{1}{c}{$\omega_0$}&
\multicolumn{1}{c}{$\lambda_0$}&
\multicolumn{1}{c}{ID}&
\multicolumn{1}{c}{Missing$^{a}$}\\
&
&
\multicolumn{1}{c}{[cm$^{-1}$]}&
\multicolumn{1}{c}{[\AA]}&
&
 \\\hline
Al{\sc \,ii}  & 1670   & 59851.976(4)$^{b}$    & 1670.78861(11)  & Al$^{2}_{1}$ & $\omega$ \\  
Al{\sc \,iii} & 1854   & 53916.5480(11)$^{b}$  & 1854.718146(39) & Al$^{3}_{1}$ & $\omega$ \\  
              & 1862   & 53682.8884(11)$^{b}$  & 1862.790974(39) & Al$^{3}_{2}$ & $\omega$ \\  
Si{\sc \,ii}  & 1260   & 79338.50$^{c}$        & 1260.4221       & Si$^{2}_{1}$ & \underline{$\omega$} \\
              & 1304   & 76665.35$^{c}$        & 1304.3702       & Si$^{2}_{3}$ & \underline{$\omega$} \\
              & 1526   & 65500.4538(7)$^{b}$   & 1526.706980(16) & Si$^{2}_{2}$ & $\omega$ \\  
              & 1808   & 55309.3404(4)$^{b}$   & 1808.012883(13) & Si$^{2}_{4}$ & $\omega$ \\  
Si{\sc \,iv}  & 1393   & 71748.355(2)$^{b}$    & 1393.760177(39) & Si$^{4}_{1}$ & $\omega$ \\  
              & 1402   & 71287.376(2)$^{b}$    & 1402.772912(39) & Si$^{4}_{2}$ & $\omega$ \\  
Ti{\sc \,ii}  & 1910.6 & 52339.240(1)$^{d}$    & 1910.61238(4)   & Ti$^{2}_{4}$ & $\omega$ \\  
              & 1910.9 & 52329.889(1)$^{d}$    & 1910.95380(4)   & Ti$^{2}_{5}$ & $\omega$ \\  
              & 3230   & 30958.5871(10)$^{e,f}$& 3230.12157(10)  & Ti$^{2}_{6}$ & $\omega$ \\  
              & 3242   & 30836.4271(10)$^{e,f}$& 3242.91785(11)  & Ti$^{2}_{2}$ & $\omega$ \\  
              & 3384   & 29544.4551(10)$^{e,f}$& 3384.72988(11)  & Ti$^{2}_{1}$ & $\omega$ \\  
Fe{\sc \,i}   & 2167   & 46137.098(11)$^{g,h}$ & 2167.4532(5)    & Fe$^{1}_{3}$ & \underline{$\omega$}, \underline{$q$} \\
              & 2298   & 43499.506(6)$^{g,h}$  & 2298.8767(3)    & Fe$^{1}_{11}$& \underline{$\omega$}, \underline{$q$} \\
              & 2463   & 40594.432(5)$^{g,h}$  & 2463.3920(3)    & Fe$^{1}_{6}$ & \underline{$\omega$} \\
              & 2484   & 40257.313(10)$^{g,h}$ & 2484.0207(6)    & Fe$^{1}_{1}$ & \underline{$\omega$} \\
              & 2501   & 39969.852(5)$^{g,h}$  & 2501.8857(3)    & Fe$^{1}_{7}$ & \underline{$\omega$}, I \\
              & 2523   & 39625.804(5)$^{g,h}$  & 2523.6081(3)    & Fe$^{1}_{2}$ & \underline{$\omega$} \\
              & 2719   & 36766.966(5)$^{g,h}$  & 2719.8328(4)    & Fe$^{1}_{4}$ & \underline{$\omega$} \\
              & 2967   & 33695.397(5)$^{g,h}$  & 2967.7644(4)    & Fe$^{1}_{8}$ & \underline{$\omega$} \\
              & 2984   & 33507.124(4)$^{g,h}$  & 2984.4400(4)    & Fe$^{1}_{10}$& \underline{$\omega$} \\
              & 3021   & 33095.942(5)$^{g,h}$  & 3021.5185(5)    & Fe$^{1}_{5}$ & \underline{$\omega$} \\
              & 3441   & 29056.323(5)$^{g,h}$  & 3441.5917(6)    & Fe$^{1}_{12}$& \underline{$\omega$}, $\Gamma$ \\
              & 3720   & 26874.550(5)$^{g,h}$  & 3720.9926(7)    & Fe$^{1}_{9}$ & \underline{$\omega$} \\
              & 3861   & 25899.989(5)$^{g,h}$  & 3861.0055(7)    & Fe$^{1}_{13}$& \underline{$\omega$} \\
Fe{\sc \,ii}  & 1260   & 79331.359(4)$^{f}$    & 1260.53557(6)   & Fe$^{2}_{7}$ & $\omega$, \underline{$q$} \\  
              & 1588   & 62945.045(2)$^{i}$    & 1588.68741(6)   & Fe$^{2}_{11}$& $\omega$, \underline{$q$} \\
              & 1608   & 62171.623(3)$^{f}$    & 1608.450852(78) & Fe$^{2}_{5}$ & $\omega$ \\  
              & 1611   & 62065.527(3)$^{f}$    & 1611.200369(78) & Fe$^{2}_{10}$& $\omega$ \\  
              & 1620   & 61726.069(3)$^{i}$    & 1620.06106(7)   & Fe$^{2}_{13}$& $\omega$, \underline{$q$} \\
              & 1901   & 52582.850(2)$^{i}$    & 1901.76075(9)   & Fe$^{2}_{12}$& $\omega$, \underline{$q$} \\
              & 2234   & 44753.818(2)$^{i}$    & 2234.44624(9)   & Fe$^{2}_{14}$& $\omega$, \underline{$q$} \\
              & 2249   & 44446.9044(19)$^{f}$  & 2249.875472(96) & Fe$^{2}_{9}$ & $\omega$ \\  
              & 2260   & 44232.5390(19)$^{e,f}$& 2260.779108(97) & Fe$^{2}_{8}$ & I \\  
              & 2367   & 42237.0563(19)$^{f}$  & 2367.58924(11)  & Fe$^{2}_{15}$& $\omega$ \\
Ni{\sc \,ii}  & 1317   & 75917.64$^{c}$        & 1317.217        & Ni$^{2}_{2}$ & \underline{$\omega$}, I \\  
              & 1370   & 72985.67$^{c}$        & 1370.132        & Ni$^{2}_{1}$ & \underline{$\omega$}, I \\  
              & 1393   & 71770.82$^{c}$        & 1393.324        & Ni$^{2}_{7}$ & \underline{$\omega$}, $f$, I \\  
              & 1454   & 68735.99$^{c}$        & 1454.842        & Ni$^{2}_{5}$ & \underline{$\omega$}, I \\  
              & 1467.2 & 68154.31$^{c}$        & 1467.259        & Ni$^{2}_{9}$ & \underline{$\omega$}, I \\  
              & 1467.7 & 68131.21$^{c}$        & 1467.756        & Ni$^{2}_{8}$ & \underline{$\omega$}, I \\  
              & 1502   & 66571.34$^{c}$        & 1502.148        & Ni$^{2}_{11}$& \underline{$\omega$}, $f$, I \\  
              & 1703   & 58705.713(15)$^{j,f}$ & 1703.41172(44)  & Ni$^{2}_{10}$& \underline{$\omega$}, I \\  
              & 1709   & 58493.0772(40)$^{j,f}$& 1709.60402(12)  & Ni$^{2}_{4}$ & $\omega$, I \\  
              & 1741   & 57420.0191(40)$^{j,f}$& 1741.55289(12)  & Ni$^{2}_{3}$ & $\omega$, I \\  
              & 1751   & 57080.3791(40)$^{j,f}$& 1751.91549(12)  & Ni$^{2}_{6}$ & $\omega$, I \vspace{0.2em}\\
\hline
\end{tabular}
}
\setlength{\extrarowheight}{0pt}
{
\footnotesize
$^{a}$Key for ``missing'' information: $\omega$\,=\,frequency meas.; $q$\,=\,$q$ coefficient calc.; $f$ = oscillator strength meas./calc.; $\Gamma$\,=\,damping constant meas./calc.; I = isotopic structure meas./calc.\\
References:
$^{b}$\citet{Griesmann:2000:L113};
$^{c}$\citet{Morton:2003:205};
$^{d}$\citet{Ruffoni:2010:424};
$^{e}$\citet{Aldenius:2009:014008};
$^{f}$\citet{Nave:2012:1570};
$^{g}$\citet{Nave:1994:221};
$^{h}$\citet{Nave:2011:737};
$^{i}$\citet{Nave:2013:1};
$^{j}$\citet{Pickering:2000:163}.}
\end{table}
\setlength{\tabcolsep}{\oldtabcolsep}

Given the now rather complete, if not completely certain,
isotopic/hyperfine structure information for most transitions studied
here, we were able to explore the likely systematic errors incurred in
quasar absorption-line analyses if these structures are ignored. For
example, differential saturation of the Mg isotopic components will
cause spurious line-shifts of up to $\sim$20\,\ms\ in the strong and
often-used \ion{Mg}{ii} doublet transitions if the isotopic structure
is not modelled. The same effect in the broadly-spaced hyperfine
structures of the \ion{Al}{iii} and \ion{Mn}{ii} transitions is much
more pronounced, particularly for the former, causing spurious shifts
of up to $\sim$120\,\ms. We also find that the Doppler broadening
parameters ($b$) for the \ion{Al}{iii} and \ion{Mn}{ii} transitions
will be systematically overestimated by $\sim$0.6 and $\sim$0.8\,\kms,
respectively, in the optically thin regime if the hyperfine structures
are not taken into account. However, the column density estimates will
be almost unaffected in such cases, assuming that more than one
transition of each species is measured simultaneously and the velocity
structure of the absorption is simple and well-resolved. These latter
conditions are often not met in real quasar absorption analyses. For
varying-$\alpha$ analyses, in which accurate absorption profile
modelling is required, ignoring the \ion{Al}{iii} and \ion{Mn}{ii}
hyperfine structures causes substantial difficulties in fitting these
transitions simultaneously with those of other species, undermining
the assumptions and reliability of the many-multiplet method.

Finally, to ensure the transparency, repeatability and ease of
updating the laboratory data and isotopic/hyperfine structure
calculations presented here, we supply a spreadsheet containing all
data details and calculations in the Supporting Information. This will
be maintained and updated at
{\urlstyle{rm}\url{https://researchdata.ands.org.au/laboratory-atomic-transition-data-for-precise-optical-quasar-absorption-spectroscopy}}
and readers are invited to alert the authors to updated information
and/or to request that new transitions be included.

\section*{Acknowledgments}

The authors thank the Australian Research Council for
\textit{Discovery Project} grant DP110100866 which supported this
work. Part of this research was undertaken on the NCI National
Facility, which is supported by the Australian Commonwealth
Government.


\begin{thebibliography}{93}
\small
\itemindent -0.48cm
\expandafter\ifx\csname natexlab\endcsname\relax\def\natexlab#1{#1}\fi

\bibitem[{Aldenius(2009)}]{Aldenius:2009:014008}
Aldenius M., 2009, Phys. Scr., 134, 014008

\bibitem[{Aldenius {et~al}\mbox{.}(2006)Aldenius, Johansson, \&
  Murphy}]{Aldenius:2006:444}
Aldenius M., Johansson S., Murphy M.T., 2006, MNRAS, 370, 444

\bibitem[{Angeli(2004)}]{Angeli:2004:185}
Angeli I., 2004, At. Data Nucl. Data Tables, 87, 185

\bibitem[{Angstmann {et~al}\mbox{.}(2004)Angstmann, Dzuba, \&
  Flambaum}]{Angstmann:2004:014102}
Angstmann E.J., Dzuba V.A., Flambaum V.V., 2004, Phys. Rev. A, 70, 014102

\bibitem[{Arbes {et~al}\mbox{.}(1994)Arbes, Benzing, Gudjons, Kurth, \&
  Werth}]{Arbes:1994:27}
Arbes F., Benzing M., Gudjons T., Kurth F., Werth G., 1994, Z. Phys. D, 31, 27

\bibitem[{Ashenfelter {et~al}\mbox{.}(2004)Ashenfelter, Mathews, \&
  Olive}]{Ashenfelter:2004:041102}
Ashenfelter T., Mathews G.J., Olive K.A., 2004, Phys. Rev. Lett., 92, 041102

\bibitem[{Asplund {et~al}\mbox{.}(2009)Asplund, Grevesse, Sauval, \&
  Scott}]{Asplund:2009:481}
Asplund M., Grevesse N., Sauval A.J., Scott P., 2009, ARA\&A, 47, 481

\bibitem[{Bahcall \& Salpeter(1965)}]{Bahcall:1965:1677}
Bahcall J.N., Salpeter E.E., 1965, ApJ, 142, 1677

\bibitem[{Balling \& K{\v r}en(2008)}]{Balling:2008:3}
Balling P., K{\v r}en P., 2008, Eur. Phys. J. D, 48, 3

\bibitem[{Batteiger {et~al}\mbox{.}(2009)Batteiger, Kn{\"u}nz, Herrmann,
  Saathoff, Sch{\"u}ssler, Bernhardt, Wilken, Holzwarth,
  {et~al.}}]{Batteiger:2009:022503}
Batteiger V. {et~al.}, 2009, Phys. Rev. A, 80, 022503

\bibitem[{Beckmann {et~al}\mbox{.}(1974)Beckmann, B{\"o}klen, \&
  Elke}]{Beckmann:1974:173}
Beckmann A., B{\"o}klen K.D., Elke D., 1974, Z. Phys., 270, 173

\bibitem[{{Berengut}(2006)}]{Berengut:2006:PhD}
{Berengut} J.C., 2006, PhD thesis, Univ. New South Wales

\bibitem[{Berengut(2011)}]{Berengut:2011:052520}
Berengut J.C., 2011, Phys. Rev. A, 84, 052520

\bibitem[{Berengut {et~al}\mbox{.}(2003)Berengut, Dzuba, \&
  Flambaum}]{Berengut:2003:022502}
Berengut J.C., Dzuba V.A., Flambaum V.V., 2003, Phys. Rev. A, 68, 022502

\bibitem[{{Berengut} {et~al}\mbox{.}(2011){Berengut}, {Dzuba}, {Flambaum},
  {King}, {Kozlov}, {Murphy}, \& {Webb}}]{Berengut:2011:9}
{Berengut} J.C., {Dzuba} V.A., {Flambaum} V.V., {King} J.A., {Kozlov}
  M.G., {Murphy} M.T., {Webb} J.K., 2011, {Atomic Transition Frequencies,
  Isotope Shifts, and Sensitivity to Variation of the Fine Structure Constant
  for Studies of Quasar Absorption Spectra}, {Martins} C., {Molaro} P., eds.,
  Astrophys. Space Sci. Proc., Springer, Berlin Heidelberg, p.9,
  arXiv:1011.4136

\bibitem[{Berengut {et~al}\mbox{.}(2004)Berengut, Dzuba, Flambaum, \&
  Marchenko}]{Berengut:2004:064101}
Berengut J.C., Dzuba V.A., Flambaum V.V., Marchenko M.V., 2004, Phys. Rev.
  A, 70, 064101

\bibitem[{Berengut {et~al}\mbox{.}(2005)Berengut, Flambaum, \&
  Kozlov}]{Berengut:2005:044501}
Berengut J.C., Flambaum V.V., Kozlov M.G., 2005, Phys. Rev. A, 72, 044501

\bibitem[{Berengut {et~al}\mbox{.}(2006)Berengut, Flambaum, \&
  Kozlov}]{Berengut:2006:012504}
Berengut J.C., Flambaum V.V., Kozlov M.G., 2006, Phys. Rev. A, 73, 012504

\bibitem[{Berengut {et~al}\mbox{.}(2008)Berengut, Flambaum, \&
  Kozlov}]{Berengut:2008:235702}
Berengut J.C., Flambaum V.V., Kozlov M.G., 2008, J. Phys. B, 41, 235702

\bibitem[{Blackwell-Whitehead {et~al}\mbox{.}(2005)Blackwell-Whitehead, Toner,
  Hibbert, Webb, \& Ivarsson}]{Blackwell-Whitehead:2005:705}
Blackwell-Whitehead R.J., Toner A., Hibbert A., Webb J., Ivarsson S., 2005,
  MNRAS, 364, 705

\bibitem[{Bouch{\'e} {et~al}\mbox{.}(2013)Bouch{\'e}, Murphy, Kacprzak,
  P{\'e}roux, Contini, Martin, \& Dessauges-Zavadsky}]{Bouche:2013:50}
Bouch{\'e} N., Murphy M.T., Kacprzak G.G., P{\'e}roux C., Contini T., Martin
  C.L., Dessauges-Zavadsky M., 2013, Science, 341, 50

\bibitem[{Campbell {et~al}\mbox{.}(1997)Campbell, Billowes, \&
  Grant}]{Campbell:1997:2351}
Campbell P., Billowes J., Grant I.S., 1997, J. Phys. B, 30, 2351

\bibitem[{Churchill \& Vogt(2001)}]{Churchill:2001:679}
Churchill C.W., Vogt S.S., 2001, AJ, 122, 679

\bibitem[{Cowie \& Songaila(1995)}]{Cowie:1995:596}
Cowie L.L., Songaila A., 1995, ApJ, 453, 596

\bibitem[{Dessauges-Zavadsky {et~al}\mbox{.}(2007)Dessauges-Zavadsky, Calura,
  Prochaska, D'Odorico, \& Matteucci}]{Dessauges-Zavadsky:2007:431}
Dessauges-Zavadsky M., Calura F., Prochaska J.X., D'Odorico S., Matteucci F.,
  2007, A\&A, 470, 431

\bibitem[{Dixit {et~al}\mbox{.}(2008)Dixit, Nataraj, Sahoo, Chaudhuri, \&
  Majumder}]{Dixit:2008:025001}
Dixit G., Nataraj H.S., Sahoo B.K., Chaudhuri R.K., Majumder S., 2008, J.
  Phys. B, 41, 025001

\bibitem[{Dzuba \& Flambaum(2008)}]{Dzuba:2008:012514}
Dzuba V.A., Flambaum V.V., 2008, Phys. Rev. A, 77, 012514

\bibitem[{Dzuba {et~al}\mbox{.}(1996)Dzuba, Flambaum, \&
  Kozlov}]{Dzuba:1996:3948}
Dzuba V.A., Flambaum V.V., Kozlov M.G., 1996, Phys. Rev. A, 54, 3948

\bibitem[{Dzuba {et~al}\mbox{.}(2002)Dzuba, Flambaum, Kozlov, \&
  Marchenko}]{Dzuba:2002:022501}
Dzuba V.A., Flambaum V.V., Kozlov M.G., Marchenko M., 2002, Phys. Rev. A,
  66, 022501

\bibitem[{Dzuba {et~al}\mbox{.}(1999{\natexlab{a}})Dzuba, Flambaum, \&
  Webb}]{Dzuba:1999:230}
Dzuba V.A., Flambaum V.V., Webb J.K., 1999{\natexlab{a}}, Phys. Rev. A, 59,
  230

\bibitem[{Dzuba {et~al}\mbox{.}(1999{\natexlab{b}})Dzuba, Flambaum, \&
  Webb}]{Dzuba:1999:888}
Dzuba V.A., Flambaum V.V., Webb J.K., 1999{\natexlab{b}}, Phys. Rev. Lett.,
  82, 888

\bibitem[{Dzuba \& Johnson(2007)}]{Dzuba:2007:062510}
Dzuba V.A., Johnson W.R., 2007, Phys. Rev. A, 76, 062510

\bibitem[{Fenner {et~al}\mbox{.}(2005)Fenner, Murphy, \&
  Gibson}]{Fenner:2005:468}
Fenner Y., Murphy M.T., Gibson B.K., 2005, MNRAS, 358, 468

\bibitem[{Griesmann \& Kling(2000)}]{Griesmann:2000:L113}
Griesmann U., Kling R., 2000, ApJL, 536, L113

\bibitem[{Gunn \& Peterson(1965)}]{Gunn:1965:1633}
Gunn J.E., Peterson B.A., 1965, ApJ, 142, 1633

\bibitem[{Hannemann {et~al}\mbox{.}(2006)Hannemann, Salumbides, Witte,
  Zinkstok, van Duijn, Eikema, \& Ubachs}]{Hannemann:2006:012505}
Hannemann S., Salumbides E.J., Witte S., Zinkstok R.T., van Duijn E.-J.,
  Eikema K. S.E., Ubachs W., 2006, Phys. Rev. A, 74, 012505

\bibitem[{Itano \& Wineland(1981)}]{Itano:1981:1364}
Itano W.M., Wineland D.J., 1981, Phys. Rev. A, 24, 1364

\bibitem[{Johnson \& Sapirstein(1986)}]{Johnson:1986:1126}
Johnson W.R., Sapirstein J., 1986, Phys. Rev. Lett., 57, 1126

\bibitem[{Jomaron {et~al}\mbox{.}(1999)Jomaron, Dworetsky, \&
  Allen}]{Jomaron:1999:555}
Jomaron C.M., Dworetsky M.M., Allen C.S., 1999, MNRAS, 303, 555

\bibitem[{Juncar {et~al}\mbox{.}(1981)Juncar, Pinard, Hamon, \&
  Chartier}]{Juncar:1981:77}
Juncar P., Pinard J., Hamon J., Chartier A., 1981, Metrologia, 17, 77

\bibitem[{King {et~al}\mbox{.}(2012)King, Webb, Murphy, Flambaum, Carswell,
  Bainbridge, Wilczynska, \& Koch}]{King:2012:3370}
King J.A., Webb J.K., Murphy M.T., Flambaum V.V., Carswell R.F.,
  Bainbridge M.B., Wilczynska M.R., Koch F.E., 2012, MNRAS, 422, 3370

\bibitem[{Kozlov {et~al}\mbox{.}(2004)Kozlov, Korol, Berengut, Dzuba, \&
  Flambaum}]{Kozlov:2004:062108}
Kozlov M.G., Korol V.A., Berengut J.C., Dzuba V.A., Flambaum V.V., 2004,
  Phys. Rev. A, 70, 062108

\bibitem[{Kozlov \& Porsev(1999)}]{Kozlov:1999:352}
Kozlov M.G., Porsev S.G., 1999, Opt. Spectrosc., 87, 352

\bibitem[{Levshakov {et~al}\mbox{.}(2006)Levshakov, Centuri{\'o}n, Molaro, \&
  Kostina}]{Levshakov:2006:L21}
Levshakov S.A., Centuri{\'o}n M., Molaro P., Kostina M.V., 2006, A\&A, 447,
  L21

\bibitem[{Lu {et~al}\mbox{.}(1993)Lu, Wolfe, Turnshek, \& Lanzetta}]{Lu:1993:1}
Lu L., Wolfe A.M., Turnshek D.A., Lanzetta K.M., 1993, ApJS, 84, 1

\bibitem[{Maleki \& Goble(1992)}]{Maleki:1992:524}
Maleki S., Goble A.T., 1992, Phys. Rev. A, 45, 524

\bibitem[{M{\aa}rtensson-Pendrill {et~al}\mbox{.}(1992)M{\aa}rtensson-Pendrill,
  Ynnerman, Warston, Vermeeren, Silverans, Klein, Neugart, Schulz,
  {et~al.}}]{Martensson-Pendrill:1992:4675}
M{\aa}rtensson-Pendrill A.-M. {et~al.}, 1992, Phys. Rev. A, 45, 4675

\bibitem[{Matsubara {et~al}\mbox{.}(2003)Matsubara, Tanaka, Imajo, Urabe, \&
  Watanabe}]{Matsubara:2003:209}
Matsubara K., Tanaka U., Imajo H., Urabe S., Watanabe M., 2003, App. Phys. B,
  76, 209

\bibitem[{Molaro {et~al}\mbox{.}(2013)Molaro, Centuri{\'o}n, Whitmore, Evans,
  Murphy, Agafonova, Bonifacio, D'Odorico, {et~al.}}]{Molaro:2013:A68}
Molaro P. {et~al.}, 2013, A\&A, 555, A68

\bibitem[{Molaro {et~al}\mbox{.}(2008)Molaro, Reimers, Agafonova, \&
  Levshakov}]{Molaro:2008:173}
Molaro P., Reimers D., Agafonova I.I., Levshakov S.A., 2008, Eur. Phys. J.
  Special Topics, 163, 173

\bibitem[{Morton(1991)}]{Morton:1991:119}
Morton D.C., 1991, ApJS, 77, 119

\bibitem[{Morton(2003)}]{Morton:2003:205}
Morton D.C., 2003, ApJS, 149, 205

\bibitem[{Murphy {et~al}\mbox{.}(2003)Murphy, Webb, \&
  Flambaum}]{Murphy:2003:609}
Murphy M.T., Webb J.K., Flambaum V.V., 2003, MNRAS, 345, 609

\bibitem[{Murphy {et~al}\mbox{.}(2001{\natexlab{a}})Murphy, Webb, Flambaum,
  Churchill, \& Prochaska}]{Murphy:2001:1223}
Murphy M.T., Webb J.K., Flambaum V.V., Churchill C.W., Prochaska J.X.,
  2001{\natexlab{a}}, MNRAS, 327, 1223

\bibitem[{Murphy {et~al}\mbox{.}(2001{\natexlab{b}})Murphy, Webb, Flambaum,
  Dzuba, Churchill, Prochaska, Barrow, \& Wolfe}]{Murphy:2001:1208}
Murphy M.T., Webb J.K., Flambaum V.V., Dzuba V.A., Churchill C.W.,
  Prochaska J.X., Barrow J.D., Wolfe A.M., 2001{\natexlab{b}}, MNRAS, 327,
  1208

\bibitem[{Murphy {et~al}\mbox{.}(2001{\natexlab{c}})Murphy, Webb, Flambaum,
  Prochaska, \& Wolfe}]{Murphy:2001:1237}
Murphy M.T., Webb J.K., Flambaum V.V., Prochaska J.X., Wolfe A.M.,
  2001{\natexlab{c}}, MNRAS, 327, 1237

\bibitem[{Nave(2012)}]{Nave:2012:1570}
Nave G., 2012, MNRAS, 420, 1570

\bibitem[{Nave \& Johansson(2013)}]{Nave:2013:1}
Nave G., Johansson S., 2013, ApJS, 204, 1

\bibitem[{Nave {et~al}\mbox{.}(1994)Nave, Johansson, Learner, Thorne, \&
  Brault}]{Nave:1994:221}
Nave G., Johansson S., Learner R. C.M., Thorne A.P., Brault J.W., 1994,
  ApJS, 94, 221

\bibitem[{Nave \& Sansonetti(2011)}]{Nave:2011:737}
Nave G., Sansonetti C.J., 2011, J. Opt. Soc. Am. B, 28, 737

\bibitem[{Norl{\'e}n(1973)}]{Norlen:1973:249}
Norl{\'e}n G., 1973, Phys. Scr., 8, 249

\bibitem[{N{\"o}rtersh{\"a}user {et~al}\mbox{.}(1998)N{\"o}rtersh{\"a}user,
  Blaum, Icker, M{\"u}ller, Schmitt, Wendt, \& Wiche}]{Nortershauser:1998:33}
N{\"o}rtersh{\"a}user W., Blaum K., Icker K., M{\"u}ller P., Schmitt A., Wendt
  K., Wiche B., 1998, Eur. Phys. J. D, 2, 33

\bibitem[{Pettini {et~al}\mbox{.}(1990)Pettini, Boksenberg, \&
  Hunstead}]{Pettini:1990:48}
Pettini M., Boksenberg A., Hunstead R.W., 1990, ApJ, 348, 48

\bibitem[{Pickering {et~al}\mbox{.}(2000)Pickering, Thorne, Murray, Litz{\'e}n,
  Johansson, Zilio, \& Webb}]{Pickering:2000:163}
Pickering J.C., Thorne A.P., Murray J.E., Litz{\'e}n U., Johansson S., Zilio
  V., Webb J.K., 2000, MNRAS, 319, 163

\bibitem[{Pickering {et~al}\mbox{.}(1998)Pickering, Thorne, \&
  Webb}]{Pickering:1998:131}
Pickering J.C., Thorne A.P., Webb J.K., 1998, MNRAS, 300, 131

\bibitem[{Porsev {et~al}\mbox{.}(2007)Porsev, Koshelev, Tupitsyn, Kozlov,
  Reimers, \& Levshakov}]{Porsev:2007:052507}
Porsev S.G., Koshelev K.V., Tupitsyn I.I., Kozlov M.G., Reimers D.,
  Levshakov S.A., 2007, Phys. Rev. A, 76, 052507

\bibitem[{Porsev {et~al}\mbox{.}(2009)Porsev, Kozlov, \&
  Reimers}]{Porsev:2009:032519}
Porsev S.G., Kozlov M.G., Reimers D., 2009, Phys. Rev. A, 79, 032519

\bibitem[{Prochaska {et~al}\mbox{.}(2008)Prochaska, Chen, Wolfe,
  Dessauges-Zavadsky, \& Bloom}]{Prochaska:2008:59}
Prochaska J.X., Chen H.-W., Wolfe A.M., Dessauges-Zavadsky M., Bloom J.S.,
  2008, ApJ, 672, 59

\bibitem[{Prochaska {et~al}\mbox{.}(2002)Prochaska, Henry, O'Meara, Tytler,
  Wolfe, Kirkman, Lubin, \& Suzuki}]{Prochaska:2002:933}
Prochaska J.X., Henry R. B.C., O'Meara J.M., Tytler D., Wolfe A.M., Kirkman
  D., Lubin D., Suzuki N., 2002, PASP, 114, 933

\bibitem[{Prochaska \& McWilliam(2000)}]{Prochaska:2000:L57}
Prochaska J.X., McWilliam A., 2000, ApJ, 537, L57

\bibitem[{Prochaska \& Wolfe(1996)}]{Prochaska:1996:403}
Prochaska J.X., Wolfe A.M., 1996, ApJ, 470, 403

\bibitem[{Rauch(1998)}]{Rauch:1998:267}
Rauch M., 1998, ARA\&A, 36, 267

\bibitem[{Richter {et~al}\mbox{.}(2011)Richter, Krause, Fechner, Charlton, \&
  Murphy}]{Richter:2011:A12}
Richter P., Krause F., Fechner C., Charlton J.C., Murphy M.T., 2011, A\&A,
  528, A12

\bibitem[{Rosman \& Taylor(1998)}]{Rosman:1998:1275}
Rosman K. J.R., Taylor P. D.P., 1998, J. Phys. Chem. Ref. Data, 27, 1275

\bibitem[{Ruffoni \& Pickering(2010)}]{Ruffoni:2010:424}
Ruffoni M.P., Pickering J.C., 2010, ApJ, 725, 424

\bibitem[{Safronova \& Johnson(2001)}]{Safronova:2001:052501}
Safronova M.S., Johnson W.R., 2001, Phys. Rev. A, 64, 052501

\bibitem[{Salumbides {et~al}\mbox{.}(2006)Salumbides, Hannemann, Eikema, \&
  Ubachs}]{Salumbides:2006:L41}
Salumbides E.J., Hannemann S., Eikema K. S.E., Ubachs W., 2006, MNRAS, 373,
  L41

\bibitem[{Savedoff(1956)}]{Savedoff:1956:688}
Savedoff M.P., 1956, Nature, 178, 688

\bibitem[{Savukov \& Dzuba(2008)}]{Savukov:2008:042501}
Savukov I.M., Dzuba V.A., 2008, Phys. Rev. A, 77, 042501

\bibitem[{Schmidt(1963)}]{Schmidt:1963:1040}
Schmidt M., 1963, Nature, 197, 1040

\bibitem[{Sur {et~al}\mbox{.}(2005)Sur, Sahoo, Chaudhuri, Das, \&
  Mukherjee}]{Sur:2005:25}
Sur C., Sahoo B.K., Chaudhuri R.K., Das B.P., Mukherjee D., 2005, Eur. Phys.
  J. D, 32, 25

\bibitem[{Tupitsyn {et~al}\mbox{.}(2003)Tupitsyn, Shabaev, L{\'o}pez-Urrutia,
  Dragani{\'c}, Orts, \& Ullrich}]{Tupitsyn:2003:022511}
Tupitsyn I.I., Shabaev V.M., L{\'o}pez-Urrutia J. R.C., Dragani{\'c} I.,
  Orts R.S., Ullrich J., 2003, Phys. Rev. A, 68, 022511

\bibitem[{van Wijngaarden \& Li(1994)}]{Wijngaarden:1994:67}
van Wijngaarden W.A., Li J., 1994, Z. Phys. D, 32, 67

\bibitem[{Varshalovich {et~al}\mbox{.}(1996)Varshalovich, Panchuk, \&
  Ivanchik}]{Varshalovich:1996:6}
Varshalovich D.A., Panchuk V.E., Ivanchik A.V., 1996, Astron. Lett., 22, 6

\bibitem[{Webb {et~al}\mbox{.}(1999)Webb, Flambaum, Churchill, Drinkwater, \&
  Barrow}]{Webb:1999:884}
Webb J.K., Flambaum V.V., Churchill C.W., Drinkwater M.J., Barrow J.D.,
  1999, Phys. Rev. Lett., 82, 884

\bibitem[{Webb {et~al}\mbox{.}(2011)Webb, King, Murphy, Flambaum, Carswell, \&
  Bainbridge}]{Webb:2011:191101}
Webb J.K., King J.A., Murphy M.T., Flambaum V.V., Carswell R.F.,
  Bainbridge M.B., 2011, Phys. Rev. Lett., 107, 191101

\bibitem[{Webb {et~al}\mbox{.}(2001)Webb, Murphy, Flambaum, Dzuba, Barrow,
  Churchill, Prochaska, \& Wolfe}]{Webb:2001:091301}
Webb J.K., Murphy M.T., Flambaum V.V., Dzuba V.A., Barrow J.D., Churchill
  C.W., Prochaska J.X., Wolfe A.M., 2001, Phys. Rev. Lett., 87, 091301

\bibitem[{Whaling {et~al}\mbox{.}(1995)Whaling, Anderson, Carle, Brault, \&
  Zarem}]{Whaling:1995:1}
Whaling W., Anderson W. H.C., Carle M.T., Brault J.W., Zarem H.A., 1995, J.
  Quant. Spectrosc. Radiat. Transfer, 53, 1

\bibitem[{Wolf {et~al}\mbox{.}(2008)Wolf, van~den Berg, Gohle, Salumbides,
  Ubachs, \& Eikema}]{Wolf:2008:032511}
Wolf A.L., van~den Berg S.A., Gohle C., Salumbides E.J., Ubachs W., Eikema
  K. S.E., 2008, Phys. Rev. A, 78, 032511

\bibitem[{Wolf {et~al}\mbox{.}(2009)Wolf, van~den Berg, Ubachs, \&
  Eikema}]{Wolf:2009:223901}
Wolf A.L., van~den Berg S.A., Ubachs W., Eikema K. S.E., 2009, Phys. Rev.
  Lett., 102, 223901

\bibitem[{Wolfe {et~al}\mbox{.}(2005)Wolfe, Gawiser, \&
  Prochaska}]{Wolfe:2005:861}
Wolfe A.M., Gawiser E., Prochaska J.X., 2005, ARA\&A, 43, 861

\bibitem[{Yei {et~al}\mbox{.}(1993)Yei, Sieradzan, \& Havey}]{Yei:1993:1909}
Yei W., Sieradzan A., Havey M.D., 1993, Phys. Rev. A, 48, 1909

\bibitem[{Zych {et~al}\mbox{.}(2009)Zych, Murphy, Hewett, \&
  Prochaska}]{Zych:2009:1429}
Zych B.J., Murphy M.T., Hewett P.C., Prochaska J.X., 2009, MNRAS, 392, 1429

\end{thebibliography}

\appendix

\section{New atomic calculations for singly-ionised silicon and titanium}\label{app:a}

New \emph{ab initio} calculations of the $q$-values and isotope shifts
presented for Si{\sc \,ii} and Ti{\sc \,ii} were performed using the
atomic structure package AMBiT, an implementation of the combination
of configuration interaction and many-body perturbation theory
(CI+MBPT) described in \citet{Berengut:2006:012504} \citep[see
also][]{Dzuba:1996:3948,Berengut:2005:044501}. Both Si{\sc \,ii} and
Ti{\sc \,ii} are well-approximated as three-valence-electron ions
above a closed-shell core. The calculations closely follow the
strategy adopted for Ti{\sc \,ii} in \citet{Berengut:2008:235702}.

Briefly, the calculations start by solving the Dirac-Fock
(relativistic Hartree-Fock) equations to obtain single-particle
orbitals for the core electrons, which are frozen at CI level. We then
diagonalize the Dirac-Fock Hamiltonian over a set of 40 b-splines
\citep{Johnson:1986:1126} spanning 40 atomic units to obtain a large
set of valence and virtual orbitals from which we select those with
the lowest eigenvalues. In this work we use a basis size 16spdf for
the configuration interaction, indicating that states 1s--16s,
2p--16p, etc., are included. This is enough to saturate the CI
energy levels in Si{\sc \,ii} and Ti{\sc \,ii}.

Valence-valence correlations are treated to all orders by the
configuration interaction. Core--valence correlations, that go beyond
the frozen-core approximation, are treated using MBPT to second order
of the residual Coulomb interaction by modifying the radial integrals
in the CI procedure. The MBPT basis can be made somewhat larger than
the CI basis, which allows for saturation of the effect of virtual
excitations from the closed-shell core (to second order). The MBPT
operator may be separated into one-, two-, and three-valence electron
parts, denoted $\Sigma^{(1)}$, $\Sigma^{(2)}$, and $\Sigma^{(3)}$. It
was shown by \citet{Berengut:2008:235702} that the effective
three-body operator, $\Sigma^{(3)}$, is important for isotope shift
calculations in Ti{\sc \,ii}, and in this work we also include it in
our Si{\sc \,ii} calculations, although it plays a smaller role in
this system.

The fine-structure constant, $\alpha$, is a free parameter in AMBiT
that must be entered by hand. To obtain the sensitivity of transitions
to $\alpha$-variation, $q$, the entire spectrum is recalculated with
smaller and larger values of $\alpha$. $q$ is then the numerical
derivative at $\alpha_0$, the current value of $\alpha$:
\[
	q = \left.\frac{d\omega}{d\alpha^2}\right|_{\alpha=\alpha_0} \,.
\]

The specific mass shift, \kSMS, is obtained from AMBiT using the
finite-field method of \citet{Berengut:2003:022502}. The SMS is a
two-body operator, proportional to $\vec{p}_1\cdot\vec{p}_2$, and it
is added to the Coulomb operator everywhere that it appears in the
energy calculation:
\[
	\tilde Q = \frac{1}{\left|\vec{r}_1 - \vec{r}_2\right|} + \lambda\,\vec{p}_1 \cdot \vec{p}_2 \,.
\]
The operator $\tilde Q$ has the same symmetry and structure as the
Coulomb operator. The scaling factor $\lambda$ is varied, and the
energy spectrum is recalculated. \kSMS\ is extracted from the
numerical derivative
\[
	\kSMS = \left.\frac{d\omega}{d\lambda}\right|_{\lambda=0} \,.
\]
Values of $\lambda$ at the level $10^{-3}$ give a large enough effect
while ensuring numerical stability.

Finally, we estimate the values of the field shift parameter, \FS,
using a finite-field method with a small CI calculation. The field can
be a scaled change in the nuclear radius, $\lambda (U_{R+\delta R} -
U_R)$, where $U_R$ is the potential of a nucleus with radius $R$. \FS\
is then extracted from $d\omega/d\lambda$
\citep{Berengut:2003:022502}. Alternatively, we can simply vary the
nuclear radius itself over a larger range and extract \FS\ via
$d\omega/dR$. Both methods agree. Since the change in potential is a
one-body operator and correlations do not strongly affect it, this
gives sufficient accuracy in \FS\ for our purposes.

\subsection{Silicon}

For the silicon calculation we begin with a $V^{N-1}$ Dirac-Fock
calculation, including 3s$^2$ in the core. These orbitals are removed
from the core and become valence orbitals for the purposes of both CI
and MBPT calculations. This necessitates including subtraction
diagrams in the MBPT, but ensures a good starting approximation for
the atomic orbitals. An alternative is to use the $V^{N-3}$
approximation which doesn't include 3s$^2$ in the core; we find that
both methods give consistent results when the CI is saturated and all
second-order MBPT diagrams are taken into account.

The CI calculation includes all single and double excitations to the
basis set 16spdf from the configurations 3s$^2$\,3p, 3s\,3p$^2$,
3s$^2$\,4p, and 3s$^2$\,3d. The MBPT basis is larger, 30spdfg; we
tested saturation of the MBPT basis by extending it to 35spdfgh
and found that this made no difference to our results. Estimates of
the uncertainty in \kSMS\ and $q$ are taken from the difference
between full CI+MBPT calculation and the CI alone. Final results are
presented in \Tref{tab:SiII_final}.

\begin{table*}
\begin{center}
\caption{Energy, $q$-values and isotope-shift constants for transitions from the Si{\sc \,ii} ground state 3s$^2$\,3p\ $^2$P$^{\rm o}_{1/2}$.}
\label{tab:SiII_final}
\begin{tabular}{lrrrrbrb}
\hline
Upper level & J & \multicolumn{2}{c}{$\omega_0$ [\cm]} & \multicolumn{1}{c}{\kNMS} & \multicolumn{1}{c}{\kSMS} & \multicolumn{1}{c}{\FS} & q \\
 & & Expt. & Theor. & \multicolumn{1}{c}{[\GHzamu]} & \multicolumn{1}{c}{[\GHzamu]} & \multicolumn{1}{c}{[\MHzfm]} & \multicolumn{1}{c}{[\cm]} \\
\hline
3s$^2$\,3p\ $^2$P$^{\rm o}$ & 3/2 &   287 &  $ 304$ & $   -5$ &    18(3)   & $   0$ & 304(9) \\ 
3s\,3p$^2$\ $^4$P   & 1/2 & 42824 & 42528 & $ -704$ & -1577(97)  & $-309$ & 470(17) \\
                   & 3/2 & 42933 & 42644 & $ -706$ & -1573(98)  & $-309$ & 586(20) \\
                   & 5/2 & 43108 & 42829 & $ -709$ & -1565(99)  & $-309$ & 773(26) \\
3s\,3p$^2$\ $^2$D   & 3/2 & 55309 & 55233 & $ -910$ & -1091(119) & $-181$ & 526(16) \\
                   & 5/2 & 55325 & 55251 & $ -910$ & -1091(121) & $-181$ & 547(16) \\
3s$^2$\,4s\ $^2$S   & 1/2 & 65500 & 66067 & $-1077$ &  1454(47)  & $ 144$ &  47(4) \\
3s\,3p$^2$\ $^2$S   & 1/2 & 76665 & 76904 & $-1261$ &  -768(126) & $-279$ & 644(24) \\
3s$^2$\,3d\ $^2$D   & 3/2 & 79339 & 79588 & $-1305$ &  1413(109) & $ -23$ & 270(8) \\
                   & 5/2 & 79355 & 79606 & $-1305$ &  1408(109) & $ -24$ & 290(8) \\
3s$^2$\,4p\ $^2$P$^{\rm o}$ & 1/2 & 81191 & 81072 & $-1335$ &  1085(53)  & $  90$ &  81(1) \\
                   & 3/2 & 81251 & 81136 & $-1336$ &  1088(54)  & $  90$ & 144(2) \\
3s\,3p$^2$\ $^2$P   & 1/2 & 83802 & 83912 & $-1378$ &   462(134) & $-303$ & 544(17) \\
                    & 3/2 & 84004 & 84127 & $-1382$ &   476(133) & $-303$ & 756(23) \\
\hline
\end{tabular}
\end{center}
\end{table*}

\subsection{Titanium}

Our titanium calculations follow \citet{Berengut:2008:235702}
closely. To obtain good starting orbitals in the titanium calculation
requires inclusion of 3d$^2$ electrons at the Dirac-Fock level, again
corresponding to the $V^{N-1}$ approximation. To include this, we
simply scale the potential due to the filled 3d$^{10}$ level by the
factor 2/10, and include it in the Dirac-Fock core. The 3d orbital
becomes a valence orbital for the CI and MBPT, and again subtraction
diagrams must be included to correct for the change in Dirac-Fock
potential.

Saturation at our accuracy is achieved in Ti{\sc \,ii} with a 16spdf
basis in the CI, taking all single and double excitations from the
leading configurations 3d$^2$\,4s, 3d$^2$\,4p, 3d\,4s\,4p, and
3d$^3$. The MBPT diagrams are calculated using the basis
33spdfg. The Ti{\sc \,ii} spectrum can be further improved by
modification of the energy denominator in second-order perturbation
theory \citep{Berengut:2008:235702}. We add a constant $\delta$ to the
energy denominator:
\[
\frac{1}{E - E_{\rm M}} \rightarrow \frac{1}{E - E_{\rm M} + \delta}
\]
where $E_{\rm M}$ is the energy of the intermediate state. By taking
$\delta$ as the difference between the CI and Dirac-Fock energies of
the ground state, $\delta = E^{\rm CI}-E^{\rm DF} \approx -0.70$, we are
correcting the ground state energy to the CI value. This shift also
restores the correct asymptotic behaviour in the Brillouin-Wigner
perturbation theory for a large number of particles
\citep{Kozlov:1999:352}.

As with our Si{\sc \,ii} calculation, we estimate uncertainties by
taking the difference between the calculations that include MBPT and
those that don't. For the $q$ values we use the difference between the
CI and CI+MBPT results that don't include modification of the energy
denominator $\delta$; these errors are usually slightly larger and are
possibly more indicative of the size of neglected contributions. Final
results are presented in \Tref{tab:TiII_final}.

\begin{table*}
\begin{center}
\caption{Energy, $q$-values and isotope-shift constants for transitions from the Ti{\sc \,ii} ground state 3d$^2$\,4s\ $^4$F$_{3/2}$.}
\label{tab:TiII_final}
\begin{tabular}{lrrrrbrb}
\hline
Upper level & J & \multicolumn{2}{c}{$\omega_0$ [\cm]} & \multicolumn{1}{c}{\kNMS} & \multicolumn{1}{c}{\kSMS} & \multicolumn{1}{c}{\FS} & q \\
 & & Expt. & Theor. & \multicolumn{1}{c}{[\GHzamu]} & \multicolumn{1}{c}{[\GHzamu]} & \multicolumn{1}{c}{[\MHzfm]} & \multicolumn{1}{c}{[\cm]} \\
\hline
3d$^2$\,4p\ $^4$G$^{\rm o}$   & 5/2 & 29544 & 29652 & $-486$ & -346(104) & $-408$ &   408(30) \\
                    & 7/2 & 29735 & 29878 & $-489$ & -329(97)  & $-408$ &   637(69) \\
                    & 9/2 & 29968 & 30145 & $-493$ & -315(95)  & $-407$ &   906(106) \\
                    &11/2 & 30241 & 30449 & $-497$ & -300(98)  & $-407$ &  1212(149) \\
3d$^2$\,4p\ $^4$F$^{\rm o}$   & 3/2 & 30836 & 30965 & $-507$ & -279(172) & $-411$ &   563(31) \\
                    & 5/2 & 30959 & 31113 & $-509$ & -265(163) & $-411$ &   721(70) \\
                    & 7/2 & 31114 & 31297 & $-512$ & -254(160) & $-411$ &   903(100) \\
                    & 9/2 & 31301 & 31516 & $-515$ & -250(166) & $-410$ &  1119(129) \\
3d$^2$\,4p\ $^2$F$^{\rm o}$   & 5/2 & 31207 & 31431 & $-513$ & -248(193) & $-399$ &   657(47) \\
                    & 7/2 & 31491 & 31769 & $-518$ & -224(178) & $-398$ &   988(110) \\
3d$^2$\,4p\ $^2$D$^{\rm o}$   & 3/2 & 31757 & 31992 & $-522$ & -194(244) & $-395$ &   655(98) \\
                    & 5/2 & 32025 & 32311 & $-527$ & -188(243) & $-398$ &   956(173) \\
3d$^2$\,4p\ $^4$D$^{\rm o}$   & 1/2 & 32532 & 32755 & $-535$ & -225(224) & $-411$ &   739(79) \\
                    & 3/2 & 32603 & 32853 & $-536$ & -228(236) & $-408$ &   855(84) \\
                    & 5/2 & 32698 & 32976 & $-538$ & -225(245) & $-403$ &   999(62) \\
                    & 7/2 & 32767 & 33063 & $-539$ & -215(225) & $-408$ &  1049(117) \\
3d\,4s\,4p\ $^4$F$^{\rm o}$ & 3/2 & 52330 & 50831 & $-861$ & 3656(510) & $ 346$ & -1689(250) \\
                    & 5/2 & 52472 & 51000 & $-863$ & 3671(512) & $ 347$ & -1482(370) \\
                    & 7/2 & 52705 & 51264 & $-867$ & 3692(509) & $ 348$ & -1198(360) \\
                    & 9/2 & 53099 & 51656 & $-873$ & 3702(517) & $ 349$ &  -910(193) \\
3d\,4s\,4p\ $^4$D$^{\rm o}$ & 1/2 & 52339 & 50870 & $-861$ & 3712(584) & $ 330$ & -1414(166) \\
                    & 3/2 & 52459 & 50977 & $-863$ & 3692(601) & $ 330$ & -1343(250) \\
                    & 5/2 & 52631 & 51155 & $-866$ & 3686(614) & $ 330$ & -1209(370) \\
                    & 7/2 & 52847 & 51401 & $-869$ & 3708(614) & $ 332$ &  -992(360) \\
3d\,4s\,4p\ $^2$D$^{\rm o}$ & 5/2 & 53555 & 52392 & $-881$ & 3263(603) & $ 235$ &  -882(182) \\
                    & 3/2 & 53597 & 52449 & $-881$ & 3264(607) & $ 236$ &  -819(191) \\
\hline
\end{tabular}
\end{center}
\end{table*}

A further complication in the calculation of $q$-values arises in the
Ti{\sc \,ii} spectrum because of the close proximity of the
3d\,4s\,4p\ $^4$F$^{\rm o}$ and $^4$D$^{\rm o}$ multiplets, which causes
mixing of their $J=3/2$, $5/2$, and $7/2$ levels. Near a level
pseudo-crossing, the states become strongly mixed, which can cause the
gradient $d\omega/dx$ to change
dramatically. \citet{Dzuba:2002:022501} found similar effects in the
\ion{Fe}{ii} spectrum, and used experimental Land\'e $g$-factors to
resolve the level of mixing.
Since experimental $g$-factors are not available for these Ti{\sc
  \,ii} lines, we estimate the mixing of the levels from the
experimental energy differences and the difference between calculated
$g$-factors and the $g$-factors expected in pure $LS$-coupling scheme.

\Tref{tab:TiII_mixing} shows the result of ``unmixing'' the
levels. For each $J$, the difference between the computed value,
$g_{\rm calc}$, and the $LS$ value, $g_{LS}$, is used to estimate the
amount of mixing of the two levels. With this mixing, it is possible
to calculate the $q$-values of the unmixed states, $q_{LS}$. Our
calculation reproduces the experimental $^4$F$^{\rm o}$--$^4$D$^{\rm
  o}$ energy difference, $\Delta$, fairly well, which means our mixing
levels are likely to be reasonable. On the other hand, the procedure
to derive $q_{LS}$ is sensitive to small changes in $g_{\rm calc}$,
and this could artificially exaggerate the differences between the
$q_{LS}$ found for the different levels, particularly for the strongly
mixed $J=5/2$ and $J=7/2$ cases.

Therefore in \Tref{tab:TiII_final} we present the calculated
$q$-values, $q_{\rm calc}$, and simply increase the errors to account
for all possible mixing coefficients. At worst, this will overestimate
our uncertainties. Better $q$-values will be possible when the
experimental $g$-factors for these levels are determined. Finally, we
note that this mixing doesn't affect the specific mass shift
calculations simply because the 3d\,4s\,4p\ $^4$F$^{\rm o}$ and
$^4$D$^{\rm o}$ levels have very similar \kSMS\ values anyway.

\begin{table}
\begin{center}
\caption{Mixing of the 3d\,4s\,4p\ $^4$F$^{\rm o}$ and $^4$D$^{\rm o}$ multiplets in Ti{\sc \,ii}. The difference between the calculated and $LS$ $g$-factors is used to estimate the proportion of mixing at $\alpha = \alpha_0$. $\Delta$ is the energy difference between the two levels for each angular momentum, $J$. $q_{\rm calc}$ is the computed $q$-value, while $q_{LS}$ is the expected $q$-value in the absence of the level pseudocrossing.}
\label{tab:TiII_mixing}
\begin{tabular}{lrrrrrrrr}
\hline
 & $g_{LS}$ & $g_{\rm calc}$ & Mixing & \multicolumn{2}{c}{$\Delta$ [\cm]} & \multicolumn{1}{c}{$q_{\rm calc}$} & \multicolumn{1}{c}{$q_{LS}$} \\
 & & & & expt. & calc. & [\cm] & [\cm]  \\
\hline
$J = 3/2$: \\
             $^4$F$^{\rm o}$ & 0.400 & 0.533 & 16\% & 129 & 147 & -1689 & -1774 \\
\vspace{1mm} $^4$D$^{\rm o}$ & 1.200 & 1.071 &      &     &     & -1343 & -1259 \\

$J = 5/2$: \\
             $^4$F$^{\rm o}$ & 1.029 & 1.137 & 32\% & 159 & 156 & -1482 & -1719 \\
\vspace{1mm} $^4$D$^{\rm o}$ & 1.371 & 1.263 &      &     &     & -1209 &  -973 \\

$J = 7/2$: \\
$^4$F$^{\rm o}$ & 1.238 & 1.306 & 36\% & 141 & 137 & -1198 & -1457 \\
$^4$D$^{\rm o}$ & 1.429 & 1.360 &      &     &     &  -992 &  -732 \\
\hline
\end{tabular}
\end{center}
\end{table}

\section*{Supporting Information}\label{sec:supp}

Additional Supporting Information may be found in the online ver-
sion of this article:\vspace{-0.5em}\newline

\noindent \textbf{Atomic Data spreadsheet.} Comprehensive record of
the measurements and theoretical estimates from the literature, plus
the calculations performed in this paper, which contribute to the
synthesized information presented in Tables \ref{tab:Mg}--\ref{tab:Zn}
(\textsl{Oxford University Press to supply URL}). An up-to-date
version of the spreadsheet is maintained by M.T.M.~and is available
through
{\urlstyle{rm}\url{https://researchdata.ands.org.au/laboratory-atomic-transition-data-for-precise-optical-quasar-absorption-spectroscopy}}. The
authors welcome updated information for inclusion in new
versions.\vspace{-0.5em}\newline

\noindent Please note: Oxford University Press are not responsible for the
content or functionality of any supporting materials supplied by
the authors. Any queries (other than missing material) should be
directed to the corresponding author for the article

\bspsmall

\label{lastpage}

\end{document}